\documentclass[a4paper,12pt]{article}
\usepackage[title]{appendix}

\usepackage{a4wide}
\usepackage{amsmath}
\usepackage{mathtools}
\usepackage{amssymb}
\usepackage{amsthm}
\usepackage{amsfonts,bm}
\usepackage{scrextend}
\usepackage{subfig}
\usepackage{graphicx}
\usepackage{wrapfig}
\usepackage{tikz}
\usepackage{xcolor}
\usepackage{placeins}

\usepackage{floatrow}
\usepackage{subfig}
\usepackage{url}
\usepackage{hyperref}
\usepackage{algpseudocode}
\usepackage{algorithm}
\usepackage{array}
\usepackage{eufrak}
\usepackage{yfonts}
\usepackage{mathrsfs}
\usepackage{enumerate}
\usepackage{gensymb}
\usepackage{sansmathaccent}
\usepackage[bottom,perpage]{footmisc}
\usepackage{authblk}
\usetikzlibrary{external}

\numberwithin{equation}{section}

\DeclarePairedDelimiter\floor{\lfloor}{\rfloor}
\title{A mathematical model of contact inhibition of locomotion: coupling contractility and focal adhesions.}
\author{Aydar Uatay}
\affil{\footnotesize Technische Universit\"at Kaiserslautern, Felix-Klein-Zentrum f\"ur Mathematik\protect\\ Paul-Ehrlich-Str.31, 67663 Kaiserslautern, Germany\protect\\(uatay at mathematik.uni-kl.de)}
\date{}

\begin{document}
\maketitle
\begin{abstract}
Cell migration is often accompanied by collisions with other cells, which can lead to cessation of movement, repolarization, and migration away from the contact site - a process termed contact inhibition of locomotion (CIL). During CIL, the coupling between actomyosin contractilityand cell-substrate adhesions is modified. However, mathematical models describing stochastic cell migration and collision outcomes as a result of the coupling remain elusive. Here, we extend our stochastic model of single cell migration \cite{2018arXiv181011435U} to include CIL. Our simulation results explain, in terms of the modified contractility and adhesion dynamics, several experimentally observed findings regarding CIL. These include response modulation in the presence of an external cue and alterations of group migration in the absence of CIL. Together with \cite{2018arXiv181011435U}, our work is able to explain a wide range of observations about single and collective cell migration.


\end{abstract}

\section{Introduction}
Cell migration is vital for the development of an organism and is required for several important processes, such as wound healing and immune response. Given its essential role, disregulation of migration can lead to progression of chronic inflammation, atherosclerosis, and cancer spread. As a migrating cell often moves in a crowded environment, it collides and interacts with other cells. One possible outcome of such interaction is cessation of movement, followed by migration away from the collision site. Abercrombie and Heaysman termed this phenomena contact inhibition of locomotion  (CIL) \cite{ABERCROMBIE1954293}. Since then, its role in many important processes, such as cancer dissemination and embryo development has been established \cite{stramer2017mechanisms}. 

%For example, following EMT (epithelial-mesenchymal transition) cells acquire ability to undergo CIL \cite{SCARPA2015421}, which may facilitate cancer cell spreading. I 

Collisions between cells of the same (homotypic) or different (heterotypic) types can result in CIL \cite{stramer2017mechanisms}. Moreover, the response of cells, exhibiting CIL, can vary: collisions can lead to adhesion, walking past each other, or chaining \cite{Desai20130717}, \cite{Scarpa901}. Collisions outcome can also be influenced by the presence of a chemotactic gradient \cite{lin2015interplay}, where the CIL signal can be overridden by a directional cue. Complicating matters even further, there is compelling evidence \cite{Desai20130717}, \cite{lin2015interplay}, \cite{Scarpa901} that the collision outcome is stochastic. Regardless of the aftermath, post-collision signaling pathways are integrated into an already intricate process of cell motility, which, along with variability of the contact outcomes, renders the elucidation of the underlying mechanisms a challenging task (see \cite{stramer2017mechanisms} for a review). For example, it has been shown that CIL is responsible for migration towards a chemoattractant of otherwise unresponsive cells \cite{theveneau2010collective}, or that heterotypic CIL is required for chase-and-run movement \cite{theveneau2013chase}. 

Dynamic interactions of cellular structures such as focal adhesions (FAs) and stress fibers (SFs), which are essential for freely migrating cells, are modified in a contact-dependent manner in order to yield CIL. It has been shown that the number of FAs is increased at the free edge of cells undergoing CIL \cite{SCARPA2015421}. Due to activation of small GTPase RhoA in the vicinity of cell-cell contacts \cite{carmona2008contact}, the contractility of SFs there increases as well \cite{roycroft2016molecular}. Thus, after collision the following events occur: a free (leading) edge protrudes forward, cell-substrate adhesions are formed at the front, rear FAs and cell-cell junctions rupture due to increased contractility there, the cell body retracts and moves forward. That is, CIL follows the stereotypical steps of a cell migration cycle, although the preceding signaling events in a colliding and freely migrating cell are different.  

Various mathematical models have been developed to address CIL specifically, and more broadly, collective behavior emerging as result of cell-cell interactions. For example, the phase-field models in \cite{kulawiak2016modeling} and \cite{lober2015collisions} were able to reproduce, respectively, experimentally observed statistical outcomes of binary interactions and emergent collective migration as a result of inelastic collisions. Particle- and agent-based models in \cite{davis2012emergence}, \cite{Desai20130717}, \cite{zimmermann2016contact} were also able to simulate outcomes in agreement with experimental observations. Cooperation of co-attraction and contact inhibition has been studied by mechanistic models in \cite{MERCHANT2018}, \cite{Szabo543}, \cite{Woods2014}. These models, however, do not describe CIL in terms of the migration cycle, which a colliding cell must follow, as described above. To do so, a model must also take into account the coupling of relevant structures (e.g. FAs, SFs), and should also be stochastic, as paths of freely migrating cell and CIL outcomes are stochastic as well \cite{Desai20130717}, \cite{lin2015interplay}, \cite{Scarpa901}. In our previous work \cite{2018arXiv181011435U}, we constructed a minimal stochastic cell motility model, which took into account the migration cycle, and the mechanochemical interaction of FAs and SFs. Encouraged by the fact that it was able to explain a variety of experimentally observed results concerning freely migrating cells, we extend and generalize the model here to include cell-cell collisions and CIL specifically. We do so by slightly modifying FA and SF dynamics in a manner described above: enhanced FA affinity away from the contact site and increased contractility in its vicinity. As in \cite{2018arXiv181011435U}, the extended model is described by a piecewise deterministic Markov process (PDMP). Unlike the original model, here we have an ``active" boundary in a sense that a jump occurs when the process hits it. In order to perform numerical simulations, we propose an efficient method for a general PDMP with ``active" boundary where solving the deterministic flow is relatively expensive. The numerical simulations themselves are able to explain several experimentally observed results regarding CIL, such as modulation of post collision outcome in the presence of a chemotactic gradient \cite{lin2015interplay} in a 1D setting, inducement of directed migration of non-chemotaxing cells due to CIL \cite{theveneau2010collective}, and invasive migration in the presence and absence of heterotypic CIL, respectively.

%Another type of models are particle-based model of motile cell clusters \cite{zimmermann2016contact}, taking into account cell-substrate adhesions and CIL was able to reproduce traction stresses, ob     

This paper is organized as follows: in Section \ref{section: single cell PDMP} we briefly overview the minimal single cell migration model developed in \cite{2018arXiv181011435U}. We then extend this model in Section \ref{section: CIL model} to include CIL mechanism. Numerical simulations are performed in Section \ref{section: numerical simulations}. Finally, a discussion and an outlook on future work are presented in Section \ref{section: discussion and outlook}.     

\section{Single cell motility model}\label{section: single cell PDMP}
As described above, cell migration occurs in a cyclical manner, which can be stereotypically divided into the following steps: 1) protrusion of the leading edge, 2) formation of focal adhesions at the cell front, 3) adhesions in cell rear rupture due to myosin generated contractile forces in stress fibers, which leads to 4) contraction of cell body and translocation \cite{Abercrombie129}. Based on this, we constructed in \cite{2018arXiv181011435U} a piecewise deterministic process of cyclical cell motility, which we briefly reintroduce here and extend it below to include cell collisions.

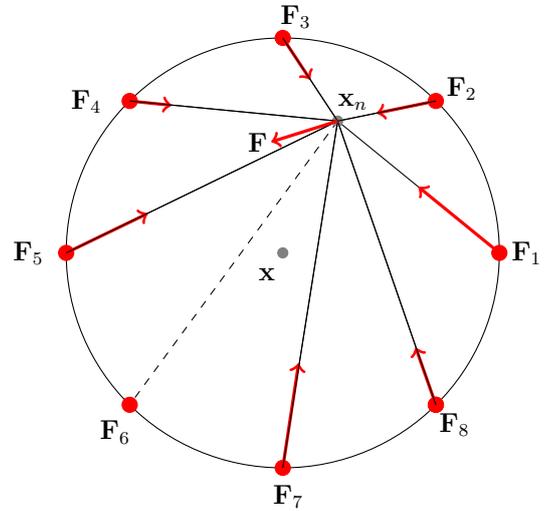
\begin{wrapfigure}[19]{hr}{0.4\textwidth} 
%\vspace*{-2cm}
\begin{tikzpicture}[scale=0.95,transform shape,every node/.style={scale=0.9}]
\draw (0.76,1.84) -- (3,0);
\draw (0.76,1.84) -- (0,3);
\draw (0.76,1.84) -- (-3,0);
\draw (0.76,1.84) -- (0,-3);
\draw (0.76,1.84) -- (2.12,2.12);
\draw (0.76,1.84) -- (-2.12,2.12);
\draw (0.76,1.84) -- (2.12,-2.12);
\draw [dashed] (0.76,1.84) -- (-2.12,-2.12);
\draw (0,0) circle [radius=3cm];
\filldraw [gray] (0.76,1.84) circle [radius=2pt];

\filldraw [red] (3,0) circle [radius=3pt];
\filldraw [red] (0,3) circle [radius=3pt];
\filldraw [red] (0,-3) circle [radius=3pt];
\filldraw [red] (-3,0) circle [radius=3pt];
\filldraw [red] (2.12,2.12) circle [radius=3pt];
\filldraw [red] (-2.12,2.12) circle [radius=3pt];
\filldraw [red] (2.12,-2.12) circle [radius=3pt];
\filldraw [red] (2.12,-2.12) circle [radius=3pt];
\filldraw [red] (-2.12,-2.12) circle [radius=3pt];

\filldraw [gray] (0,0) circle [radius=2pt];
\draw (-0.3,-0.3) node[text width = 2pt] {$\mathbf{x}$};

\draw [very thick, red][->](3,0) -- (1.88,0.92);
\draw (3.2,0) node[text width = 2pt] {$\mathbf{F}_1$};
\draw [very thick, red][->](2.12,2.12) -- (1.3,1.95);
\draw (2.3,2.3) node[text width = 2pt] {$\mathbf{F}_2$};
\draw [very thick, red][->](0,3) -- (0.38,2.42);
\draw (0,3.3) node[text width = 2pt] {$\mathbf{F}_3$};
\draw [very thick, red][->](-2.12,2.12) -- (-1.54,2.06);
\draw (-2.9,2.12) node[text width = 2pt] {$\mathbf{F}_4$};
\draw [very thick, red][->](-3,0) -- (-1.87,0.55);
\draw (-3.7,0) node[text width = 2pt] {$\mathbf{F}_5$};
\draw [very thick, red][->](0,-3) -- (0.22,-1.54);
\draw (-2.5,-2.5) node[text width = 2pt] {$\mathbf{F}_6$};
\draw (-0.1,-3.4) node[text width = 2pt] {$\mathbf{F}_7$};
\draw [very thick, red][->](2.12,-2.12) -- (1.84,-1.32);
\draw (2.2,-2.4) node[text width = 2pt] {$\mathbf{F}_8$};

\draw (0.76,1.84) -- (0,3);
\draw (0.76,1.84) -- (-3,0);
\draw (0.76,1.84) -- (0,-3);
\draw (0.76,1.84) -- (2.12,2.12);
\draw (0.76,1.84) -- (-2.12,2.12);
\draw (0.76,1.84) -- (2.12,-2.12);

\draw (0.8,2.11) node[text width = 2pt] {$\mathbf{x}_n$};

\draw[very thick, red] [->](0.76,1.84) -- (-0.16,1.55);
\draw (-0.45,1.55) node[text width = 2pt] {\textbf{F}};
\end{tikzpicture}
\caption{Schematic representation of a cell. Solid lines and the corresponding red circles represent stress fibers and bound focal adhesions, respectively. The dashed line corresponds to an unbound FA and an absent SF.}
\label{fig: cell representation}
\end{wrapfigure}

Figure \ref{fig: cell representation} depicts a cell as a disk of radius $R_{cell}$. Let $\mathbf{x}(t)\in\mathbb{R}^2$ denote the cell centroid at time $t$. Suppose there are $M$ equally spaced FAs on the cell circumference, such that their relative distance is constant. Let $\mathbf{Y}(t)\in\lbrace0,1\rbrace^M$ denote the state of FAs at time $t$ and suppose one end of SFs is anchored at an FA and the other at a node $\mathbf{x}_n(t)\in\Omega_{cell}:=\left\lbrace(x,y)\in\mathbb{R}^2\phantom{,}|\phantom{,}x^2+y^2\leq R^2_{cell} \right\rbrace$ (in the cell reference frame). Let $\theta(t)$ denote the polar position of the first FA. Since the relative distance of FAs is constant, then their polar position is uniquely determined by $\theta$. Then, the force $\mathbf{F}_j=\mathbf{F}_j(\mathbf{x}_n,\theta)$ is given by:
\begin{align}\label{eq: Fi}
\mathbf{F}_j = 
\begin{cases}
\left(T_j + EA\frac{L_j-L_0}{L_0}\right)\mathbf{e}_j,\phantom{abc} L_0<L_j\\
T_j\mathbf{e}_j,\phantom{asdasdasdaabcsaa}L_c\leq L_j\leq L_0\\
\frac{L_j-L_c+\delta}{\delta}T_j\mathbf{e}_j \phantom{sdaasdssda} L_c-\delta\leq L_j<L_c\\
0 \phantom{sdaasdAADSADAdA} L_i<L_c-\delta,  
\end{cases}
\end{align}
where $T_j$ is the magnitude of contractile force due to myosin motors, $EA$ is the one-dimensional Young's modulus, $L_0$ and $L_c$ are, respectively, rest and critical lengths, $L_j(\mathbf{x}_n,\theta)$ and $\mathbf{e}_j (\mathbf{x}_n,\theta)$ are the length of SF and the unit vector along the $j^{\text{th}}$ SF, respectively, and $\delta$ is a small positive constant\footnote{Introduced here purely for technical reasons (continuity of $\mathbf{F}_j$). See \cite{2018arXiv181011435U} for details.}. Then the net force at $\mathbf{x}_n$ is given by:
\begin{align} \label{eq: total F}
\mathbf{F}(\mathbf{x}_n,\theta,\mathbf{Y}):=-\sum_{i=j}^{M}Y_{j}\mathbf{F}_j(\mathbf{x}_n,\theta).
\end{align}

Let $\mu\in\lbrace0,1\rbrace$ denote the motility state of a cell: $\mu=0$ and $\mu=1$ correspond to a stationary and a moving cell, respectively. In \cite{2018arXiv181011435U}, considering the cell migration cycle, these values of $\mu$ also indicate the type of the last FA event: $\mu=0$ and $\mu=1$ correspond to binding and unbinding events, respectively. Then we have:
\begin{align}\label{eq: ss ODE}
\dot{\mathbf{x}} &=\mu\beta_{ECM}^{-1}\mathbf{F}(\mathbf{x}_n,\theta,\mathbf{Y})\cdot\hat{\mathbf{r}}(\mathbf{x}_n)\hat{\mathbf{r}}(\mathbf{x}_n)\nonumber \\ 
\dot{\mathbf{x}}_n &= \beta_{cell}^{-1}\mathbf{F}(\mathbf{x}_n,\theta,\mathbf{Y})\nonumber\\
\dot{\theta} &=  \mu\beta_{rot}^{-1}\lVert\mathbf{x}_n\rVert\mathbf{F}(\mathbf{x}_n,\theta,\mathbf{Y})\cdot\hat{\boldsymbol{\varphi}}(\mathbf{x}_n),
\end{align}         
for $t\in[0,\tau)$, where $\tau$ is the (random) time of the next (random) FA event; $\beta_{ECM}$, $\beta_{rot}$ are the translational and rotational drag coefficients, and $\beta_{cell}$ is the drag coefficient inside cytoplasm; $\hat{\mathbf{r}}(\mathbf{x}_n)$ and $\hat{\boldsymbol{\varphi}}(\mathbf{x}_n)$ are radial and angular unit vectors at $\mathbf{x}_n$, respectively. Note that within the context of the migration cycle, the cell body movement occurs after an FA ruptures.   

Let $a^\pm_j(\mathbf{Y},\mathbf{X})dt$ be the probability of $j^{\text{th}}$ FA binding/unbinding in time interval $[t,t+dt)$, given $\mathbf{Y}(t)$ and $\mathbf{X}(t):=(\mathbf{x}(t),\mathbf{x}_n(t),\theta(t))$. Then we have \cite{2018arXiv181011435U}:
\begin{align*}
\mathbb{P}\left(\mathcal{T}_{k+1}-\mathcal{T}_k>\tau|\mathbf{Y}\left(\mathcal{T}_k\right),\mathbf{X}\left(\mathcal{T}_k\right)\right) = \exp\left(-\int_0^\tau a_0\left(\mathbf{Y}\left(\mathcal{T}_k\right),\mathbf{X}\left(\mathcal{T}_k+s\right)\right)ds\right),
\end{align*}
where $\mathcal{T}_{k}$ is the time of $k^\text{th}$ event and $a_0(\mathbf{Y},\mathbf{X})=\sum_{j=1}^{M}a^+_j(\mathbf{Y},\mathbf{X})+a^-_j(\mathbf{Y},\mathbf{X})$. That is, the FA event interarrival time is distributed according to the survival function above. The distribution of the next FA event, given that an event occurred at time $\mathcal{T}_{k+1}$, is then:
\begin{align*}
\mathbb{P}\left(j^\pm|\mathcal{T}_{k+1}\right) = \frac{a^\pm_j\left(\mathbf{Y}\left(\mathcal{T}^-_{k+1}\right),\mathbf{X}\left(\mathcal{T}^-_{k+1}\right)\right)}{a_0\left(\mathbf{Y}\left(\mathcal{T}^-_{k+1}\right),\mathbf{X}\left(\mathcal{T}^-_{k+1}\right)\right)},
\end{align*}
where $j^\pm$ indicates binding/unbinding of $j^{\text{th}}$ FA\footnote{Here, when referring to time, $t^-=\lim\limits_{\epsilon\uparrow0}{t-\epsilon}$}. Note that $\mathbf{Y}(t)=Const.$ for $t\in[\mathcal{T}_k,\mathcal{T}_{k+1})$ and $\mathbf{Y}$ jumps to a new state at $t=\mathcal{T}_{k+1}$. Depending on the event occurred, $\mu$ changes accordingly and between the events $\mathbf{X}$ evolves according to equation \eqref{eq: ss ODE}. A more detailed treatment of the model is given in \cite{2018arXiv181011435U}.

\section{Modeling Contact inhibition of locomotion}\label{section: CIL model}
\begin{figure}[H]
	\begin{tikzpicture}[scale=0.65,transform shape]
	\draw [xshift=0.3cm][yshift = -0.3cm](0.5,1.22) -- (2,0);
	\draw [xshift=0.3cm][yshift = -0.3cm](0.5,1.22) -- (0,2);
	\draw [xshift=0.3cm][yshift = -0.3cm](0.5,1.22) -- (-2,0);
	\draw [xshift=0.3cm][yshift = -0.3cm](0.5,1.22) -- (0,-2);
	%\draw (0.5,1.22) -- (1.41,1.43);
	\draw [xshift=0.3cm][yshift = -0.3cm](0.5,1.22) -- (-1.43,1.43);
	\draw [xshift=0.3cm][yshift = -0.3cm](0.5,1.22) -- (1.43,-1.43);
	\draw [xshift=0.3cm][yshift = -0.3cm](0.5,1.22) -- (-1.43,-1.43);
	\draw [xshift=0.3cm][yshift = -0.3cm](0,0) circle [radius=2cm];
	\filldraw [black] [xshift=0.3cm][yshift = -0.3cm](0.5,1.22) circle [radius=2pt];
	
	\filldraw [xshift=0.3cm][yshift = -0.3cm][red] (2,0) circle [radius=2pt];
	\filldraw [xshift=0.3cm][yshift = -0.3cm][red] (0,2) circle [radius=2pt];
	\filldraw [xshift=0.3cm][yshift = -0.3cm][blue] (0,-2) circle [radius=2pt];
	\filldraw [xshift=0.3cm][yshift = -0.3cm][blue] (-2,0) circle [radius=2pt];
	\filldraw [xshift=0.3cm][yshift = -0.3cm][red] (1.43,1.43) circle [radius=2pt];
	\filldraw [xshift=0.3cm][yshift = -0.3cm][red] (-1.43,1.43) circle [radius=2pt];
	\filldraw [xshift=0.3cm][yshift = -0.3cm][blue] (1.43,-1.43) circle [radius=2pt];
	\filldraw [xshift=0.3cm][yshift = -0.3cm][blue] (-1.43,-1.43) circle [radius=2pt];
	
	\filldraw [xshift=0.3cm][yshift = -0.3cm][black] (0,0) circle [radius=1.33pt];
%		\draw [xshift=0.3cm][yshift = -0.3cm][thick, red][->](0.5,1.22) -- (-0.28,-0.28);
	%\draw (0.48,1.5) node[text width = 1.33pt] {$\mathbf{x}_n$};
	
	% Second 
	\draw [xshift=-2.82cm][yshift = -2.82cm](-0.5,-1.22) -- (2,0);
	\draw [xshift=-2.82cm][yshift = -2.82cm](-0.5,-1.22) -- (0,2);
	\draw [xshift=-2.82cm][yshift = -2.82cm](-0.5,-1.22) -- (-2,0);
	\draw [xshift=-2.82cm][yshift = -2.82cm](-0.5,-1.22) -- (0,-2);
	%\draw [xshift=-2.82cm][yshift = -2.82cm](-0.5,-1.22) -- (-1.41,-1.43);
	\draw [xshift=-2.82cm][yshift = -2.82cm](-0.5,-1.22) -- (-1.43,1.43);
	\draw [xshift=-2.82cm][yshift = -2.82cm](-0.5,-1.22) -- (1.43,-1.43);
	\draw [xshift=-2.82cm][yshift = -2.82cm](-0.5,-1.22) -- (1.43,1.43);
	\draw [xshift=-2.82cm][yshift = -2.82cm](0,0) circle [radius=2cm];
	\filldraw [xshift=-2.82cm][yshift = -2.82cm][black] (-0.5,-1.22) circle [radius=2pt];
	
	\filldraw [xshift=-2.82cm][yshift = -2.82cm][blue] (2,0) circle [radius=2pt];
	\filldraw [xshift=-2.82cm][yshift = -2.82cm][blue] (0,2) circle [radius=2pt];
	\filldraw [xshift=-2.82cm][yshift = -2.82cm][red] (0,-2) circle [radius=2pt];
	\filldraw [xshift=-2.82cm][yshift = -2.82cm][red] (-2,0) circle [radius=2pt];
	\filldraw [xshift=-2.82cm][yshift = -2.82cm][blue] (1.43,1.43) circle [radius=2pt];
	\filldraw [xshift=-2.82cm][yshift = -2.82cm][blue] (-1.43,1.43) circle [radius=2pt];
	\filldraw [xshift=-2.82cm][yshift = -2.82cm][red] (1.43,-1.43) circle [radius=2pt];
	\filldraw [xshift=-2.82cm][yshift = -2.82cm][red] (-1.43,-1.43) circle [radius=2pt];
	
	\filldraw [xshift=-2.82cm][yshift = -2.82cm][black] (0,0) circle [radius=1.33pt];
%		\draw [xshift=-2.82cm][yshift = -2.82cm][thick, red][->](-0.5,-1.22) -- (0.28,0.28);
	
	\draw [xshift=3.9cm][very thick, black][->](-1.43,-1.43) -- (-0.8,-1.43);
	
	%Aftermath
	\draw [xshift=8.3cm][yshift = -0.3cm](-0.5,-1.22) -- (2,0);
	\draw [xshift=8.3cm][yshift = -0.3cm](-0.5,-1.22) -- (0,2);
	\draw [xshift=8.3cm][yshift = -0.3cm](-0.5,-1.22) -- (-2,0);
	\draw [xshift=8.3cm][yshift = -0.3cm](-0.5,-1.22) -- (0,-2);
%	\draw [xshift=8.3cm][yshift = -0.3cm](-0.5,-1.22) -- (1.41,1.43);
	\draw [xshift=8.3cm][yshift = -0.3cm](-0.5,-1.22) -- (-1.43,1.43);
	\draw [xshift=8.3cm][yshift = -0.3cm](-0.5,-1.22) -- (1.43,-1.43);
	\draw [xshift=8.3cm][yshift = -0.3cm](-0.5,-1.22) -- (-1.43,-1.43);
	%\draw (0.5,1.22) -- (-1.43,-1.43);
	\draw [xshift=8.3cm][yshift = -0.3cm](0,0) circle [radius=2cm];
	\filldraw [xshift=8.3cm][yshift = -0.3cm][black] (-0.5,-1.22) circle [radius=2pt];
	
	\filldraw [xshift=8.3cm][yshift = -0.3cm][blue] (2,0) circle [radius=2pt];
	\filldraw [xshift=8.3cm][yshift = -0.3cm][blue] (0,2) circle [radius=2pt];
	\filldraw [xshift=8.3cm][yshift = -0.3cm][red] (0,-2) circle [radius=2pt];
	\filldraw [xshift=8.3cm][yshift = -0.3cm][red] (-2,0) circle [radius=2pt];
	\filldraw [xshift=8.3cm][yshift = -0.3cm][blue] (1.43,1.43) circle [radius=2pt];
	\filldraw [xshift=8.3cm][yshift = -0.3cm][blue] (-1.43,1.43) circle [radius=2pt];
	\filldraw [xshift=8.3cm][yshift = -0.3cm][red] (1.43,-1.43) circle [radius=2pt];
	\filldraw [xshift=8.3cm][yshift = -0.3cm][red] (-1.43,-1.43) circle [radius=2pt];
	
	\filldraw [xshift=8.3cm][yshift = -0.3cm][black] (0,0) circle [radius=1.33pt];
%		\draw [xshift=8.3cm][yshift = -0.3cm][thick, red][->](-0.5,-1.22) -- (0.28,0.28);
	%\draw (0.48,1.5) node[text width = 1.33pt] {$\mathbf{x}_n$};
	
	% Second 
	\draw [xshift=5.18cm][yshift = -2.82cm](0.5,1.22) -- (2,0);
	\draw [xshift=5.18cm][yshift = -2.82cm](0.5,1.22) -- (0,2);
	\draw [xshift=5.18cm][yshift = -2.82cm](0.5,1.22) -- (-2,0);
	\draw [xshift=5.18cm][yshift = -2.82cm](0.5,1.22) -- (0,-2);
%		\draw [xshift=5.18cm][yshift = -2.82cm](-0.5,-1.22) -- (-1.41,-1.43);
	\draw [xshift=5.18cm][yshift = -2.82cm](0.5,1.22) -- (-1.43,1.43);
	\draw [xshift=5.18cm][yshift = -2.82cm](0.5,1.22) -- (1.43,-1.43);
	\draw [xshift=5.18cm][yshift = -2.82cm](0.5,1.22) -- (1.43,1.43);
	\draw [xshift=5.18cm][yshift = -2.82cm](0,0) circle [radius=2cm];
	\filldraw [xshift=5.18cm][yshift = -2.82cm][black] (0.5,1.22) circle [radius=2pt];
	
	\filldraw [xshift=5.18cm][yshift = -2.82cm][red] (2,0) circle [radius=2pt];
	\filldraw [xshift=5.18cm][yshift = -2.82cm][red] (0,2) circle [radius=2pt];
	\filldraw [xshift=5.18cm][yshift = -2.82cm][blue] (0,-2) circle [radius=2pt];
	\filldraw [xshift=5.18cm][yshift = -2.82cm][blue] (-2,0) circle [radius=2pt];
	\filldraw [xshift=5.18cm][yshift = -2.82cm][red] (1.43,1.43) circle [radius=2pt];
	\filldraw [xshift=5.18cm][yshift = -2.82cm][red] (-1.43,1.43) circle [radius=2pt];
	\filldraw [xshift=5.18cm][yshift = -2.82cm][blue] (1.43,-1.43) circle [radius=2pt];
	\filldraw [xshift=5.18cm][yshift = -2.82cm][blue] (-1.43,-1.43) circle [radius=2pt];
	
	\filldraw [xshift=5.18cm][yshift = -2.82cm][black] (0,0) circle [radius=1.33pt];
%		\draw [xshift=5.18cm][yshift = -2.82cm][thick, red][->](0.5,1.22) -- (-0.28,-0.28);

	\draw [xshift=12cm][very thick, black][->](-1.43,-1.43) -- (-0.8,-1.43);
	
	%Aftermath II
	\draw [xshift=16.6cm][yshift = -0.3cm](-0.5,-1.22) -- (2,0);
	\draw [xshift=16.6cm][yshift = -0.3cm](-0.5,-1.22) -- (0,2);
	\draw [xshift=16.6cm][yshift = -0.3cm](-0.5,-1.22) -- (-2,0);
	\draw [xshift=16.6cm][yshift = -0.3cm](-0.5,-1.22) -- (0,-2);
	\draw [xshift=16.6cm][yshift = -0.3cm](-0.5,-1.22) -- (1.41,1.43);
	\draw [xshift=16.6cm][yshift = -0.3cm](-0.5,-1.22) -- (-1.43,1.43);
	\draw [xshift=16.6cm][yshift = -0.3cm](-0.5,-1.22) -- (1.43,-1.43);
	\draw [xshift=16.6cm][yshift = -0.3cm](-0.5,-1.22) -- (-1.43,-1.43);
	%\draw (0.5,1.22) -- (-1.43,-1.43);
	\draw [xshift=16.6cm][yshift = -0.3cm](0,0) circle [radius=2cm];
	\filldraw [xshift=16.6cm][yshift = -0.3cm][black] (-0.5,-1.22) circle [radius=2pt];
	
	\filldraw [xshift=16.6cm][yshift = -0.3cm][blue] (2,0) circle [radius=2pt];
	\filldraw [xshift=16.6cm][yshift = -0.3cm][blue] (0,2) circle [radius=2pt];
	\filldraw [xshift=16.6cm][yshift = -0.3cm][red] (0,-2) circle [radius=2pt];
	\filldraw [xshift=16.6cm][yshift = -0.3cm][red] (-2,0) circle [radius=2pt];
	\filldraw [xshift=16.6cm][yshift = -0.3cm][blue] (1.43,1.43) circle [radius=2pt];
	\filldraw [xshift=16.6cm][yshift = -0.3cm][blue] (-1.43,1.43) circle [radius=2pt];
	\filldraw [xshift=16.6cm][yshift = -0.3cm][red] (1.43,-1.43) circle [radius=2pt];
	\filldraw [xshift=16.6cm][yshift = -0.3cm][red] (-1.43,-1.43) circle [radius=2pt];
	
	\filldraw [xshift=16.6cm][yshift = -0.3cm][black] (0,0) circle [radius=1.33pt];
	\draw [xshift=16.6cm][yshift = -0.3cm][thick,red][->](-0.5,-1.22) -- (0.28,0.28);
	%\draw (0.48,1.5) node[text width = 1.33pt] {$\mathbf{x}_n$};
	
	% Second 
	\draw [xshift=13.48cm][yshift = -2.82cm](0.5,1.22) -- (2,0);
	\draw [xshift=13.48cm][yshift = -2.82cm](0.5,1.22) -- (0,2);
	\draw [xshift=13.48cm][yshift = -2.82cm](0.5,1.22) -- (-2,0);
	\draw [xshift=13.48cm][yshift = -2.82cm](0.5,1.22) -- (0,-2);
%		\draw [xshift=5.18cm][yshift = -2.82cm](-0.5,-1.22) -- (-1.41,-1.43);
	\draw [xshift=13.48cm][yshift = -2.82cm](0.5,1.22) -- (-1.43,1.43);
	\draw [xshift=13.48cm][yshift = -2.82cm](0.5,1.22) -- (1.43,-1.43);
%		\draw [xshift=13.48cm][yshift = -2.82cm](0.5,1.22) -- (1.43,1.43);
	\draw [xshift=13.48cm][yshift = -2.82cm][dashed](0.5,1.22) -- (1.43,1.43);
	\draw [xshift=13.48cm][yshift = -2.82cm](0,0) circle [radius=2cm];
	\filldraw [xshift=13.48cm][yshift = -2.82cm][black] (0.5,1.22) circle [radius=2pt];
	
	\filldraw [xshift=13.48cm][yshift = -2.82cm][red] (2,0) circle [radius=2pt];
	\filldraw [xshift=13.48cm][yshift = -2.82cm][red] (0,2) circle [radius=2pt];
	\filldraw [xshift=13.48cm][yshift = -2.82cm][blue] (0,-2) circle [radius=2pt];
	\filldraw [xshift=13.48cm][yshift = -2.82cm][blue] (-2,0) circle [radius=2pt];
	\filldraw [xshift=13.48cm][yshift = -2.82cm][red] (1.43,1.43) circle [radius=2pt];
	\filldraw [xshift=13.48cm][yshift = -2.82cm][red] (-1.43,1.43) circle [radius=2pt];
	\filldraw [xshift=13.48cm][yshift = -2.82cm][blue] (1.43,-1.43) circle [radius=2pt];
	\filldraw [xshift=13.48cm][yshift = -2.82cm][blue] (-1.43,-1.43) circle [radius=2pt];
	
	\filldraw [xshift=13.48cm][yshift = -2.82cm][black] (0,0) circle [radius=1.33pt];
	\draw [xshift=13.48cm][yshift = -2.82cm][thick, red][->](0.5,1.22) -- (-0.7,0);
	
	\draw [xshift=14.0cm][yshift = -2.5cm][thick, black,->] (-2,-2)--(-3,-3);
	
	\node[ xshift = -3.3cm,yshift = 0.5cm] at (0,0) {\Large High RhoA};
	\filldraw [xshift=-5cm][yshift = 0.5cm][red] (0,0) circle [radius=2pt];
			
	\node[ xshift = -3.4cm,yshift = 1.5cm] at (0,0) {\Large High Rac1};
	\filldraw [xshift=-5cm][yshift = 1.5cm][blue] (0,0) circle [radius=2pt];
	
	\draw [xshift=0.3cm][yshift = -0.3cm][dashed](-1.43,-1.43) -- (0,0);
	\draw [xshift=0.3cm][yshift = -0.3cm][dashed](-1.5,1.31) -- (0,0);
	\draw [xshift=0.3cm][yshift = -0.3cm][dashed](1.5,-1.31) -- (0,0);
	\end{tikzpicture}
	\caption{Schematic representation of CIL stages. (Left) Cells moving towards each other collide. Blue circles indicate regions of higher protrusion and FA binding activity, characteristic of cell front. Red circles indicate regions of increased contractility, characteristic for the cell rear. A diametric dashed line indicates cell half from the point of contact, which is shown by the radial dashed line. (Middle) After the collision, both cells cease to move and repolarize, such that polarity is reflected along the former diametric dashed line. (Right) The migration cycle restarts with modified affinities for adhesion formation/rupturing. Dashed line indicates a ruptured FA. The cell on the left starts moving (ruptured FA), while that on the right remains stationary (newly formed FA).}
	\label{figure: CIL schematics}
\end{figure}
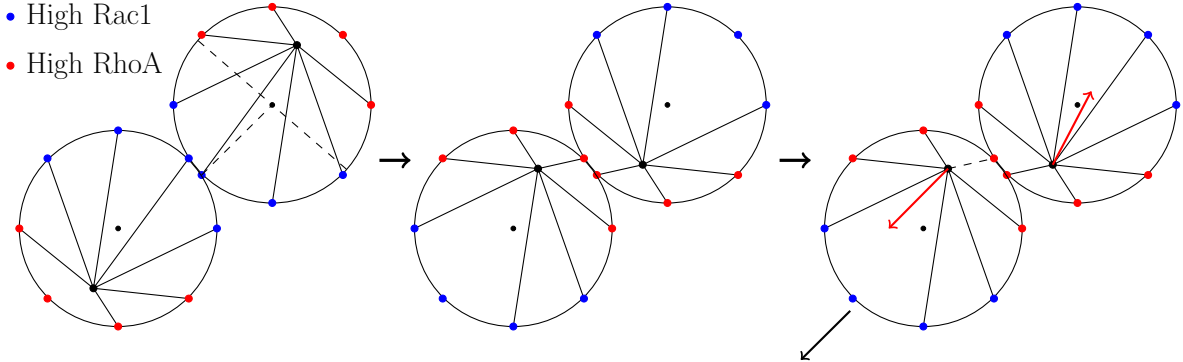

Contact inhibition of locomotion can be divided into the following sequence of stages (Figure \ref{figure: CIL schematics}). First, after collision, the movement ceases and cadherin mediated cell-cell contacts are formed. Second, in the vicinity of the contact protrusions collapse and actomyosin contractility is enhanced, as a result of Rac1 inhibition and RhoA activation. Their activity away from the collision site are altered in the opposite manner \cite{roycroft2016molecular}. Finally, the cells move away from each other.   

Within the context of our mesenchymal cell motility model described in Section \ref{section: single cell PDMP}, accounting for cell collisions has the following consequences: first, the collision causes the cells to jump into a non-motile state. Second, activation of Rac1 leads to increased FA binding affinity away from cell-cell contacts \cite{SCARPA2015421} and activation of RhoA enhances myosin generated contractile forces in SFs around the collision site \cite{roycroft2016molecular}. In the following we will consider a system of two cells, corresponding to the experimental settings in \cite{Desai20130717}, \cite{lin2015interplay}, \cite{Scarpa901}. See Appendix \ref{appendix: general CIL} for a general case and a mathematical treatment.   

Let $\bar{C}^i(t)\in\lbrace0,1\rbrace$ denote the collision state\footnote{By collision state we mean that a cell is in contact with some other cell: $\bar{C}(t)^i=1$ if it is in contact, and $\bar{C}(t)^i=0$ if it is not.} at time $t$  and $\bar{\Phi}^i(t)\in[0,2\pi)$ be the polar angle where the last contact of cell $i$ occurred\footnote{$\bar{\Phi}^i$ is constant until the next collision occurs.}, $i\in\lbrace1,2\rbrace$. Let the variables $\mu^i,\mathbf{Y}^i, \mathbf{X}^i$, corresponding to cell $i\in\lbrace1,2\rbrace$ be defined as before. Let $u_j:[0,2\pi)\times[0,2\pi)\times\lbrace0,1\rbrace\rightarrow\lbrace0,1\rbrace$, $j=1,\ldots, M$, be given by:
\begin{align}\label{eq: simple u_j}
u_j(\theta^i,\bar{\Phi}^i,\bar{C}^i) = 
\begin{cases*}
1, \phantom{abc}\bar{\Phi}^i-\frac{\pi}{2}\leq\theta^i+(j-1)\frac{2\pi}{M}\leq\bar{\Phi}^i+\frac{\pi}{2}\text{ and }\bar{C}^i=1\\
0, \phantom{abc}\text{else}.
\end{cases*}
\end{align}   
This function indicates whether $j^\text{th}$ FA is in the vicinity\footnote{By vicinity we simply mean within $\frac{\pi}{2}$ angle from the contact angle $\bar{\Phi}^i$. Here we assumed that the RhoGTPases activity is modified in half of a cell.} of the cell-cell contact site, provided there is one. 

As mentioned above, collisions lead to increased actomyosin contractility around the collision site. Thus, recalling equation \eqref{eq: Fi}, the tension due to myosin motors $T_j$ is modified as follows:
\begin{align}\label{eq: Ti modification}
T_j\rightarrow T_j(1+\delta_{myo}u_j(\theta^i,\bar{\Phi}^i,\bar{C}^i)), \phantom{abc} j=1,\ldots,M,\phantom{abc} i=1,2,
\end{align} 
where $\delta_{myo}>0$ is a parameter that signifies the increase in myosin generated force due to increased RhoA activity.  We then have $\mathbf{F}_j\rightarrow \mathbf{F}_j(\mathbf{X}^i,\bar{\Phi}^i,\bar{C}^i)$ and $\mathbf{F}\rightarrow\mathbf{F}(\mathbf{Y}^i,\mathbf{X}^i,\bar{\Phi}^i,\bar{C}^i)$ (see \eqref{eq: Fi} and \eqref{eq: total F}).  
The propensity function $a_j^+$ is modified as follows:
\begin{align}\label{eq: a+ modification}
a_j^+(\mathbf{Y}^i,\mathbf{X}^i)\rightarrow a_j^+(\mathbf{Y}^i,\mathbf{X}^i,\bar{\Phi}^i,\bar{C}^i)(1+\delta_+(1-u_j(\theta^i,\bar{\Phi}^i,\bar{C}^i))),
%a_j^-(\mathbf{Y}^i,\mathbf{X}^i,\bar{C}^i,\bar{\Phi}^i)&= k^0_{off}e^{\lVert \mathbf{F}_j(\mathbf{X}^i,\bar{C}^i,\bar{\Phi}^i)\rVert/F_b}y_j, 
\end{align}  
%\begin{align*}
%a_j^+(\mathbf{Y}^i,\mathbf{X}^i,\bar{C}^i,\bar{\Phi}^i) &= q(Q_{cue}(\mathbf{x}^i+\mathbf{x}_j))k_{force}(\lVert \mathbf{F}_j(\mathbf{X}^i,\bar{C}^i,\bar{\Phi}^i)\rVert)(1-Y_j)(1+\delta_+(1-u_j(\bar{C}^i,\bar{\Phi}^i,\theta^i)))
%%a_j^-(\mathbf{Y}^i,\mathbf{X}^i,\bar{C}^i,\bar{\Phi}^i)&= k^0_{off}e^{\lVert \mathbf{F}_j(\mathbf{X}^i,\bar{C}^i,\bar{\Phi}^i)\rVert/F_b}y_j, 
%\end{align*}  
%\footnote{In \cite{2018arXiv181011435U}, we showed that asymmetric increase of $T_j$ in prospective rear led to directed movement forward. There we also showed that preferential FA association at the prospective front leads to directed forward movement.}.
where $\delta_+>0$ is a parameter that signifies the increase in FA association rate due to increased Rac1 activity away from a contact site. Similarly, we also modify $a_j^-$:
\begin{align*}
a_j^-(\mathbf{Y}^i,\mathbf{X}^i)\rightarrow a_j^-(\mathbf{Y}^i,\mathbf{X}^i,\bar{\Phi}^i,\bar{C}^i)(1-\delta_-(1-u_j(\theta^i,\bar{\Phi}^i,\bar{C}^i))),
\end{align*}
where $\delta_-\in[0,1]$. Note that the dependence of $a_j^\pm$ on $\bar{C}^i$, $\bar{\Phi}^i$ is due to its dependence on $\mathbf{F}_j$ (see \cite{2018arXiv181011435U} for the form of $a_j^\pm$). If $\delta_-=1$, this implies that FAs away from a contact site do not disassociate. Thus, if a cell moves, it does so necessarily away from a collision site. That is, for $\delta_-=1$ cells do not crawl on top of one another.

For clarity, we introduce the following shorthand notation: 
\begin{align*}
a^{\pm,i}_j(\cdot) &= a^\pm_j(\mathbf{Y}^i(\cdot),\mathbf{X}^i(\cdot),\bar{\Phi}^i(\cdot),\bar{C}^i(\cdot))\\
a^2_0(\cdot) &= \sum_{i=1}^{2}a_0(\mathbf{Y}^i(\cdot),\mathbf{X}^i(\cdot),\bar{\Phi}^i(\cdot),\bar{C}^i(\cdot)).
\end{align*} Then, if $\mathcal{T}_k$ is the time of $k^\text{th}$ event, we have (see Appendix \ref{appendix: general CIL} for the derivation):

\begin{align}\label{eq: survival collisions}
\mathbb{P}\left(\mathcal{T}_{k+1}-\mathcal{T}_k>\tau|\lbrace\mathbf{Y}^i\left(\mathcal{T}_k\right),\mathbf{X}^i\left(\mathcal{T}_k\right),\bar{\Phi}^i\left(\mathcal{T}_k\right),\bar{C}^i\left(\mathcal{T}_k\right)\rbrace_{i=1,2}\right) = \exp\left(-\int_0^\tau a^2_0\left(\mathcal{T}_k+s\right)ds\right),
\end{align}
and
\begin{align}\label{eq: event collisions}
\mathbb{P}\left(j^{\pm,i}|\mathcal{T}_{k+1}\right) = \frac{a^{\pm,i}_j\left(\mathcal{T}^-_{k+1}\right)}{a^2_0\left(\mathcal{T}^-_{k+1}\right)},
\end{align} 
where $\mathbb{P}\left(j^{\pm,i}|\mathcal{T}_{k+1}\right)$ is the probability of binding/unbinding of $j^\text{th}$ FA of cell $i$, given the FA event time $\mathcal{T}_{k+1}$. Note that between two events, $\mathbf{X}^i$ evolves according to equation \eqref{eq: ss ODE}.  
%For example, let $t=T_{k-1}$ and suppose the next FA event time $T^*_k$ is given by \eqref{eq: survival collisions}. Suppose $t^*$ is the time of the     

%There are several salient points that need to be addressed with regards to outcomes $(IV,IV')$ and $(V,V')$. 

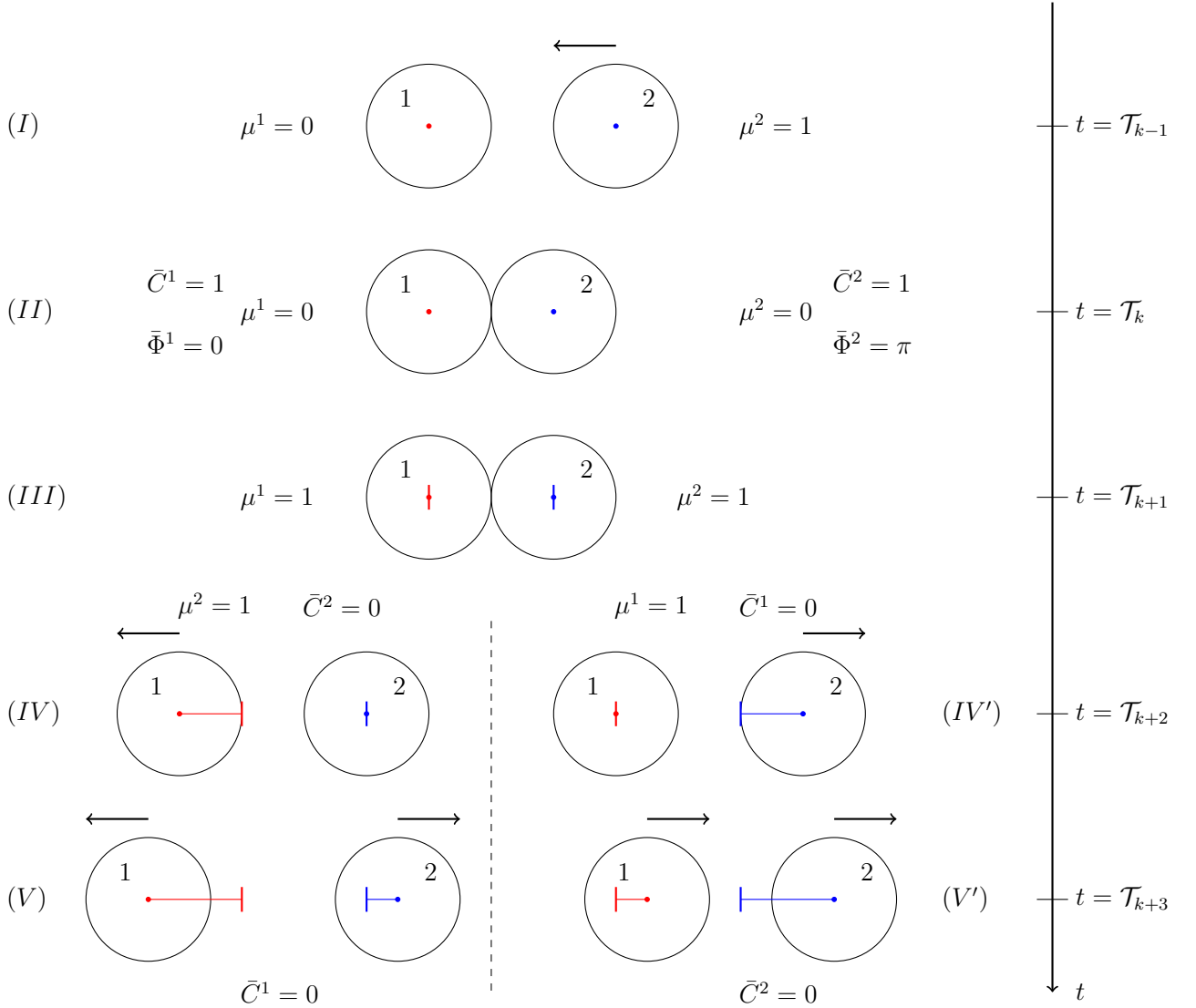
\begin{figure}[H]
\vspace*{9mm}
\begin{tikzpicture}[scale=0.9,transform shape,every node/.style={scale=1}]
\draw [thick,black,->](2,1.3)--(1,1.3);
%First
\draw (-1,0) circle [radius=1cm];
\draw (2,0) circle [radius=1cm];

\draw (-1.45,0.45) node[text width = 1.33pt] {\text{1}};
\draw (2.45,0.45) node[text width = 1.33pt] {\text{2}};

\draw (-4,0) node[text width = 1.33pt] {\text{$\mu^1=0$}};
\draw (4,0) node[text width = 1.33pt] {\text{$\mu^2=1$}};
\filldraw [blue](2,0) circle [radius=1pt];
\filldraw [red](-1,0) circle [radius=1pt];
\draw (-7.75,0) node[text width = 1.33pt] {\text{$(I)$}};
%Second
\draw [yshift = -3cm](-1,0) circle [radius=1cm];
\draw [yshift = -3cm](1,0) circle [radius=1cm];

\draw [yshift = -3cm](-1.45,0.45) node[text width = 1.33pt] {\text{1}};
\draw [yshift = -3cm](1.45,0.45) node[text width = 1.33pt] {\text{2}};

\draw [yshift = -3cm](-4,0) node[text width = 1.33pt] {\text{$\mu^1=0$}};
\draw [yshift = -3cm](4,0) node[text width = 1.33pt] {\text{$\mu^2=0$}};
\draw [yshift = -3cm](-5.5,0.5) node[text width = 1.33pt] {\text{$\bar{C}^1=1$}};
\draw [yshift = -3cm](-5.5,-0.5) node[text width = 1.33pt] {\text{$\bar{\Phi}^1=0$}};
\draw [yshift = -3cm](5.5,0.5) node[text width = 1.33pt] {\text{$\bar{C}^2=1$}};
\draw [yshift = -3cm](5.5,-0.5) node[text width = 1.33pt] {\text{$\bar{\Phi}^2=\pi$}};

\draw [yshift = -3cm](-7.75,0) node[text width = 1.33pt] {\text{$(II)$}};
\filldraw [yshift = -3cm][blue](1,0) circle [radius=1pt];
\filldraw [yshift = -3cm][red](-1,0) circle [radius=1pt];
%Third
\draw [yshift = -6cm](-1,0) circle [radius=1cm];
\draw [yshift = -6cm](1,0) circle [radius=1cm];

\draw [yshift = -6cm](-1.45,0.45) node[text width = 1.33pt] {\text{1}};
\draw [yshift = -6cm](1.45,0.45) node[text width = 1.33pt] {\text{2}};

\draw [yshift = -6cm][red,thick](-1,0.2) -- (-1,-0.2);
\draw [yshift = -6cm][blue,thick](1,0.2) -- (1,-0.2);
\draw [yshift = -6cm](-7.75,0) node[text width = 1.33pt] {\text{$(III)$}};
\filldraw [yshift = -6cm][blue](1,0) circle [radius=1pt];
\filldraw [yshift = -6cm][red](-1,0) circle [radius=1pt];

\draw [yshift = -6cm](3,0) node[text width = 1.33pt] {\text{$\mu^2=1$}};
\draw [yshift = -6cm](-4,0) node[text width = 1.33pt] {\text{$\mu^1=1$}};
%Fourth
%Right
\draw [xshift=3cm][yshift = -9.5cm](-1,0) circle [radius=1cm];
\draw [xshift=3cm][yshift = -9.5cm](2,0) circle [radius=1cm];
\draw [xshift=3cm][yshift = -9.5cm](-1.45,0.45) node[text width = 1.33pt] {\text{1}};
\draw [xshift=3cm][yshift = -9.5cm](2.45,0.45) node[text width = 1.33pt] {\text{2}};
\draw [xshift=3cm][yshift = -9.5cm][->,thick](2,1.3)--(3,1.3);

\draw [yshift = -8cm][dashed](0,0)--(0,-6);

\draw [xshift=3cm][yshift = -9.5cm][red,thick](-1,0.2) -- (-1,-0.2);
\draw [xshift=3cm][yshift = -9.5cm][blue,thick](1,0.2) -- (1,-0.2);
\draw [xshift=3cm][yshift = -9.5cm][blue] (1,0)--(2,0);
\filldraw [xshift=3cm][yshift = -9.5cm][blue](2,0) circle [radius=1pt];
\filldraw [xshift=3cm][yshift = -9.5cm][red](-1,0) circle [radius=1pt];
\draw [xshift=3cm][yshift = -9.5cm](4.25,0) node[text width = 1.33pt] {\text{$(IV')$}};

\draw [yshift = -7.75cm](2,0) node[text width = 1.33pt] {\text{$\mu^1=1$}};
\draw [yshift = -7.75cm](4,0) node[text width = 1.33pt] {\text{$\bar{C}^1=0$}};
%Left
\draw [xshift=-5cm][yshift = -9.5cm](0,0) circle [radius=1cm];
\draw [xshift=-5cm][yshift = -9.5cm](3,0) circle [radius=1cm];
\draw [xshift=-5cm][yshift = -9.5cm](-0.45,0.45) node[text width = 1.33pt] {\text{1}};
\draw [xshift=-5cm][yshift = -9.5cm](3.45,0.45) node[text width = 1.33pt] {\text{2}};

\draw [xshift=-5cm][yshift = -9.5cm][<-,thick](-1,1.3)--(0,1.3);

\draw [xshift=-5cm][yshift = -9.5cm][blue,thick](3,0.2)--(3,-0.2);
\draw [xshift=-5cm][yshift = -9.5cm][red,thick](1,0.2)--(1,-0.2);
\draw [xshift=-5cm][yshift = -9.5cm][red] (1,0)--(0,0);
\filldraw [xshift=-5cm][yshift = -9.5cm][red](0,0) circle [radius=1pt];
\filldraw [xshift=-5cm][yshift = -9.5cm][blue](3,0) circle [radius=1pt];
\draw [yshift = -9.5cm](-7.75,0) node[text width = 1.33pt] {\text{$(IV)$}};

\draw [yshift = -7.75cm](-5,0) node[text width = 1.33pt] {\text{$\mu^2=1$}};
\draw [yshift = -7.75cm](-3,0) node[text width = 1.33pt] {\text{$\bar{C}^2=0$}};
%Fifth
%Right
\draw [xshift=3cm][yshift = -12.5cm](-0.5,0) circle [radius=1cm];
\draw [xshift=3cm][yshift = -12.5cm](2.5,0) circle [radius=1cm];
\draw [xshift=3cm][yshift = -12.5cm](-0.95,0.45) node[text width = 1.33pt] {\text{1}};
\draw [xshift=3cm][yshift = -12.5cm](2.95,0.45) node[text width = 1.33pt] {\text{2}};
\draw [xshift=3cm][yshift = -12.5cm][->,thick](2.5,1.3)--(3.5,1.3);
\draw [xshift=3cm][yshift = -12.5cm][->,thick](-0.5,1.3)--(0.5,1.3);

\draw [xshift=3cm][yshift = -12.5cm][red,thick](-1,0.2) -- (-1,-0.2);
\draw [xshift=3cm][yshift = -12.5cm][red](-1,0) -- (-0.5,0);
\draw [xshift=3cm][yshift = -12.5cm][blue,thick](1,0.2) -- (1,-0.2);
\draw [xshift=3cm][yshift = -12.5cm][blue] (1,0)--(2.5,0);
\filldraw [xshift=3cm][yshift = -12.5cm][blue](2.5,0) circle [radius=1pt];
\filldraw [xshift=3cm][yshift = -12.5cm][red](-0.5,0) circle [radius=1pt];
\draw [xshift=3cm][yshift = -12.5cm](4.25,0) node[text width = 1.33pt] {\text{$(V')$}};

\draw [yshift = -14cm](4,0) node[text width = 1.33pt] {\text{$\bar{C}^2=0$}};
%Left
\draw [xshift=-5cm][yshift = -12.5cm](-0.5,0) circle [radius=1cm];
\draw [xshift=-5cm][yshift = -12.5cm](3.5,0) circle [radius=1cm];
\draw [xshift=-5cm][yshift = -12.5cm](-0.95,0.45) node[text width = 1.33pt] {\text{1}};
\draw [xshift=-5cm][yshift = -12.5cm](3.95,0.45) node[text width = 1.33pt] {\text{2}};
\draw [xshift=-5cm][yshift = -12.5cm][<-,thick](-1.5,1.3)--(-0.5,1.3);
\draw [xshift=-5cm][yshift = -12.5cm][->,thick](3.5,1.3)--(4.5,1.3);

\draw [yshift = -12.5cm](-7.75,0) node[text width = 1.33pt] {\text{$(V)$}};

\draw [xshift=-5cm][yshift = -12.5cm][red,thick](1,0.2) -- (1,-0.2);
\draw [xshift=-5cm][yshift = -12.5cm][red](1,0) -- (-0.5,0);
\draw [xshift=-5cm][yshift = -12.5cm][blue,thick](3,0.2) -- (3,-0.2);
\draw [xshift=-5cm][yshift = -12.5cm][blue] (3,0)--(3.5,0);
\filldraw [xshift=-5cm][yshift = -12.5cm][blue](3.5,0) circle [radius=1pt];
\filldraw [xshift=-5cm][yshift = -12.5cm][red](-0.5,0) circle [radius=1pt];

\draw [yshift = -14cm](-4,0) node[text width = 1.33pt] {\text{$\bar{C}^1=0$}};
%Right axis
\draw [thick,black][->](9,2) -- (9,-14);
%t=-1
\draw (8.75,0)--(9.25,0);
\draw (9.4,0) node[text width = 1.33pt] {\text{$t=\mathcal{T}_{k-1}$}};
%t=0
\draw [yshift=-3cm](8.75,0)--(9.25,0);
\draw [yshift=-3cm](9.4,0) node[text width = 1.33pt] {\text{$t=\mathcal{T}_k$}};
%t=tau1
\draw [yshift=-6cm](8.75,0)--(9.25,0);
\draw [yshift=-6cm](9.4,0) node[text width = 1.33pt] {\text{$t=\mathcal{T}_{k+1}$}};
%t=tau2
\draw [yshift=-9.5cm](8.75,0)--(9.25,0);
\draw [yshift=-9.5cm](9.4,0) node[text width = 1.33pt] {\text{$t=\mathcal{T}_{k+2}$}};
%t=tau3
\draw [yshift=-12.5cm](8.75,0)--(9.25,0);
\draw [yshift=-12.5cm](9.4,0) node[text width = 1.33pt] {\text{$t=\mathcal{T}_{k+3}$}};
\draw [yshift = -14cm](9.4,0) node[text width = 1.33pt] {\text{$t$}};
\end{tikzpicture}

\caption{Schematic representation of binary collisions. $(I)$ Cell 2 moves in the direction of cell 1. The centroids $\mathbf{x}^i$ are indicated by blue and red dots, respectively. $(II)$ Collided cells become stationary. $(III)$ An event occurs at time $t=\mathcal{T}_{k+1}$. Vertical bars indicate centroid positions at the collision time. Following an FA unbinding event at $t=\mathcal{T}_{k+1}$ in cell 1 or 2, the motility state in the corresponding cell switches and the system proceeds to configurations $(IV)$ or $(IV')$, respectively. An adhesion event leads back to $(III)$. $(IV,IV')$ Another event occurs at time $t=\mathcal{T}_{k+2}$. FA rupturing in cell 2 or 1 leads to the motility and collision state switches in $(IV)$ or $(IV')$, respectively. Cells move in the opposite or the same directions, and the proceed to configurations $(V)$ or $(V')$, respectively, until the next FA event occurs at time $t=\mathcal{T}_{k+3}$.}
\label{fig: collision schematics}
\end{figure}
Also, the event time $\mathcal{T}_k$ needs not be the time when an FA reaction occurred. It is possible that at time $\mathcal{T}_k$ a collision occurred. In this case $\mathbf{Y}^i$ is unchanged, but $\mu^i,\bar{C}^i,\bar{\Phi}^i$ jump to new values. Figure \ref{fig: collision schematics} illustrates how cell collisions are incorporated into the cell motility model.
\begin{itemize}
\item $(I)$ Suppose an FA event occurred at time $t=\mathcal{T}_{k-1}$ and cell 1 is stationary ($\mu^1=0$), while cell 2 is moving ($\mu^2=1$). Suppose $\mathcal{T}^*_k$ is given according to \eqref{eq: survival collisions}. The evolution of $\mathbf{X}^i$ is given by \eqref{eq: ss ODE} until a collision occurs at time $t=t^*<\mathcal{T}^*_{k}$. Then $\mathcal{T}_k=t^*$.
\item $(II)$ Due to the collision, both cells become stationary ($\mu^i=0$), the collision states and the contact angles jump to new values: $\bar{C}^1=\bar{C}^2=1$ and $\bar{\Phi}^1=0$, $\bar{\Phi}^2=\pi$ for cell 1 and 2, respectively. Then, $\mathbf{X}^i$ follows \eqref{eq: ss ODE} until time $t=\mathcal{T}_{k+1}$, given by \eqref{eq: survival collisions}.
\item $(III)$  At time $t=\mathcal{T}_{k+1}$ an FA event, determined by \eqref{eq: event collisions}, occurs. If an adhesion event occurs in cell $i$, $\mathbf{Y}^i$ changes accordingly, a new FA event time is found, the ODE system proceeds until this time and we are back at the same stage $(III)$. Suppose a deadhesion event occurs. If it was in cell 1(2), then $\mu^1$($\mu^2$) jumps to a new value, cell 1(2) moves until time $t=\mathcal{T}_{k+2}$ of the next event and the system proceeds to the configuration  in $(IV)$(or $(IV')$). 
\item $(IV,IV')$ Following FA rupturing in cell 2, $\mu^2=1$ and $\bar{C}_2=0$, corresponding to scenario $(IV)$. Likewise, for an FA rupturing in cell 2, $\mu^2=1$ and $\bar{C}_2=0$, corresponding to scenario $(IV')$. The collision state switches since the cells are no longer in contact. In both cases, the other cell is unaffected and continues its motion.
\item $(V,V')$ Suppose the next FA event at time $t=\mathcal{T}_{k+3}$ occurred in the previously unaffected cell. Then, its collision state $\bar{C}^i$ jumps to a new value, which is zero in this case.   
\end{itemize}
There are two implicit assumptions we made. First, a cell state changes only when a collision or an FA event occurs. Second, an FA event in a cell only changes the state of a cell in which it occurred. Thus, cell 1 and 2 continue their motion away from the collision site in $(IV)$ and $(IV')$, respectively, unaffected by what happened in the other cell. In particular $\bar{C}^1$ and $\bar{C}^2$ in $(IV)$ and $(IV')$, respectively, remain the same, since at the onset of post-collision motion in $(III)$ the cells are still in contact. When an event occurs in $(V)$ and $(V')$, the corresponding collision states are switched as cells are no longer in contact, while the other cells continue their motion. Note that whether cells move in the same or opposite directions after collisions is determined stochastically in our model, which is in line with \cite{Desai20130717}, \cite{lin2015interplay}, \cite{Scarpa901}. 

\textbf{Remark.} We assume, more generally, that cell interactions occur solely by collisions and that there is no coupling of cells before or after they interact. That is, neither the equations of motion \eqref{eq: ss ODE} between the events, nor the probabilities \eqref{eq: event collisions} of FA events in a cell depend on the state of another cell. This can be justified by the results in \cite{Desai20130717}, where it was found that CIL response in cells is statistically independent. It is, of course, possible that cells ``stick" and move together \cite{theveneau2010collective}. In this paper, however, the focus is solely on repulsive CIL. 
\section{Numerical Simulations}\label{section: numerical simulations}
\subsection{Confinement to one-dimensional lanes}
The illustration in Figure \ref{fig: collision schematics} depicts binary collisions in one-dimensional tracks. As noted in \cite{Desai20130717}, \cite{lin2015interplay}, \cite{Scarpa901}, this setup allows for a more efficient study of the CIL mechanism. In particular, it allows for unambiguous quantification of collision outcomes for measuring the CIL response. As in \cite{Desai20130717}, \cite{lin2015interplay}, we classify the outcomes into two categories. Namely, outcome 1 and 2 leading to cells moving in the opposite and the same directions, respectively, as illustrated in Figure \ref{fig: collision schematics} $(V,V')$. In order to investigate these outcomes, we introduce the following quantities:
\begin{itemize}
\item The distance between the cell centroids $d(t):=\lvert x_1^1(t+t_*)-x_1^2(t+t_*)\rvert$ at time $t$ after the first cell collision, where $x_1^i$ is the $x$-component of $\mathbf{x}^i$, $i=1,2$ and $t_*$ is the time of the collision.
\item Define $d^i(t):=x^i_1(t+t_*)-x_1^i(t_{*})$, illustrated in Figure \ref{fig: collision schematics} as the difference between the red (blue) dot and red (blue) vertical bar.
\end{itemize}
Note that restriction to movement in lanes implies that the first equation in \eqref{eq: ss ODE} is modified as follows:
\begin{align*}
\dot{\mathbf{x}} &=\mu\beta_{ECM}^{-1}\hat{\mathbf{e}}_1\cdot\left(\mathbf{F}\cdot\hat{\mathbf{r}}\hat{\mathbf{r}}\right),
\end{align*}
where $\hat{\mathbf{e}}_1=(1,0)^T$, i.e. the cells move in horizontal direction only.
 
Consider Figure \ref{fig: collision schematics} $(V,V')$. If the cells are moving in opposite directions (Figure \ref{fig: collision schematics} $(V)$), then $d^1(t)$ and $d^2(t)$ have opposite signs - negative and positive, respectively.  If the two cells are moving in the positive (negative) $x$-direction, then $d^i(t)>0$ ($d^i(t)<0$), for $i=1,2$. Note that while $d(t)$ is used as a readout of CIL in \cite{Scarpa901}\footnote{In \cite{Scarpa901} the distance between cell nuclei, rather than cell centroids, was measured.}, where its increase with time was used as an indication that cells are moving in opposite directions and hence undergoing CIL, $d^i(t)$ allows to distinguish between outcome 1 and 2. Moreover, increasing $d(t)$ might simply indicate that one cell is faster than the other, while both are moving in the same direction.  

In Section \ref{section: CIL model} we introduced three new parameters in addition to the cell motility model in \cite{2018arXiv181011435U}, namely, $\delta_{myo}$, $\delta_+$, and $\delta_-$. Their magnitude indicates the strength of CIL repolarization signal upon collision. Below we perform numerical simulations with varying values of $\delta_{myo}$, $\delta_+$ in the absence of an external cue, and in the presence of a chemotactic gradient with varying strength (mimicking the experimental setup in \cite{lin2015interplay}) and fixed $\delta_{myo}$, $\delta_+$. For each scenario we simulate 64 pairs of cells for 20 hours of simulation time. Initially, the distance between the cell centroids is $2.4R_{cell}$, where we take $R_{cell}=25\mu m$, and the initial values for the $x$-components of the centroids are $x^1_1(0)=1$ and $x^2_1(0)=3.4$ for cell 1 and 2, respectively\footnote{As in Figure \ref{fig: collision schematics}, cell 1 and 2 refers to the cells on the left and right, respectively}. Here, we also set $\delta_-=0$, as we would like to explore hallmarks of CIL (contraction of the leading edge and FA activation away from it) specifically in the absence of volume exclusion. The initial conditions for other variables and parameter values are taken as in \cite{2018arXiv181011435U}.

\textbf{Remark.} Among other factors, the collision outcome depends on whether it was a head-to-tail or a head-to-head collision \cite{Desai20130717}, \cite{lin2015interplay}. Here, we analyze the outcomes in terms of effects CIL has on FA dynamics and SF contractility.  

\subsubsection{Absence of an external cue}
Here we investigate three scenarios corresponding to three pairs of values for $\delta_{myo}$ and $\delta_+$. Similar values were used in \cite{2018arXiv181011435U} to simulate directed movement. 
\begin{table}[h]
\centering
\def\arraystretch{1}
\begin{tabular}{|c|c c c|}
\hline
Parameters & S1 & S2 & S3\\
\hline
$\delta_{myo}$ & 0.3 & 0.4 & 0.5\\
$\delta_{+}$  & 0.1 & 0.2 & 0.3\\
\hline
\end{tabular}
 \caption{Parameter values corresponding to three scenarios S1-S3.}
 \label{table: uniform parameters}
\end{table}

\begin{figure}[h]
\subfloat[]
{
	\includegraphics[width=52mm,height=50mm]{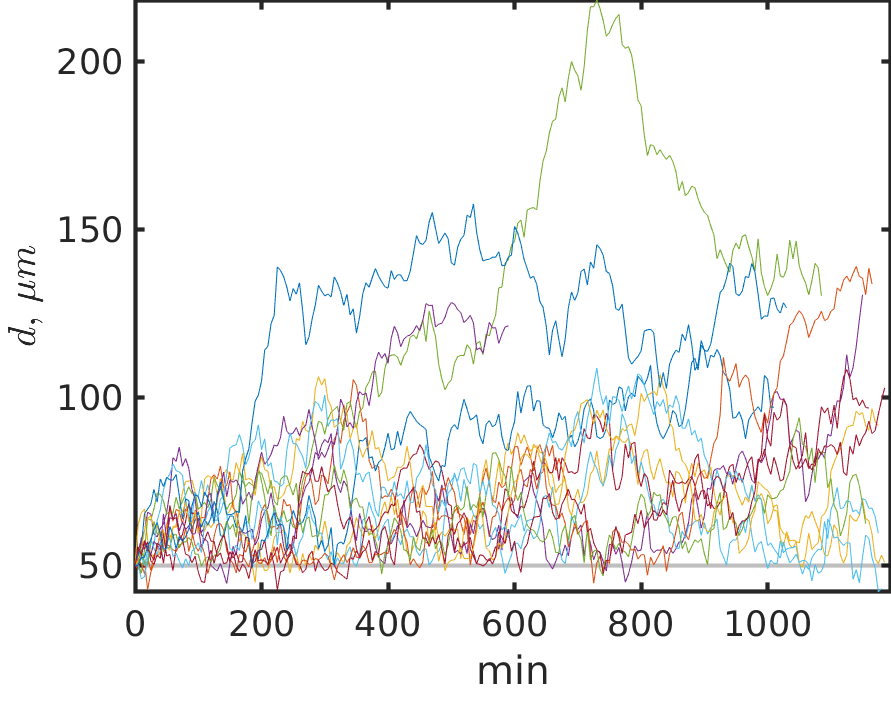}
}
\subfloat[]
{
	\includegraphics[width=52mm,height=50mm]{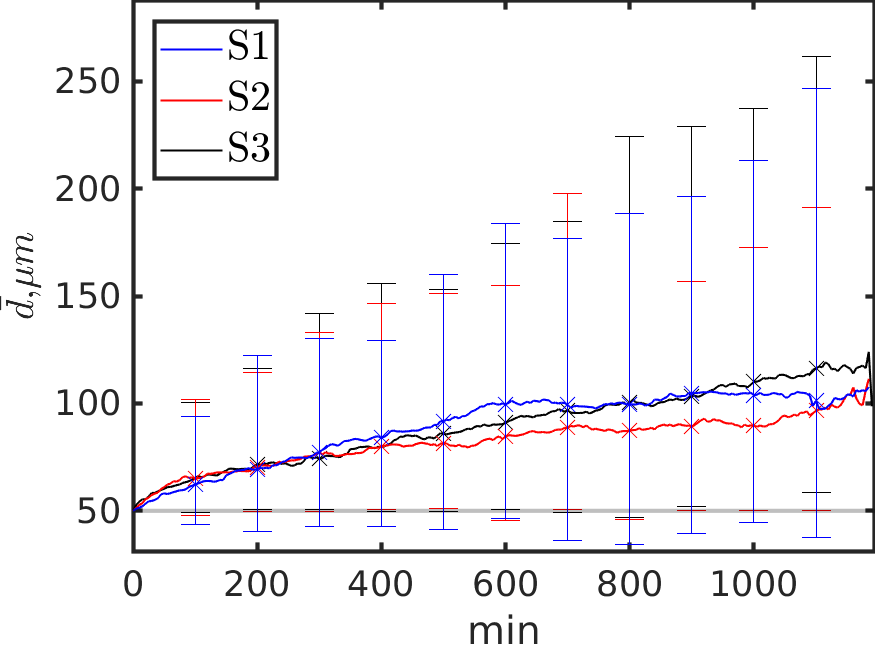}
}
\subfloat[]
{
	\includegraphics[width=52mm,height=50mm]{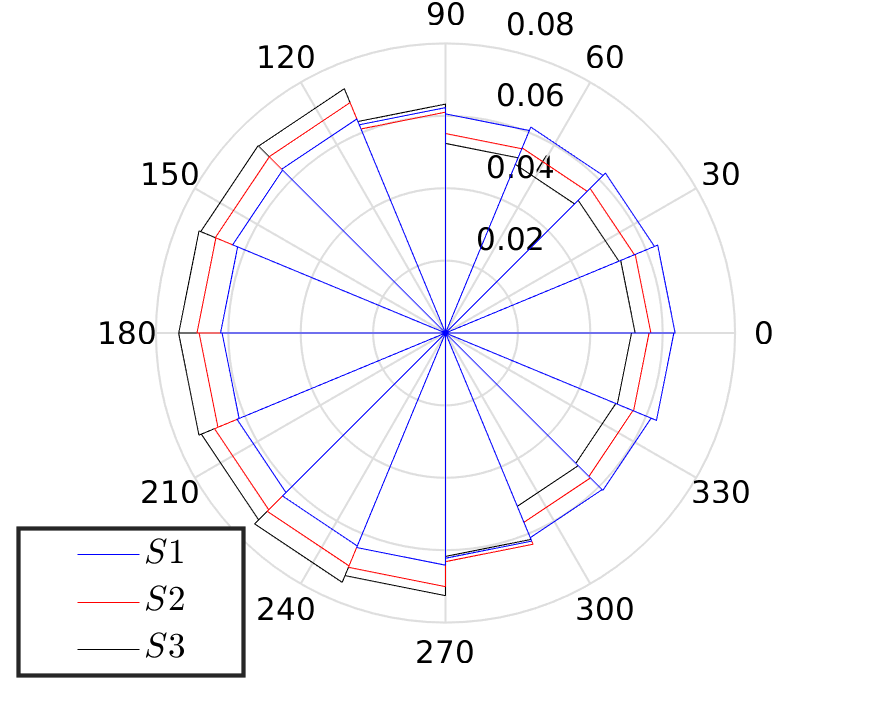}
}
\caption{The opaque horizontal line indicates the distance of $2R_{cell}$, i.e. the cells in contact. (a) Cell centroid distance $d$ of 14 cell pairs corresponding to scenario S2. (b-c) Scenarios S1 in blue, S2 in red, S3 in black. (b) Ensemble averages $\bar{d}$ for each scenario. The corresponding error bars indicate ensemble minimum and maximum. (c) Relative frequency of binding events of cells with $\bar{C}^i=1$ and colliding at $0\degree$ for each scenario. Each sector corresponds to a single FA counting counterclockwise. Cell 2 is accounted for by reflection about south-north axis.}
\label{fig: c_distances uniform}
\end{figure}
Since in our model the cells are not treated as hard spheres, it is possible that some overlaps may occur (Figure \ref{fig: c_distances uniform}a,b). However, the slight overlap is followed by an increase in $d$ and separation (Figure \ref{fig: c_distances uniform}a). Although the average distances $\bar{d}$ are similar (Figure \ref{fig: c_distances uniform}b), increasing $\delta_{myo}$ and $\delta_+$ leads to a stronger response: the minimum of $d$ is consistently lower for S1 compared to S3 (Figure \ref{fig: c_distances uniform}b) and FA formation away from the collision site is more frequent for S3 (Figure \ref{fig: c_distances uniform}c). Notice that the cells need not obey the volume exclusion principle for eventual separation to occur and the stronger response in S3 implies that the separation can be modulated by modifying contractility and FA formation. 
\begin{figure}[h]
\subfloat[]
{
	\includegraphics[width=52mm,height=50mm]{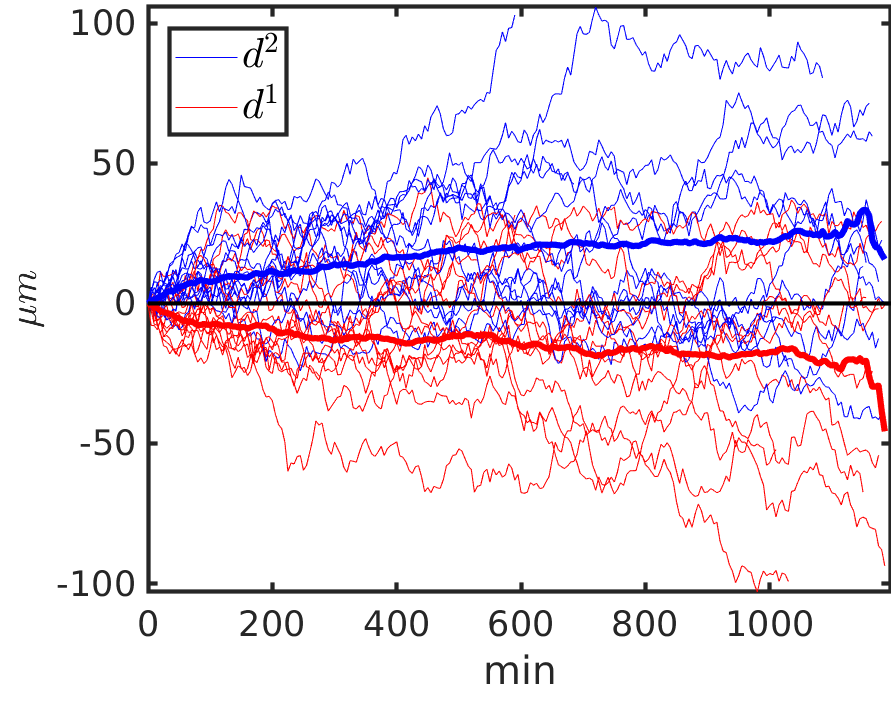}
}
\subfloat[]
{
	\includegraphics[width=52mm,height=50mm]{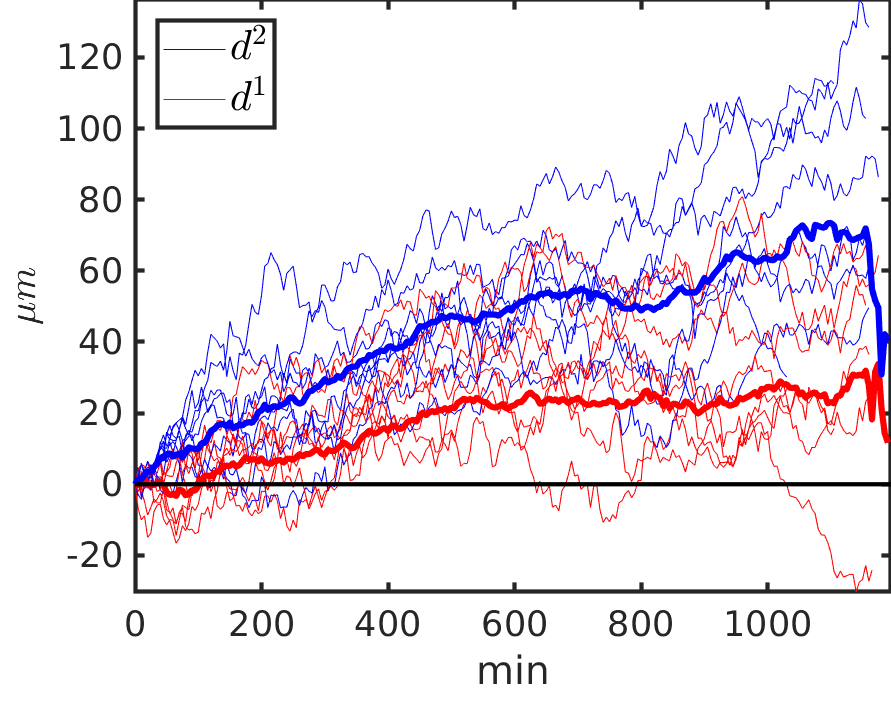}
}
\subfloat[]
{
	\includegraphics[width=52mm,height=50mm]{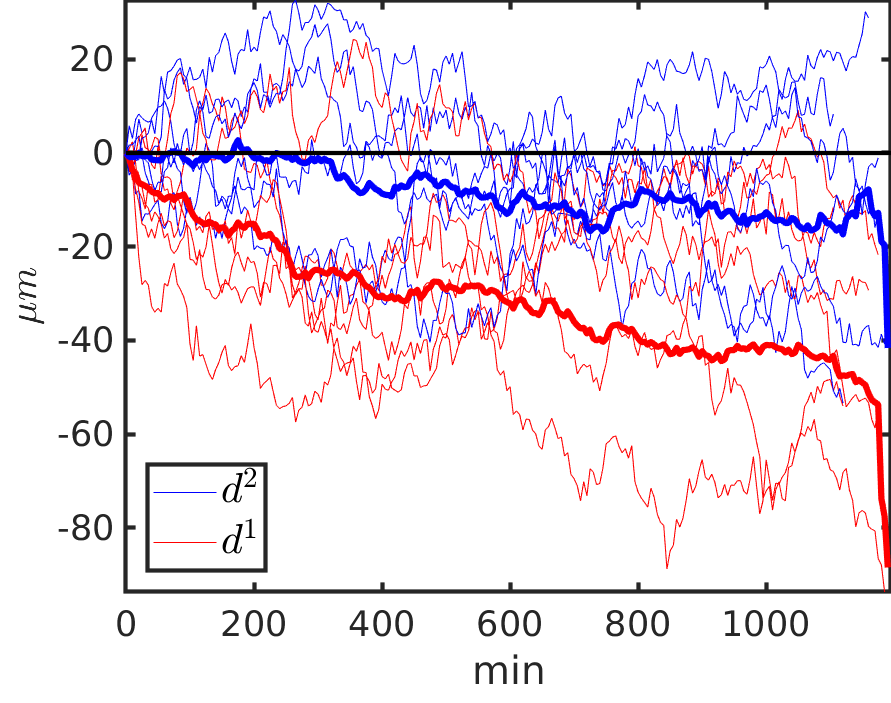}
}
\hfill
\subfloat[]
{
	\includegraphics[width=52mm,height=50mm]{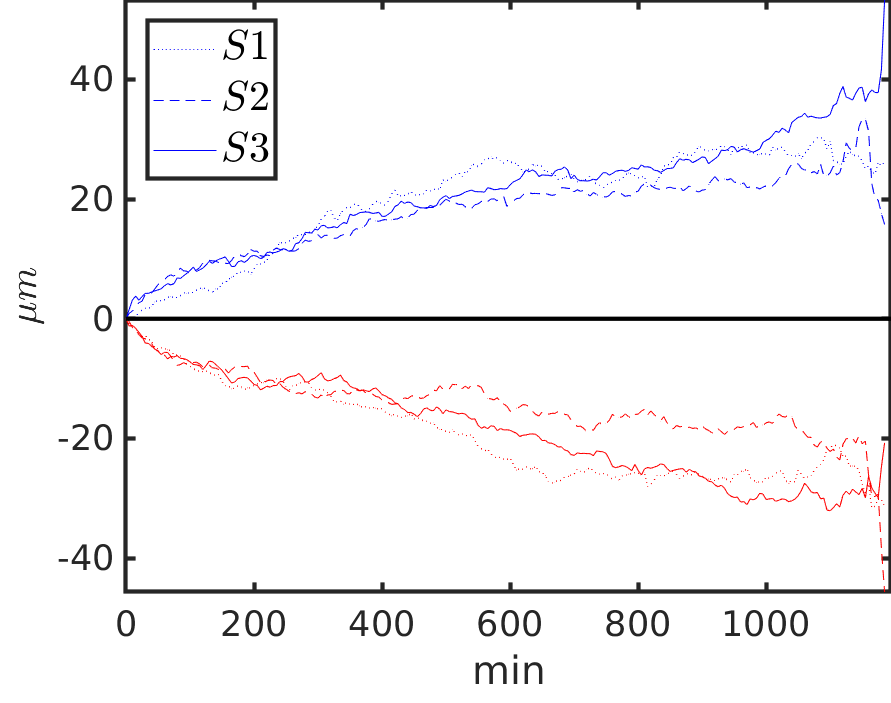}
}
\subfloat[]
{
	\includegraphics[width=52mm,height=50mm]{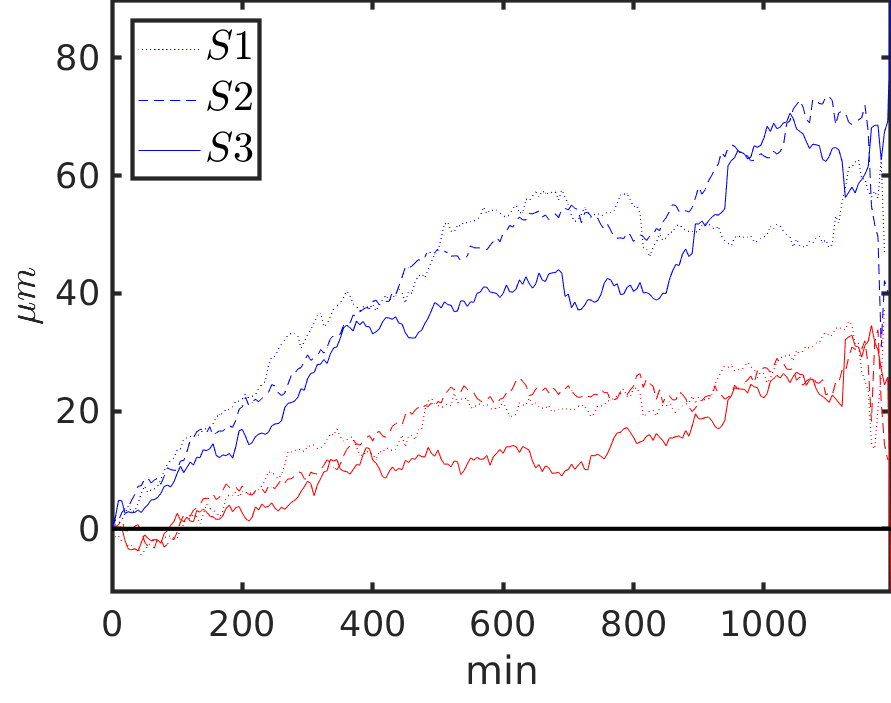}
}
\subfloat[]
{
	\includegraphics[width=52mm,height=50mm]{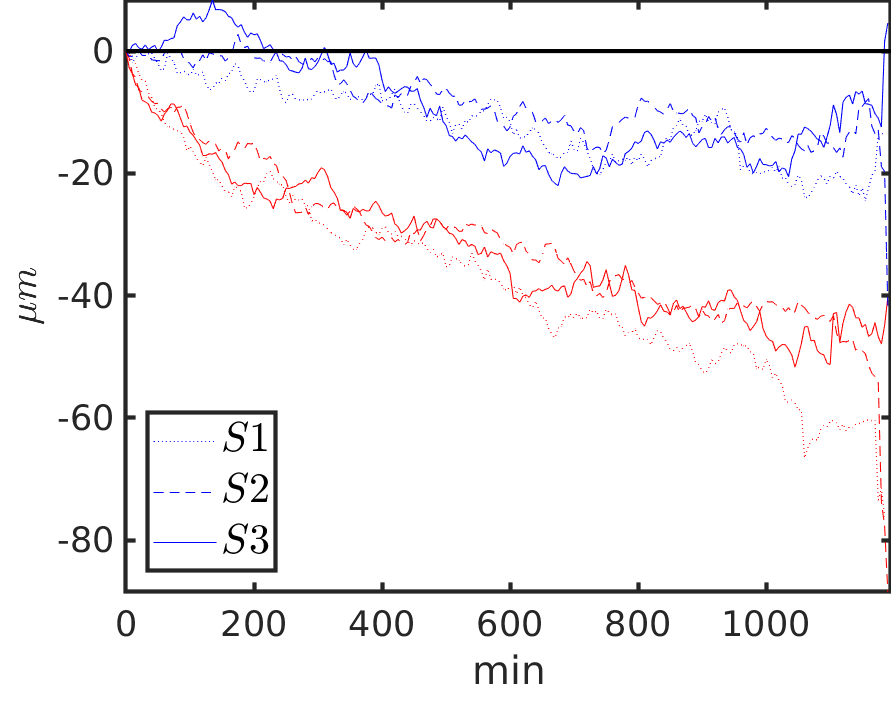}
}
\caption{(a-c) The differences $d^i$ corresponding to scenario S2. Thick lines represent the corresponding ensemble averages. $d^2$ is colored in blue, $d^1$ in red. (a) A sample of 14 pairs. (b) and (c) The differences $d^i$, whose averages over time are positive and negative, respectively. Samples of 8 and 6 pairs are shown, respectively. (d-f) Ensemble averages of $d^i$ for scenarios S1 (dot), S2 (dash), and S3 (solid). Blue and red colored plots correspond to cells 1 and 2, respectively.(e) and (f) $d^i$ with positive and negative times averages.}
\label{fig: c_point_distances uniform}
\end{figure}

Since increasing $d$ only suggests that the cells are separating, we examined their relative direction of motion after collision (Figure \ref{fig: c_point_distances uniform}). Ensemble averages of $d^i$ in Figure \ref{fig: c_point_distances uniform}d show that following collisions, the movement in the opposite directions is prevalent, which is in line with results in \cite{Desai20130717}, \cite{lin2015interplay}. It may also occur that cells follow one another after collision, as indicated by positive and negative time averages of $d^1$ and $d^2$ (Figure \ref{fig: c_point_distances uniform}b,c). The ensemble averages in Figure \ref{fig: c_point_distances uniform}(d-f) do not show a strong difference between the scenarios S1-S3. This suggests that varying the strength of cell response to collision does not have a significant effect on the relative direction of migration after the collision. In our simulations, $56\%$ of collided pairs moved in the opposite directions, compared to $\sim65\%$ in \cite{lin2015interplay}.  

\textbf{Remark.} Note that the collision times $t_*$ for each simulated pair are different. Thus, the number of cells at time $t$ after the collision time $t_*$ varies, and reduces towards the terminal time. This skews the values for ensemble averages and causes the abrupt changes in Figure \ref{fig: c_point_distances uniform}.

\begin{figure}[H]
\vspace*{-7mm}
\subfloat[]
{
	\includegraphics[width=55mm,height=50mm]{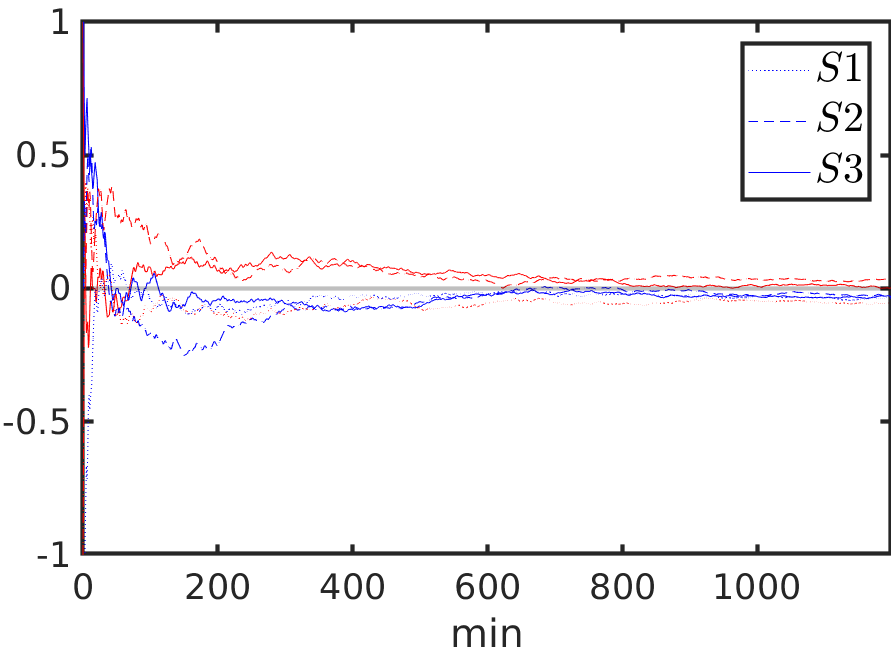}
}
\subfloat[]
{
	\includegraphics[width=55mm,height=50mm]{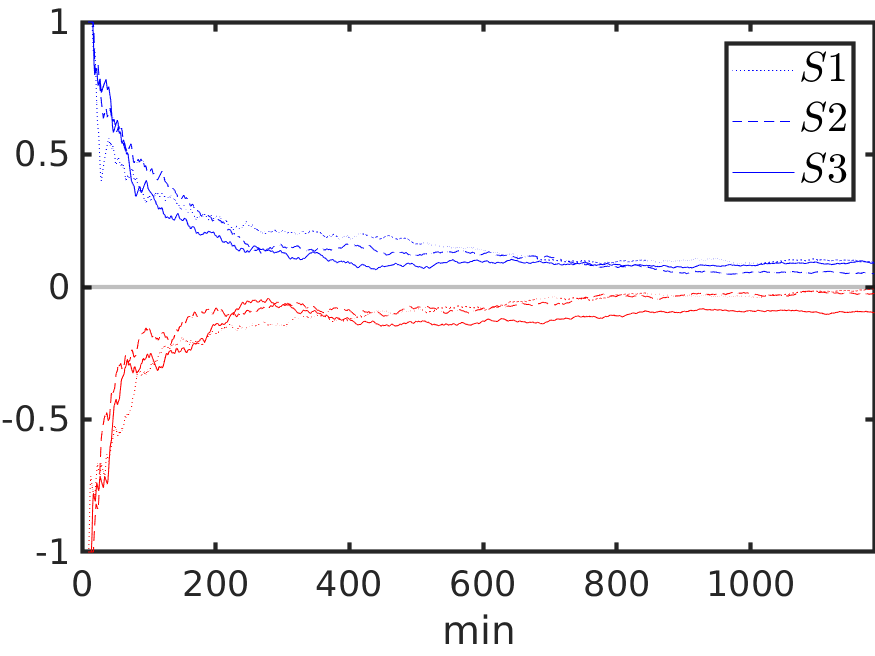}
}

\caption{(a) and (b) Ensemble averages of normalized velocities before and after collisions, respectively, for scenarios S1 (dot), S2 (dash), and S3 (solid). Plots in red and blue correspond to cells 1 and 2, respectively. In (a) non-colliding pairs have also been accounted for.}
\label{fig: velocities uniform}
\end{figure}
Note that a freely migrating cell before collision is equally likely to move in either direction, as indicated by a rapid decay of normalized velocities to zero in Figure \ref{fig: velocities uniform}a. How fast does a cell become freely migrating after a collision? Figure \ref{fig: velocities uniform}b shows a much slower decay of the normalized velocities for the three scenarios. This suggests that either there are frequent follow up collisions after the first one, resulting in cell 1(2) moving left(right), or collisions lead to persistent movement in the opposite direction. It must be the latter, since in light of our results, cells separate (Figure \ref{fig: c_distances uniform}b) and move away from each other (Figure \ref{fig: c_point_distances uniform}d). Thus, in our model transient perturbations in cell motility lead to persistent, but decaying, alterations in migration dynamics. This is unexpected, since the collision state $\bar{C}^i$ of a cell is switched off after separation, i.e. the cell migrates freely. However, studies in \cite{lin2015interplay} and \cite{Scarpa901} indicate that cells continue to move in opposite directions even after separation occurs.

\subsubsection{Presence of a chemotactic gradient}
We now explore how collision outcomes are affected in the presence of a chemotactic gradient, as experimental evidence in \cite{lin2015interplay} suggest that CIL response is modulated by the strength of the external signal. As in \cite{2018arXiv181011435U}, we suppose that $a^+_j\propto Q_{cue}$, i.e. the binding probability of the $j^{\text{th}}$ FA is proportional the (local) concentration of chemoattractant $Q_{cue}$ at the position of the FA.

\begin{figure}[H]
\vspace*{-7mm}
\subfloat[]
{
	\includegraphics[width=60mm,height=55mm]{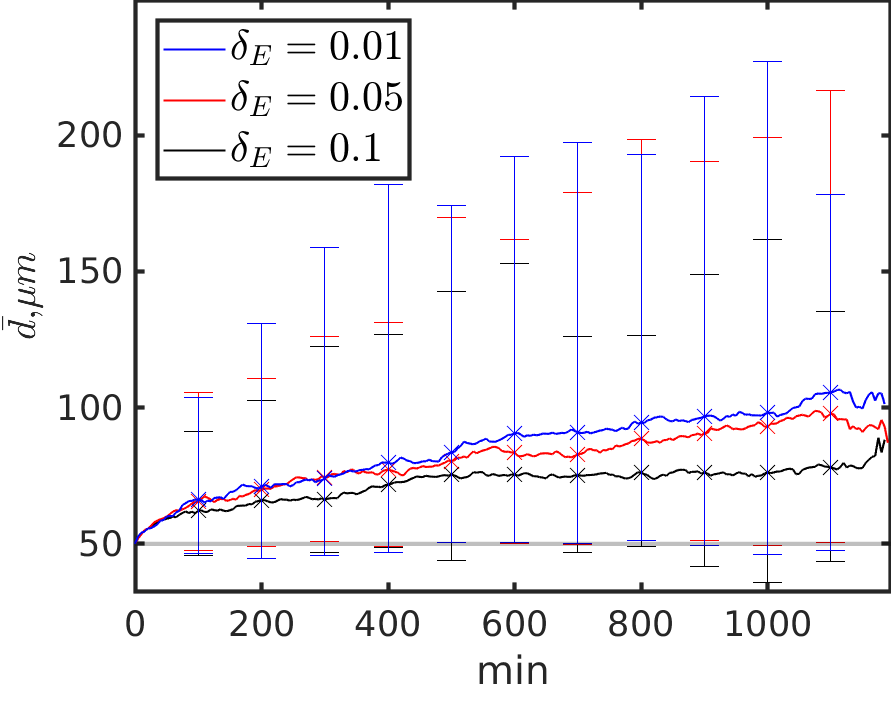}
}
\subfloat[]
{
	\includegraphics[width=60mm,height=55mm]{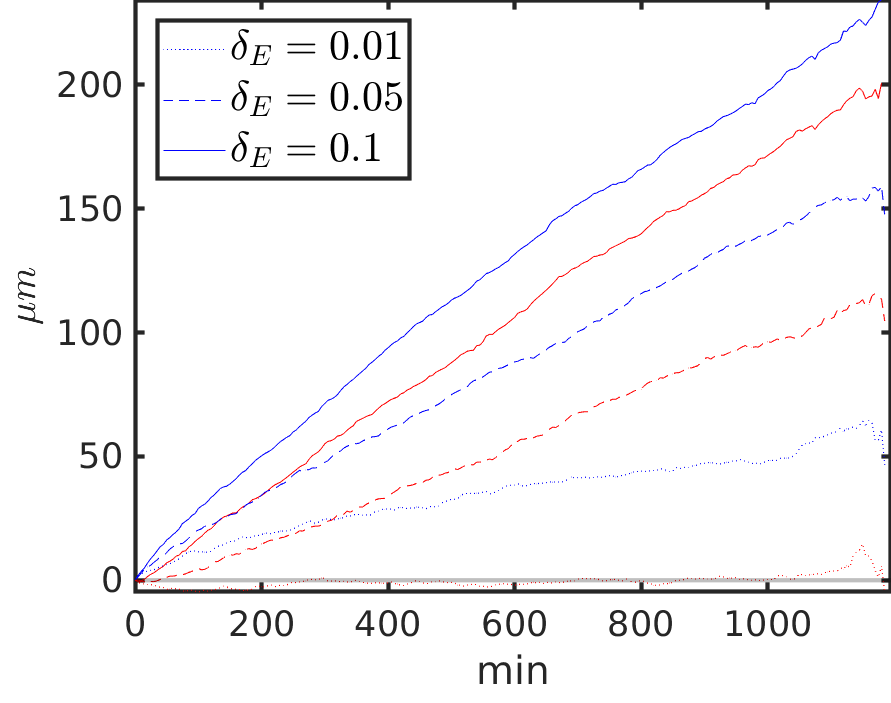}
}
\hfill
\subfloat[]
{
	\includegraphics[width=60mm,height=55mm]{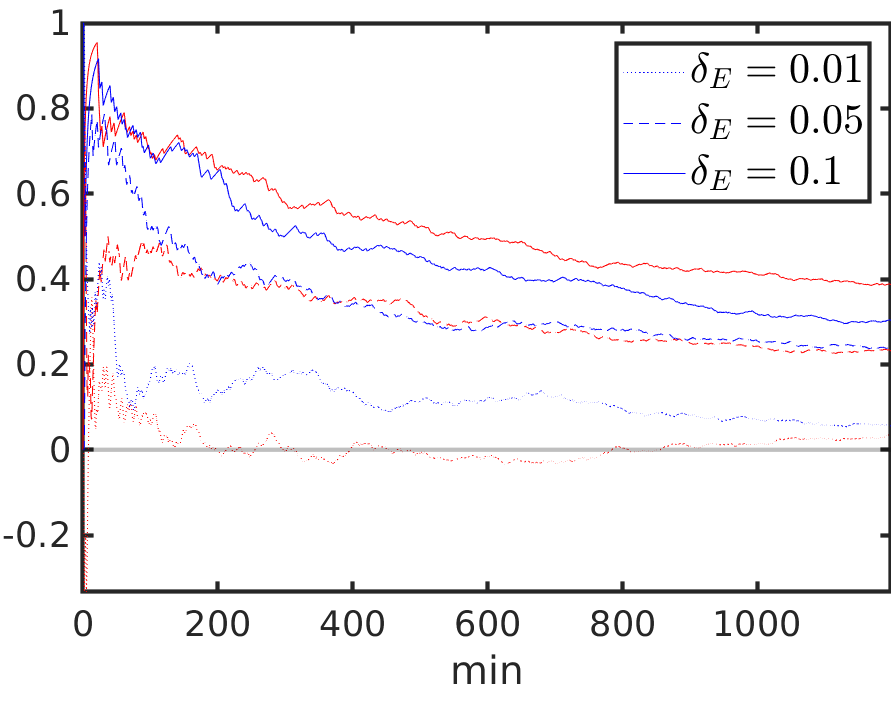}
}
\subfloat[]
{
	\includegraphics[width=60mm,height=55mm]{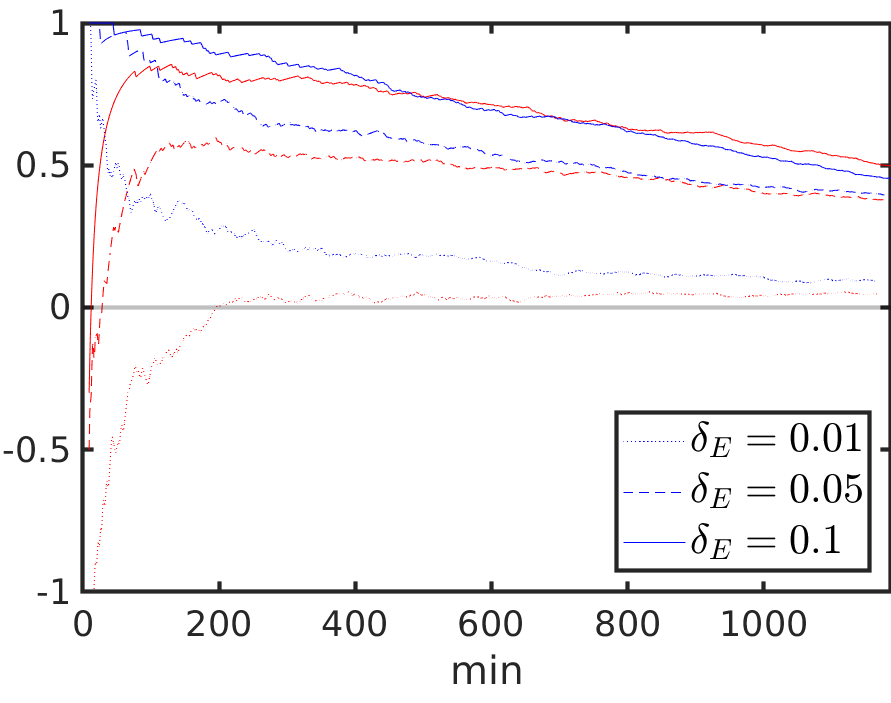}
}
\hfill
\subfloat[]
{
	\includegraphics[width=60mm,height=55mm]{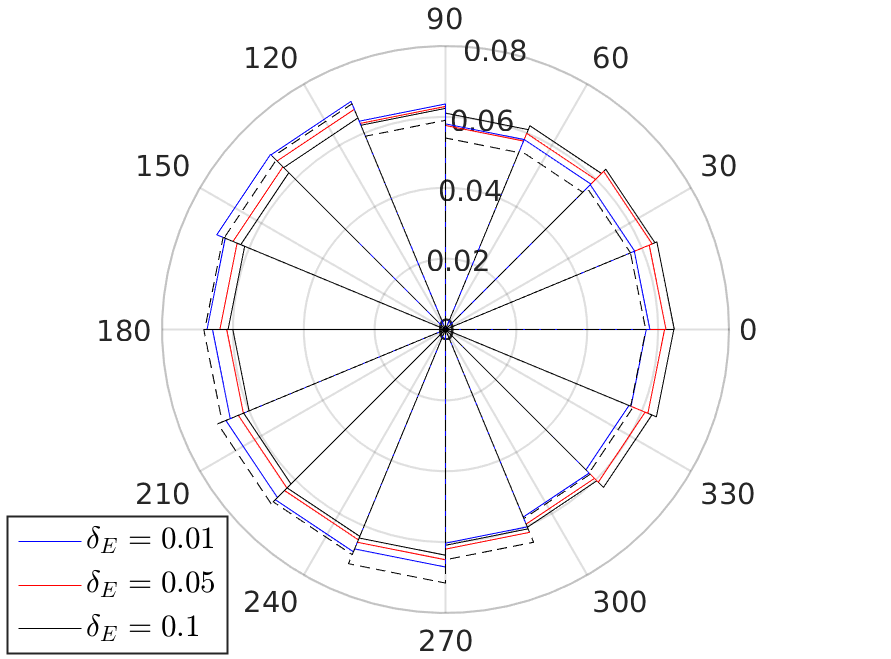}
}
\subfloat[]
{
	\includegraphics[width=60mm,height=55mm]{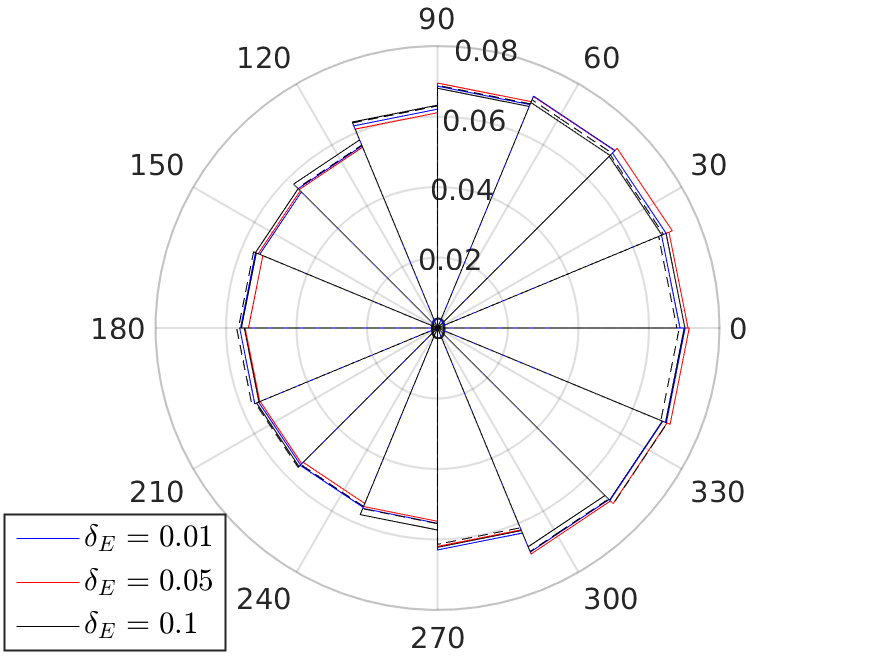}
}
\caption{The effect of varying chemotactic signal strength $\delta_E$. (a,e,f) $\delta_E=0.01$ (blue), $0.05$ (red), $0.1$ (black). (a) Ensemble averages of cell-cell distances. (b,c,d) $\delta_E=0.01$ (dot), $0.05$ (dash), $0.1$ (solid). (b) Ensemble averages of $d_1$ (red) and $d_2$ (blue). (c) and (d) Ensemble averages of normalized velocities before and after collisions, respectively, of cells 1 (red) and 2 (blue). (e) and (f) Relative frequency of binding events after collision of cells 1 and 2. Dashed lines correspond to $\delta_E=0$. }
\label{fig: results taxis}
\end{figure}
We assume that $Q_{cue}$ has the following form:
\begin{align*}
Q_{cue}(\mathbf{x}) = 
\begin{cases*}
1+\delta_Ex_1, \text{ if $x_1>0$}\\
1,\phantom{abcsasd}\text{else}
\end{cases*},
\end{align*}
where $\mathbf{x}$ is the position of an FA (in units of $R_{cell}$) in the lab reference frame, and $\delta_E>0$ indicates strength of the signal, i.e. there is a chemotactic gradient in the positive $x$-direction\footnote{The relative difference of $Q_{cue}$ between diametrically opposite FAs is always less than or equal to $\delta_E$. As in \cite{2018arXiv181011435U}, we note that chemotaxis occurs solely due to biased FA formation in the direction of the chemoattractant, and not due to the gradient taken as an input.}. We also take $\delta_{myo} = 0.4$, $\delta_+=0.2$.

The influence of a chemotactic signal on CIL can be seen in Figure \ref{fig: results taxis}. We see that increasing the signal strength reduces average cell-cell separation (Figure \ref{fig: results taxis}a). Although the difference between the averages is slight (relative to cell size), the variance (as indicated by the error bars) of cell-cell distances is noticeably smaller for the case of the strongest signal. Moreover, after the collision, cells tend to move in the same direction following the signal, as shown in Figure \ref{fig: results taxis}b. This, together with what appears to be a plateauing of cell-cell distance (Figure \ref{fig: results taxis}a), suggests emergence of collective movement. Observe that reducing the signal strength leads to reduced propensity of cells to move in the same direction, in line with the results reported in \cite{lin2015interplay}. Note that in \cite{lin2015interplay} three scenarios with different EGF concentrations were explored. There, the gradients of EGF concentration were kept constant at $3.3nM$ per length of the lane. However, the relative changes in EGF concentrations were $\frac{5.5nM-2.2nM}{2.2nM}=1.6$, $\frac{9.9nM-6.6nM}{6.6nM}=0.5$, $\frac{14.1nM-10.8nM}{10.8nM}=0.3$ and reduced relative changes led to diminished alteration of a typical CIL response, which our simulations show as well.

Motion alignment is not immediate, as the amount of time during which cells move in the opposite directions after collision depends on the magnitude of the gradient (Figure \ref{fig: results taxis}d), compared to a rapid velocity alignment of uncollided pairs (Figure \ref{fig: results taxis}c). 

We also see that the effect on adhesion dynamics of cells to the left and to the right of a collision point is different (Figure \ref{fig: results taxis}e,f). If the CIL signal in a cell and the chemotactic gradient are in the opposite directions, the affinity of FA association away from the contact reduces with increasing gradient strength (Figure \ref{fig: results taxis}e). However, if the signals are aligned, the FA binding dynamics does not appear to be significantly modified (Figure \ref{fig: results taxis}f). This suggests that in relation to adhesion dynamics, the chemotactic cue either reduces CIL response or has little to no effect. Interestingly, in \cite{theveneau2010collective} it was shown that elevated Rac1 activity (and hence enhanced adhesion to a substrate) away from the contact site (and in a free edge) is primarily due to cell-cell contacts, rather than to a chemoattractant.
\FloatBarrier
\subsection{Unconfined 2D setting}   
\begin{figure}[h]
	\begin{tikzpicture}[scale=0.65,transform shape]
%	Collision
	\draw [xshift=-5.64cm](-1.5,1.22) -- (1,1);
	\draw [xshift=-5.64cm](-1.5,1.22) -- (-1,3);
	\draw [xshift=-5.64cm](-1.5,1.22) -- (-3,1);
	\draw [xshift=-5.64cm](-1.5,1.22) -- (-1,-1);
	\draw [xshift=-5.64cm](-1.5,1.22) -- (0.43,2.43);
%	\draw [xshift=-5.64cm](-1.5,1.22) -- (-2.43,2.43);
	\draw [xshift=-5.64cm](-1.5,1.22) -- (0.43,-0.43);
	\draw [xshift=-5.64cm](-1.5,1.22) -- (-2.43,-0.43);
	%\draw (0.5,1.22) -- (-1.43,-1.43);
	\draw [xshift=-5.64cm](-1,1) circle [radius=2cm];
	\filldraw [xshift=-5.64cm][black](-1.5,1.22) circle [radius=2pt];
	
	\filldraw [xshift=-5.64cm][black] (1,1) circle [radius=2pt];
	\filldraw [xshift=-5.64cm][black] (-1,3) circle [radius=2pt];
	\filldraw [xshift=-5.64cm][black] (-1,-1) circle [radius=2pt];
	\filldraw [xshift=-5.64cm][black] (-3,1) circle [radius=2pt];
	\filldraw [xshift=-5.64cm][black] (0.43,2.43) circle [radius=2pt];
	\filldraw [xshift=-5.64cm][black] (-2.43,2.43) circle [radius=2pt];
	\filldraw [xshift=-5.64cm][black] (0.43,-0.43) circle [radius=2pt];
	\filldraw [xshift=-5.64cm][black] (-2.43,-0.43) circle [radius=2pt];
	
	\filldraw [xshift=-5.64cm][black] (-1,1) circle [radius=1.33pt];
	\draw [xshift=-5.64cm][thick,->](0.75,-0.75)--(1.15,-1.15);
	% %
	\draw (-0.5,-1.22) -- (2,0);
	\draw (-0.5,-1.22) -- (0,2);
	\draw (-0.5,-1.22) -- (-2,0);
	\draw (-0.5,-1.22) -- (0,-2);
%	\draw [xshift=8.3cm][yshift = -0.3cm](-0.5,-1.22) -- (1.41,1.43);
	\draw (-0.5,-1.22) -- (-1.43,1.43);
	\draw (-0.5,-1.22) -- (1.43,-1.43);
	\draw (-0.5,-1.22) -- (-1.43,-1.43);
	%\draw (0.5,1.22) -- (-1.43,-1.43);
	\draw (0,0) circle [radius=2cm];
	\filldraw [black] (-0.5,-1.22) circle [radius=2pt];
	
	\filldraw [blue] (2,0) circle [radius=2pt];
	\filldraw [blue] (0,2) circle [radius=2pt];
	\filldraw [red] (0,-2) circle [radius=2pt];
	\filldraw [red] (-2,0) circle [radius=2pt];
	\filldraw [blue] (1.43,1.43) circle [radius=2pt];
	\filldraw [blue] (-1.43,1.43) circle [radius=2pt];
	\filldraw [red] (1.43,-1.43) circle [radius=2pt];
	\filldraw [red] (-1.43,-1.43) circle [radius=2pt];
	
	\filldraw [black] (0,0) circle [radius=1.33pt];
%		\draw [xshift=8.3cm][yshift = -0.3cm][thick, red][->](-0.5,-1.22) -- (0.28,0.28);
	%\draw (0.48,1.5) node[text width = 1.33pt] {$\mathbf{x}_n$};
	
	% Second 
	\draw [xshift=-2.82cm][yshift=-2.82cm](0.5,1.22) -- (2,0);
	\draw [xshift=-2.82cm][yshift=-2.82cm](0.5,1.22) -- (0,2);
	\draw [xshift=-2.82cm][yshift=-2.82cm](0.5,1.22) -- (-2,0);
	\draw [xshift=-2.82cm][yshift=-2.82cm](0.5,1.22) -- (0,-2);
	\draw [xshift=-2.82cm][yshift = -2.82cm](0.5,1.22) -- (-1.41,-1.43);
	\draw [xshift=-2.82cm][yshift=-2.82cm](0.5,1.22) -- (-1.43,1.43);
%	\draw [xshift=-2.82cm][yshift=-2.82cm](0,1.22) -- (1.43,-1.43);
	\draw [xshift=-2.82cm][yshift=-2.82cm](0.5,1.22) -- (1.43,1.43);
	\draw [xshift=-2.82cm][yshift=-2.82cm](0,0) circle [radius=2cm];
	\filldraw [xshift=-2.82cm][yshift=-2.82cm][black] (0.5,1.22) circle [radius=2pt];
	
	\filldraw [xshift=-2.82cm][yshift=-2.82cm][red] (2,0) circle [radius=2pt];
	\filldraw [xshift=-2.82cm][yshift=-2.82cm][red] (0,2) circle [radius=2pt];
	\filldraw [xshift=-2.82cm][yshift=-2.82cm][blue] (0,-2) circle [radius=2pt];
	\filldraw [xshift=-2.82cm][yshift=-2.82cm][blue] (-2,0) circle [radius=2pt];
	\filldraw [xshift=-2.82cm][yshift=-2.82cm][red] (1.43,1.43) circle [radius=2pt];
	\filldraw [xshift=-2.82cm][yshift=-2.82cm][red] (-1.43,1.43) circle [radius=2pt];
	\filldraw [xshift=-2.82cm][yshift=-2.82cm][blue] (1.43,-1.43) circle [radius=2pt];
	\filldraw [xshift=-2.82cm][yshift=-2.82cm][blue] (-1.43,-1.43) circle [radius=2pt];
	
	\filldraw [xshift=-2.82cm][yshift=-2.82cm][black] (0,0) circle [radius=1.33pt];
%		\draw [xshift=5.18cm][yshift = -2.82cm][thick, red][->](0.5,1.22) -- (-0.28,-0.28);
	%Aftermath
	\draw [xshift=5.94cm](0.5,-1.22) -- (2,0);
	\draw [xshift=5.94cm](0.5,-1.22) -- (0,2);
	\draw [xshift=5.94cm](0.5,-1.22) -- (-2,0);
	\draw [xshift=5.94cm](0.5,-1.22) -- (0,-2);
	\draw [xshift=5.94cm](0.5,-1.22) -- (1.43,1.43);
%	\draw [xshift=7.96cm](-0.5,0.22) -- (-2.43,2.43);
	\draw [xshift=5.94cm](0.5,-1.22) -- (1.43,-1.43);
	\draw [xshift=5.94cm](0.5,-1.22) -- (-1.43,-1.43);
	%\draw (0.5,1.22) -- (-1.43,-1.43);
	\draw [xshift=5.94cm](0,0) circle [radius=2cm];
	\filldraw [xshift=5.94cm][black](0.5,-1.22) circle [radius=2pt];
	
	\filldraw [xshift=5.94cm][red] (2,0) circle [radius=2pt];
	\filldraw [xshift=5.94cm][blue] (0,2) circle [radius=2pt];
	\filldraw [xshift=5.94cm][red] (0,-2) circle [radius=2pt];
	\filldraw [xshift=5.94cm][blue] (-2,0) circle [radius=2pt];
	\filldraw [xshift=5.94cm][red] (1.43,1.43) circle [radius=2pt];
	\filldraw [xshift=5.94cm][blue] (-1.43,1.43) circle [radius=2pt];
	\filldraw [xshift=5.94cm][red] (1.43,-1.43) circle [radius=2pt];
	\filldraw [xshift=5.94cm][red] (-1.43,-1.43) circle [radius=2pt];
	
	\filldraw [xshift=5.94cm][black] (0,0) circle [radius=1.33pt];
	% %
	\draw [xshift=11.64cm](-0.5,-1.22) -- (2,0);
	\draw [xshift=11.64cm](-0.5,-1.22) -- (0,2);
	\draw [xshift=11.64cm](-0.5,-1.22) -- (-2,0);
	\draw [xshift=11.64cm](-0.5,-1.22) -- (0,-2);
%	\draw [xshift=8.3cm][yshift = -0.3cm](-0.5,-1.22) -- (1.41,1.43);
	\draw [xshift=11.64cm](-0.5,-1.22) -- (-1.43,1.43);
	\draw [xshift=11.64cm](-0.5,-1.22) -- (1.43,-1.43);
	\draw [xshift=11.64cm](-0.5,-1.22) -- (-1.43,-1.43);
	%\draw (0.5,1.22) -- (-1.43,-1.43);
	\draw [xshift=11.64cm](0,0) circle [radius=2cm];
	\filldraw [black] [xshift=11.64cm](-0.5,-1.22) circle [radius=2pt];
	
	\filldraw [xshift=11.64cm][blue] (2,0) circle [radius=2pt];
	\filldraw [xshift=11.64cm][blue] (0,2) circle [radius=2pt];
	\filldraw [xshift=11.64cm][red] (0,-2) circle [radius=2pt];
	\filldraw [xshift=11.64cm][red] (-2,0) circle [radius=2pt];
	\filldraw [xshift=11.64cm][blue] (1.43,1.43) circle [radius=2pt];
	\filldraw [xshift=11.64cm][blue] (-1.43,1.43) circle [radius=2pt];
	\filldraw [xshift=11.64cm][red] (1.43,-1.43) circle [radius=2pt];
	\filldraw [xshift=11.64cm][red] (-1.43,-1.43) circle [radius=2pt];
	
	\filldraw [xshift=11.64cm][black] (0,0) circle [radius=1.33pt];
%		\draw [xshift=8.3cm][yshift = -0.3cm][thick, red][->](-0.5,-1.22) -- (0.28,0.28);
	%\draw (0.48,1.5) node[text width = 1.33pt] {$\mathbf{x}_n$};
	
	% Second 
	\draw [xshift=8.82cm][yshift=-2.82cm](-0.5,1.22) -- (2,0);
	\draw [xshift=8.82cm][yshift=-2.82cm](-0.5,1.22) -- (0,2);
	\draw [xshift=8.82cm][yshift=-2.82cm](-0.5,1.22) -- (-2,0);
	\draw [xshift=8.82cm][yshift=-2.82cm](-0.5,1.22) -- (0,-2);
	\draw [xshift=8.82cm][yshift = -2.82cm](-0.5,1.22) -- (-1.41,-1.43);
	\draw [xshift=8.82cm][yshift=-2.82cm](-0.5,1.22) -- (-1.43,1.43);
%	\draw [xshift=-2.82cm][yshift=-2.82cm](0,1.22) -- (1.43,-1.43);
	\draw [xshift=8.82cm][yshift=-2.82cm](-0.5,1.22) -- (1.43,1.43);
	\draw [xshift=8.82cm][yshift=-2.82cm](-0,0) circle [radius=2cm];
	\filldraw [xshift=8.82cm][yshift=-2.82cm][black] (-0.5,1.22) circle [radius=2pt];
	
	\filldraw [xshift=8.82cm][yshift=-2.82cm][red] (2,0) circle [radius=2pt];
	\filldraw [xshift=8.82cm][yshift=-2.82cm][red] (0,2) circle [radius=2pt];
	\filldraw [xshift=8.82cm][yshift=-2.82cm][blue] (0,-2) circle [radius=2pt];
	\filldraw [xshift=8.82cm][yshift=-2.82cm][red] (-2,0) circle [radius=2pt];
	\filldraw [xshift=8.82cm][yshift=-2.82cm][red] (1.43,1.43) circle [radius=2pt];
	\filldraw [xshift=8.82cm][yshift=-2.82cm][red] (-1.43,1.43) circle [radius=2pt];
	\filldraw [xshift=8.82cm][yshift=-2.82cm][blue] (1.43,-1.43) circle [radius=2pt];
	\filldraw [xshift=8.82cm][yshift=-2.82cm][red] (-1.43,-1.43) circle [radius=2pt];
	
	\filldraw [xshift=8.82cm][yshift=-2.82cm][black] (0,0) circle [radius=1.33pt];
%		\draw [xshift=5.18cm][yshift = -2.82cm][thick, red][->](0.5,1.22) -- (-0.28,-0.28);
	
	\node[xshift=-8.17cm][yshift=-3cm] at (0,0) {\Large High RhoA};
	\filldraw[xshift=-10cm][yshift=-3cm] [red] (0,0) circle [radius=2pt];
			
	\node[xshift=-8.25cm][yshift=-4cm] at (0,0) {\Large High Rac1};
	\filldraw[xshift=-10cm][yshift=-4cm] [blue] (0,0) circle [radius=2pt];
	
	\draw [xshift=3cm][yshift=-2cm][thick,black][->](0,0)--(1,0);
	\end{tikzpicture}
	\caption{Schematic representation of CIL for non-binary collisions.}
	\label{figure: CIL schematics 2D}
\end{figure}
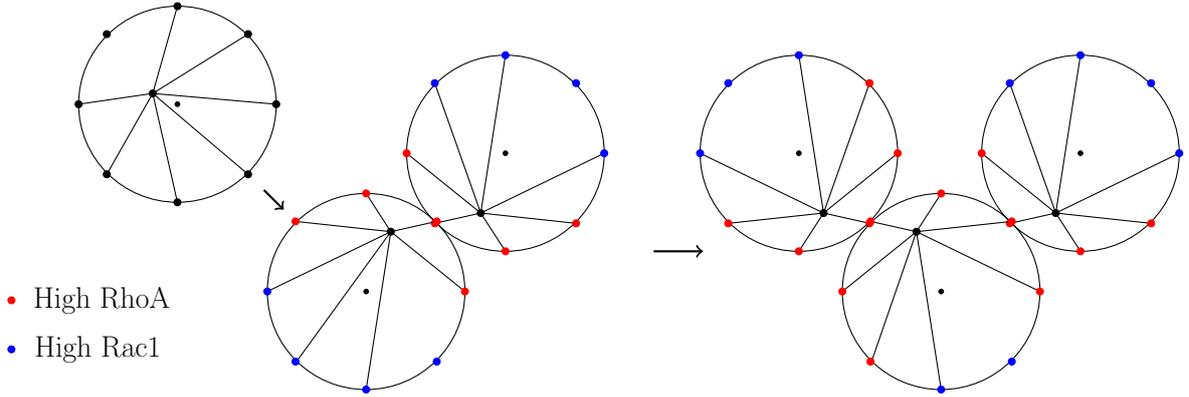

We now simulate our model in an unconfined 2D setting (see Appendix \ref{appendix: general CIL} for a general, non-binary system of cells) and investigate the effect of CIL on chemotaxing and non-chemotaxing cells. As was shown above, taking $\delta_-=0$ may lead to overlapping cells. Since in a general 2D setting a cell might have a contact with multiple cells at the same time (see Figure \ref{figure: CIL schematics 2D}), it is possible (for arbitrary values of $\delta_{myo},\delta_+$, and $\delta_-$) that multiple cells overlap each other. Note that cells undergoing CIL do not crawl on top of each other. Thus, for cells undergoing CIL we take $\delta_{myo}=0.4,\delta_{+}=0.2,\delta_{-}=1$, and for cells not exhibiting it we take $\delta_{myo}=\delta_{+}=\delta_{-}=0$.

We also explore the interplay between CIL and chemotaxis in a heterogeneous population of cells. Namely, we investigate the effect of CIL on a mix of cells responsive and non-responsive to an external cue. For chemotaxing cells we take $\delta_E=0.05$.

\begin{figure}[h]
\subfloat[]
{
	\includegraphics[width=43mm,height=39mm]{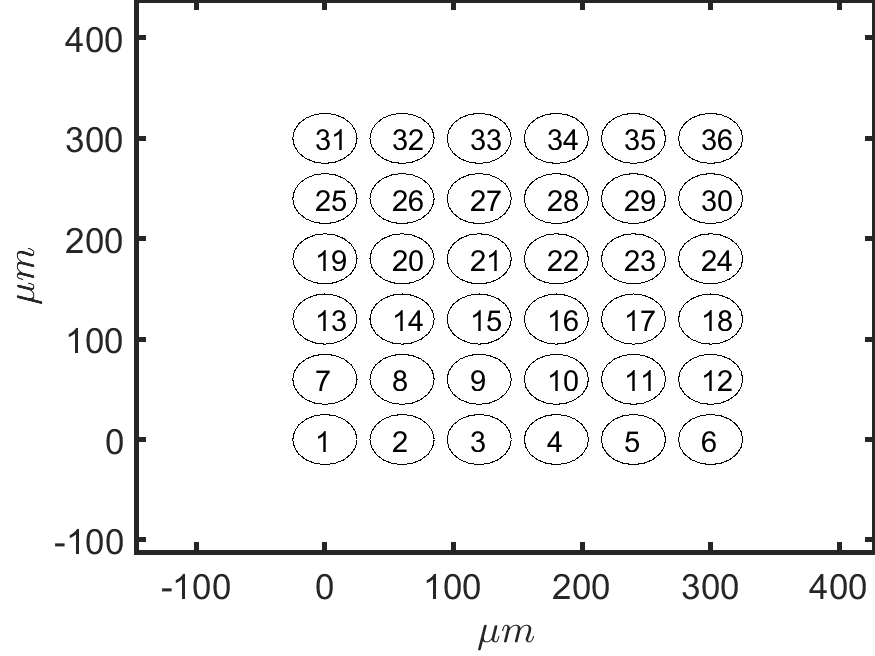}
}
\subfloat[]
{
	\includegraphics[width=43mm,height=39mm]{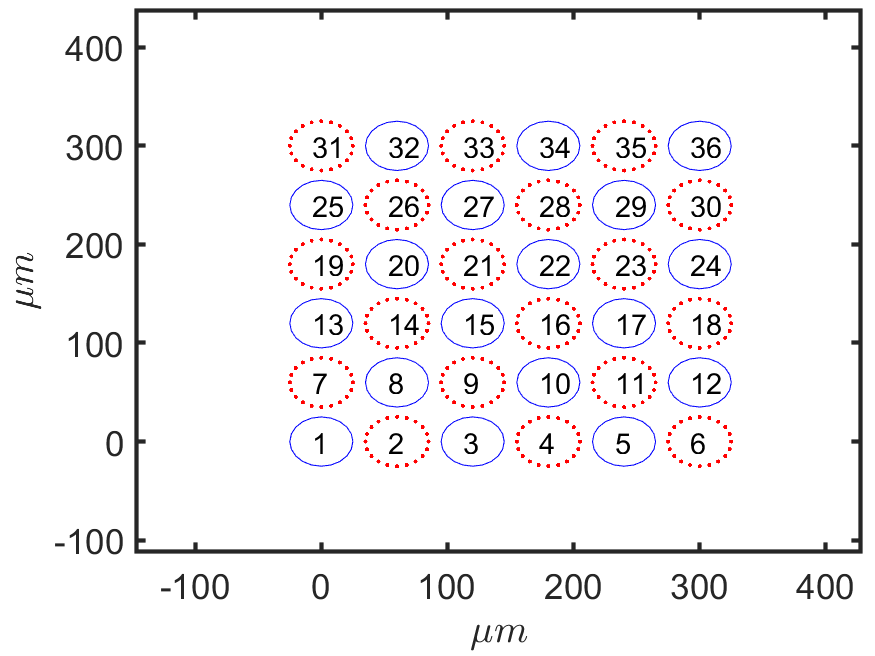}
}
\subfloat[]
{
	\includegraphics[width=43mm,height=39mm]{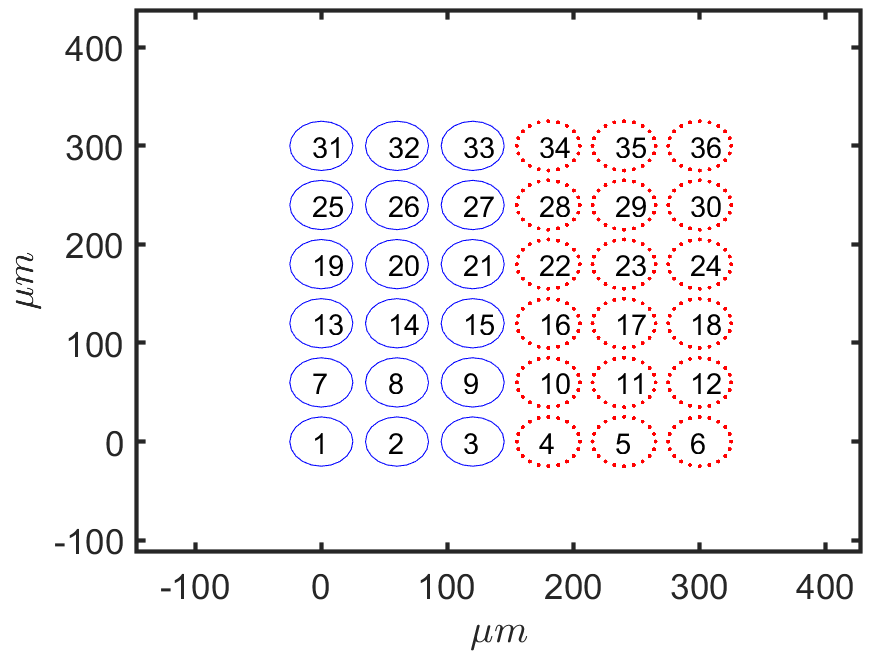}
}
\caption{Initial configuration of cells. (a) Homogeneous population. (b,c) Heterogeneous populations of chemotaxing (blue, solid) and non-chemotaxing (red, dotted) cells.}
\label{fig: initial positions populations}
\end{figure}

In the following, we simulate 36 cells and evolve them for 20 hours, such that initially the cells are positioned as in Figure \ref{fig: initial positions populations}, and the distance between the centroids of neighboring cells is $2.4R_{cell}$. 
%Here we explore what effect CIL has on cells responsive and non-responsive to an external cue. In particular, we would like to see if CIL promotes dispersion \cite{CARMONAFONTAINE20111026}, \cite{SCARPA2015421}, enhances chemotactic response \cite{theveneau2010collective}, or leads to chase-and-run behavior \cite{theveneau2013chase}. , and for cells exhibiting CIL we take $\delta_{myo}=0.4, \delta_+0.2$.    

\subsubsection{Homogeneous population}

\begin{figure}[H]
\subfloat[]
{
	\includegraphics[width=43mm,height=39mm]{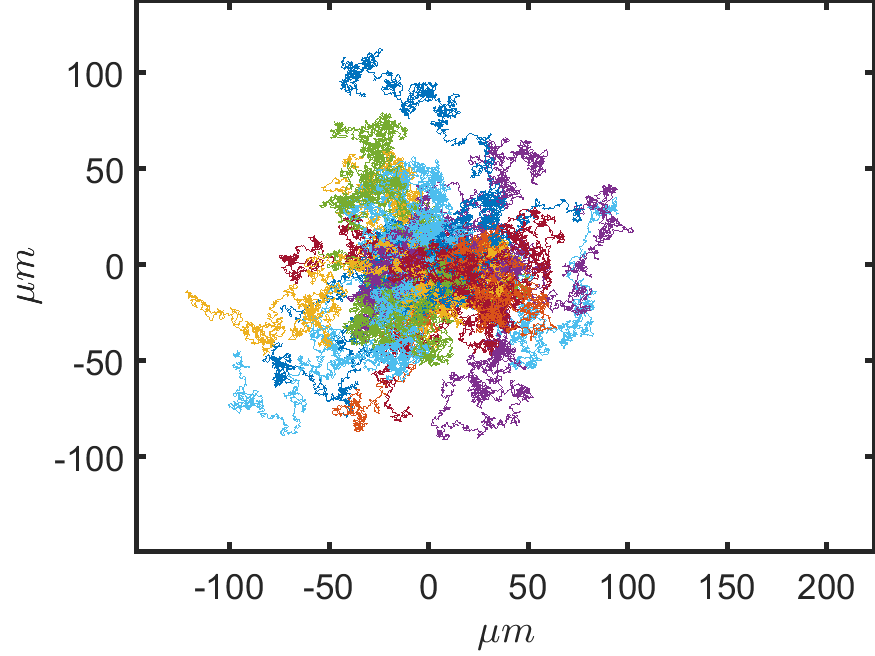}
}
\subfloat[]
{
	\includegraphics[width=43mm,height=39mm]{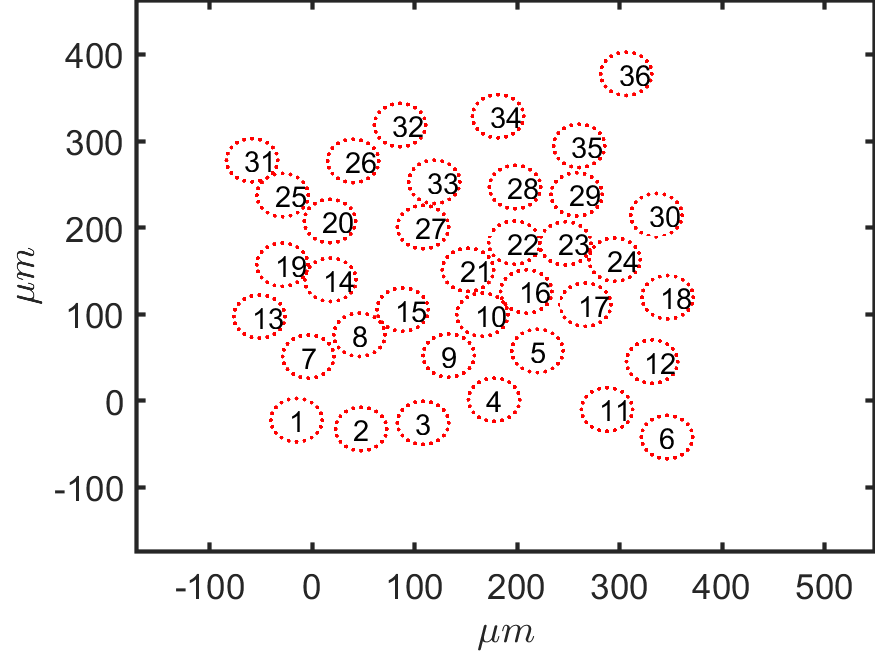}
}
\subfloat[]
{
	\includegraphics[width=43mm,height=39mm]{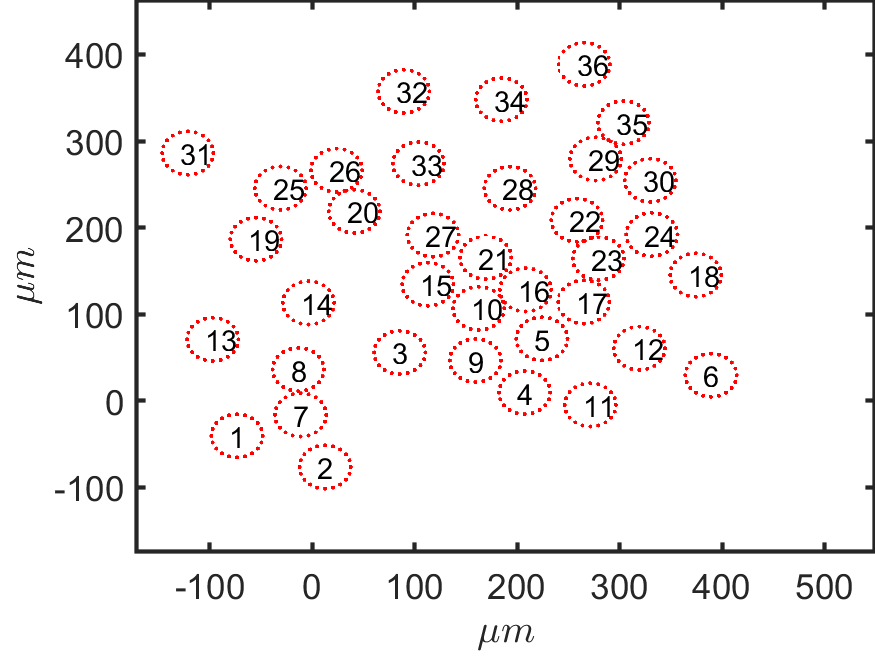}
}
\hfill
\subfloat[]
{
	\includegraphics[width=43mm,height=39mm]{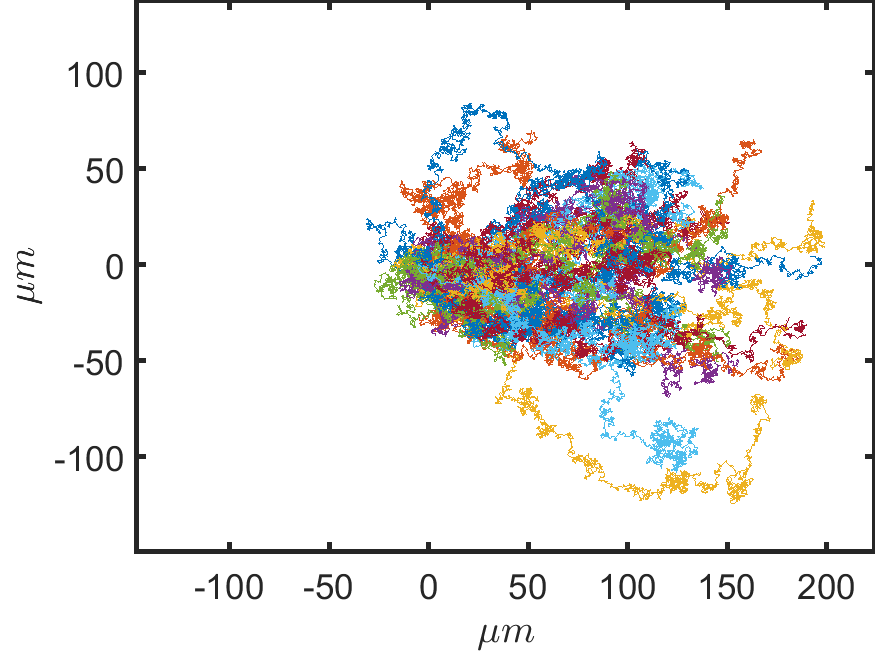}
}
\subfloat[]
{
	\includegraphics[width=43mm,height=39mm]{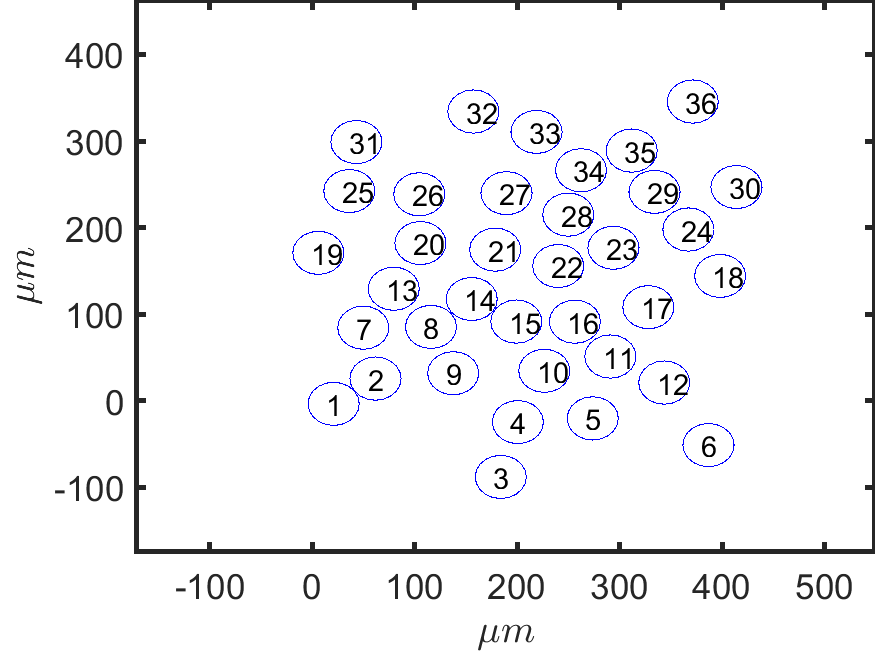}
}
\subfloat[]
{
	\includegraphics[width=43mm,height=39mm]{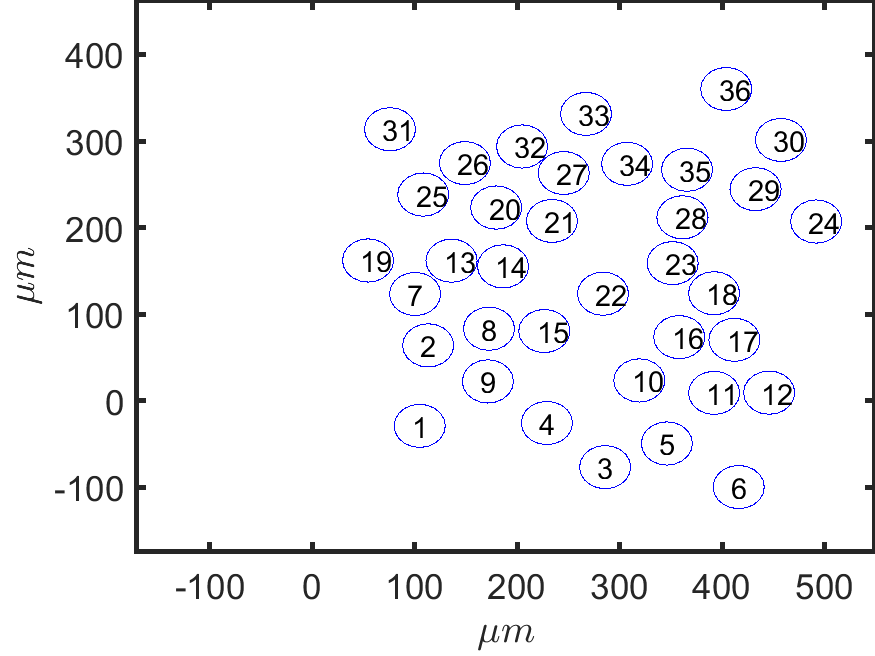}
}
\hfill
\subfloat[]
{
	\includegraphics[width=43mm,height=40mm]{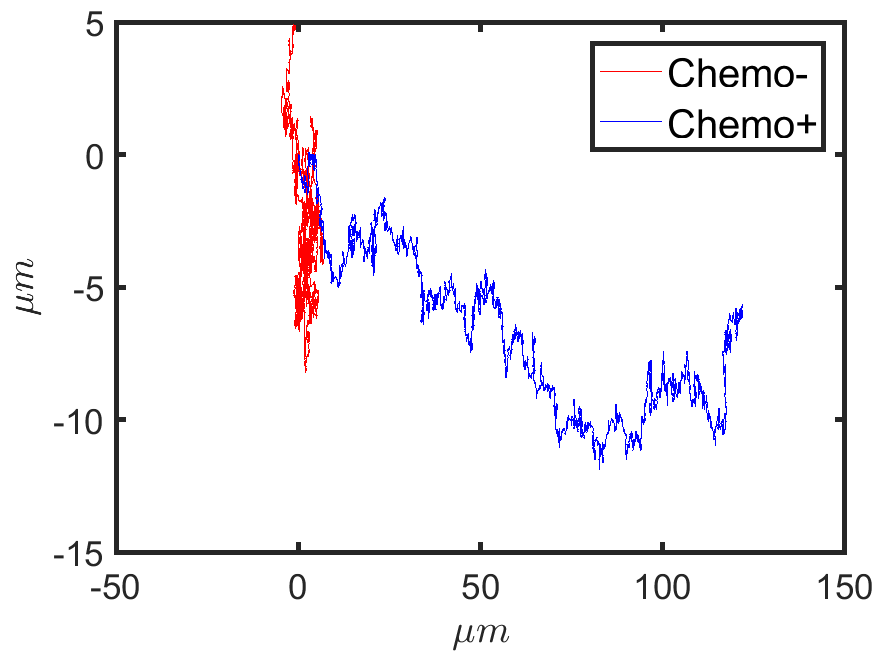}
}
\subfloat[]
{
	\includegraphics[width=43mm,height=39mm]{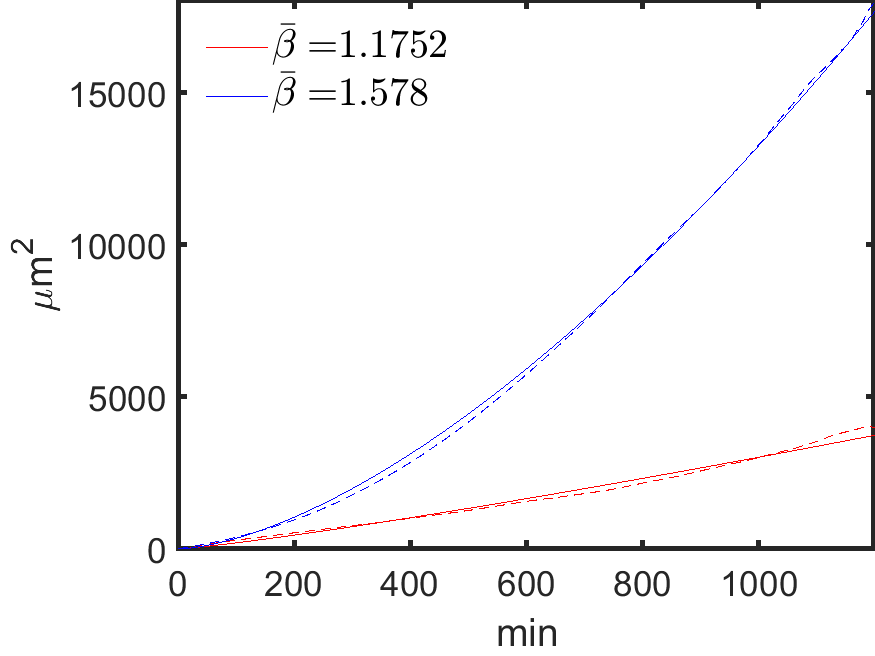}
}
%\subfloat[]
%{
%	\includegraphics[width=43mm,height=39mm]{images/vel_dirs_COM}
%}
%\subfloat[]
%{
%	\includegraphics[width=48mm,height=44mm]{images/vel_dirs_COM}
%}
\caption{Simulation results for a homogeneous population of non-chemotaxing (top row) and chemotaxing cells (middle row). (a,d) Centered trajectories. (b,e) and (c,f) Cell positions at $t=600min,1200min$, respectively. (g) Centered trajectories of the cluster centers of mass. (h) Mean-squared displacement (dashed) of chemotaxing (blue) and non-chemotaxing (red) cells, and the fit $\widehat{msd}(t)$ (solid).}
\label{fig: homogeneous 2D}
\end{figure}
Simulation results for a homogeneous population of chemotaxing and non-chemotaxing cells are shown in Figure \ref{fig: homogeneous 2D}. We see that the biased migration of chemotaxing cells occurs in a cluster-like manner. In contrast, we see that the non-chemotaxing cells disperse randomly, such that the center of mass deviates very little as compared to cell dimensions ($R_{cell}=25\mu m$). Note that the motion of randomly migrating cells exhibits a superdiffusive character (Figure \ref{fig: homogeneous 2D}h), as indicated by fitting the mean-squared displacement to the curve $\widehat{msd}(t)=\beta_0t^{\bar{\beta}}$ (see Appendix in \cite{2018arXiv181011435U} for details). In \cite{2018arXiv181011435U}, it was shown that non-interacting cells\footnote{But otherwise identical, as the parameter values are the same.} exhibit normal diffusive behavior ($\bar{\beta}\sim 1$) in the absence of any source of asymmetry affecting FA dynamics. Here, since the exponent $\bar{\beta}$  corresponding to non-chemotaxing cells is larger than one, we see that cell-cell collisions also lead to anomalous diffusion as $\bar{\beta}>1$. Comparing chemotaxing cells, we also see that $\bar{\beta}$ increases if cells collide with one another (in \cite{2018arXiv181011435U} $\bar{\beta}=1.34$ for the same value of $\delta_E$). Thus, we see that the average displacement increases due to CIL, despite the fact that motion ceases upon contact. 

It has been hypothesized that superdiffusive motion is optimal for searching a target source, that itself diffuses \cite{PhysRevLett.88.097901}, \cite{superdiffusionFood}. Thus, cancer cells that acquire ability to undergo homotypic CIL can find a diffusing source (e.g. VEGF) more efficiently and hence facilitate tumor progression. Interestingly, it has also been hypothesized that homotypic CIL facilitates dispersion of cancer cells \cite{MAYOR2010319}, \cite{stramer2017mechanisms}.    
\subsection{Inhomogeneous populations}
We now explore the effects of heterotypic CIL between populations of chemotaxing and non-chemotaxing cells (Figure \ref{fig: initial positions populations}b,c). Here, cells always exhibit CIL when they collide with the members of the same group (see Appendix \ref{appendix: homo and hetero CIL}). 

\begin{figure}[H]
\subfloat[]
{
	\includegraphics[width=43mm,height=39mm]{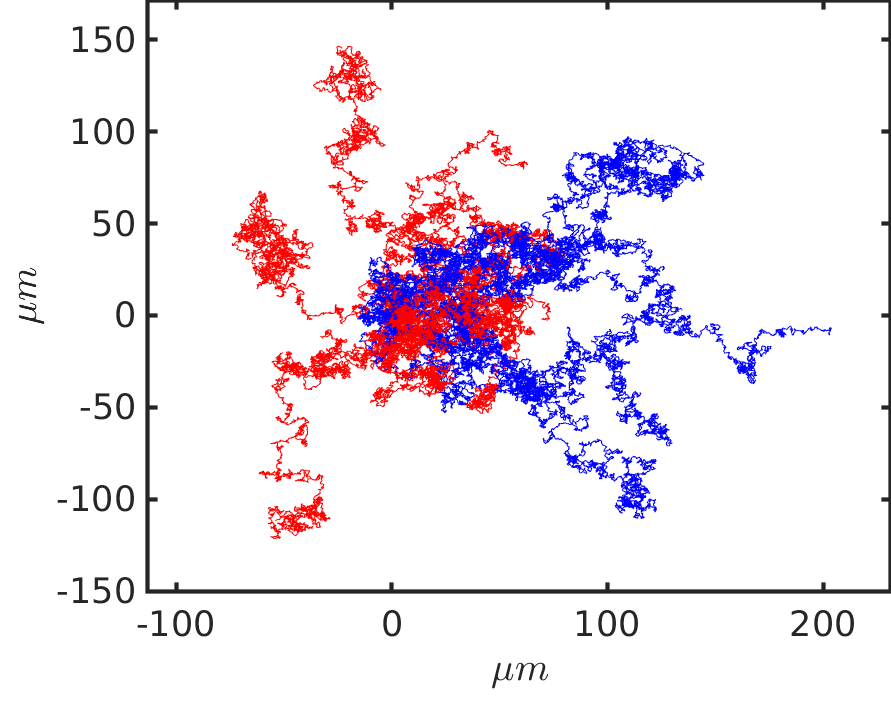}
}
\subfloat[]
{
	\includegraphics[width=43mm,height=39mm]{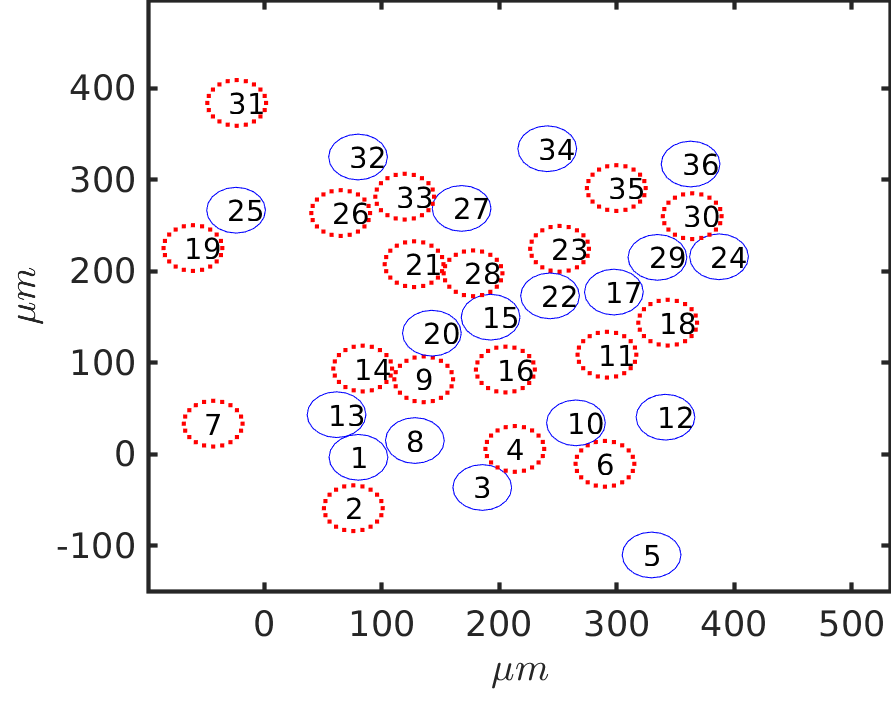}
}
\subfloat[]
{
	\includegraphics[width=43mm,height=39mm]{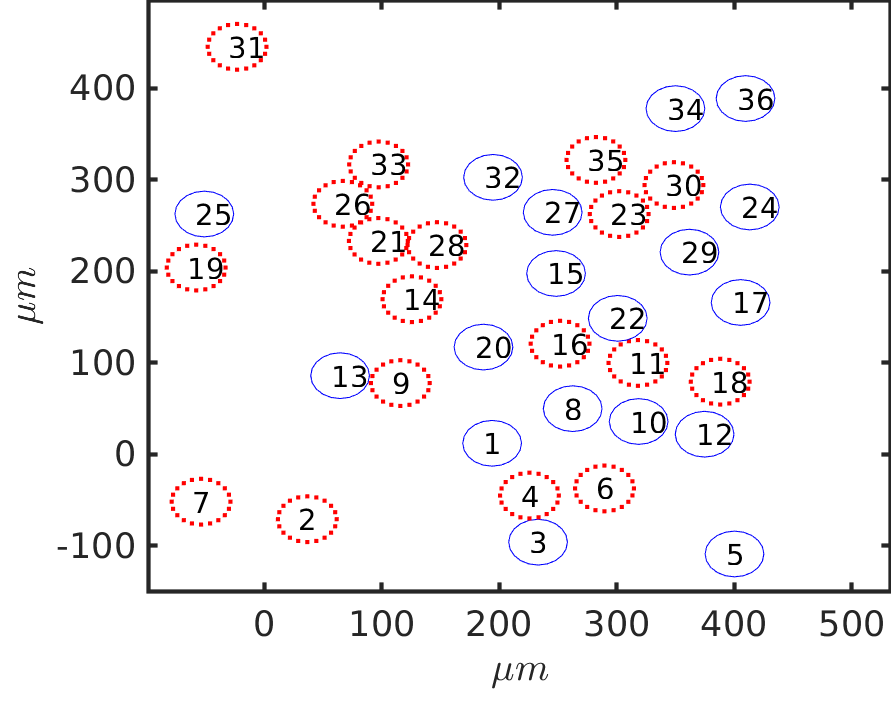}
}
\hfill
\subfloat[]
{
	\includegraphics[width=43mm,height=39mm]{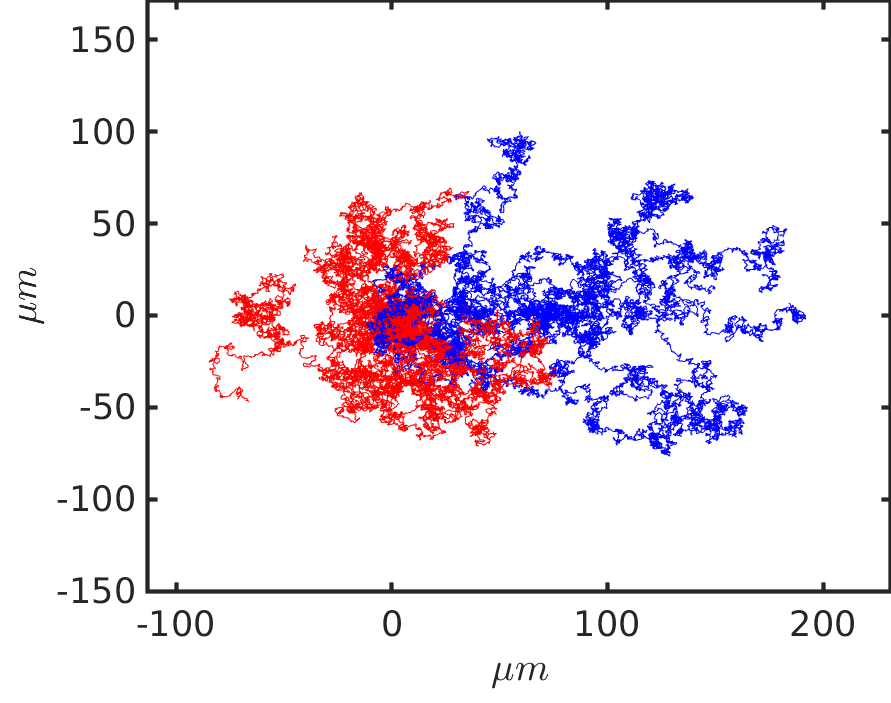}
}
\subfloat[]
{
	\includegraphics[width=43mm,height=39mm]{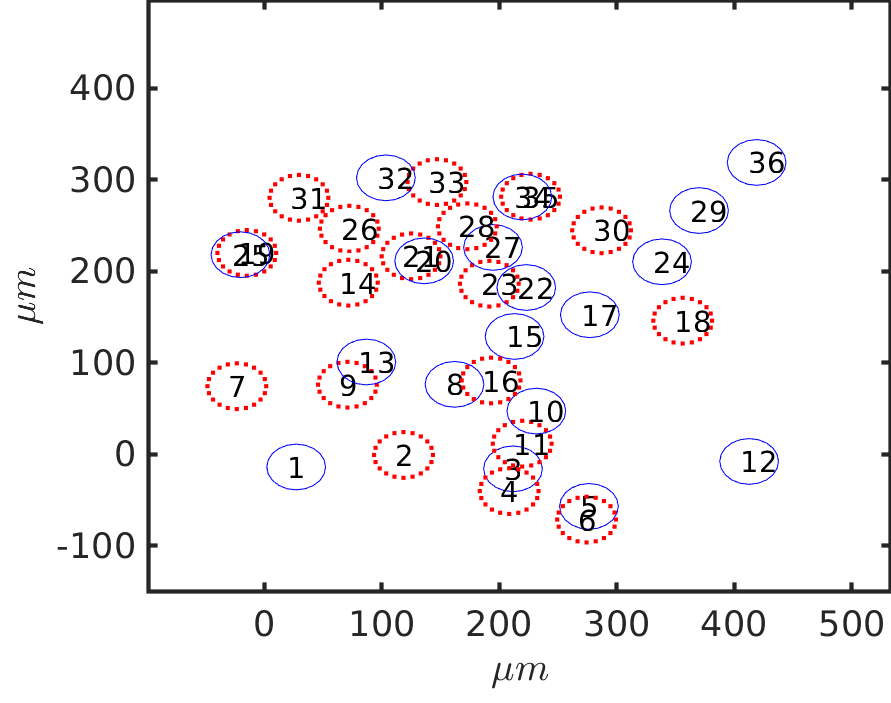}
}
\subfloat[]
{
	\includegraphics[width=43mm,height=39mm]{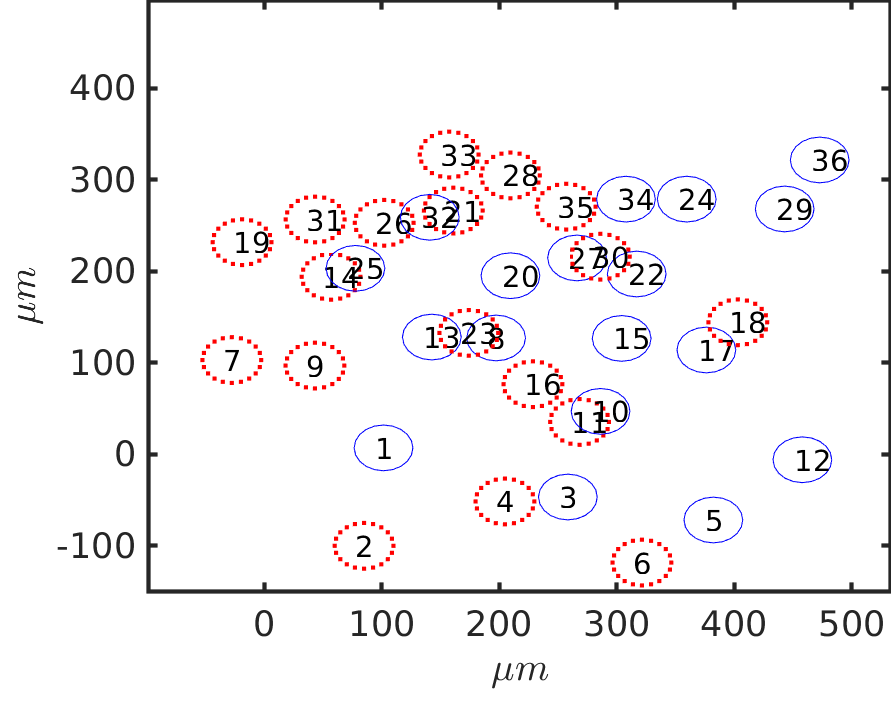}
}
\hfill
\subfloat[]
{
	\includegraphics[width=43mm,height=39mm]{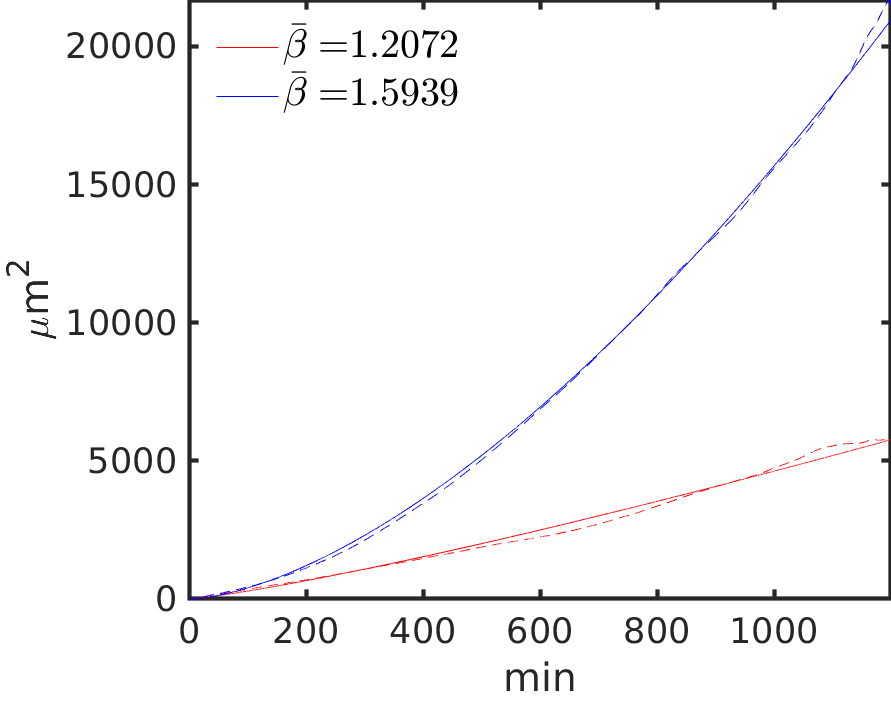}
}
\subfloat[]
{
	\includegraphics[width=43mm,height=39mm]{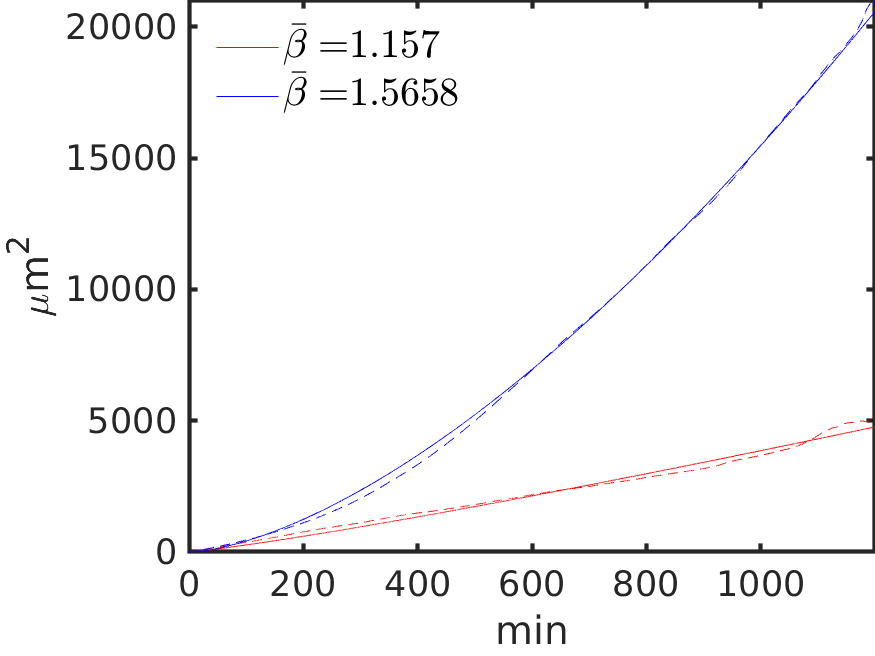}
}
%\vspace{-5mm}
%\subfloat[]
%{
%	\includegraphics[width=40mm,height=39mm]{images/vel_dirs_COM_mixing}
%}
%\subfloat[]
%{
%	\includegraphics[width=48mm,height=44mm]{images/vel_dirs_COM}
%}
\caption{Simulation results for the mixed population with (top row) and without (middle row) heterogeneous CIL. Initially, cells are positioned as in Figure \ref{fig: initial positions populations}b. (a,d) Centered trajectories of 9 chemotaxing (blue) and non-chemotaxing (red) cells. (b,e) and (c,f) Positions of chemotaxing (blue, solid) and non-chemotaxing (red, dotted) cells at $t=600min$ in (b,e) and at $t=1200min$ in (c,f). (g,h) Mean squared displacements of chemotaxing (blue) and non-chemotaxing (red) cells with (g) and without heterotypic CIL (h).}
\label{fig: mixing}
\end{figure}

\begin{figure}[H]
\subfloat[]
{
	\includegraphics[width=43mm,height=39mm]{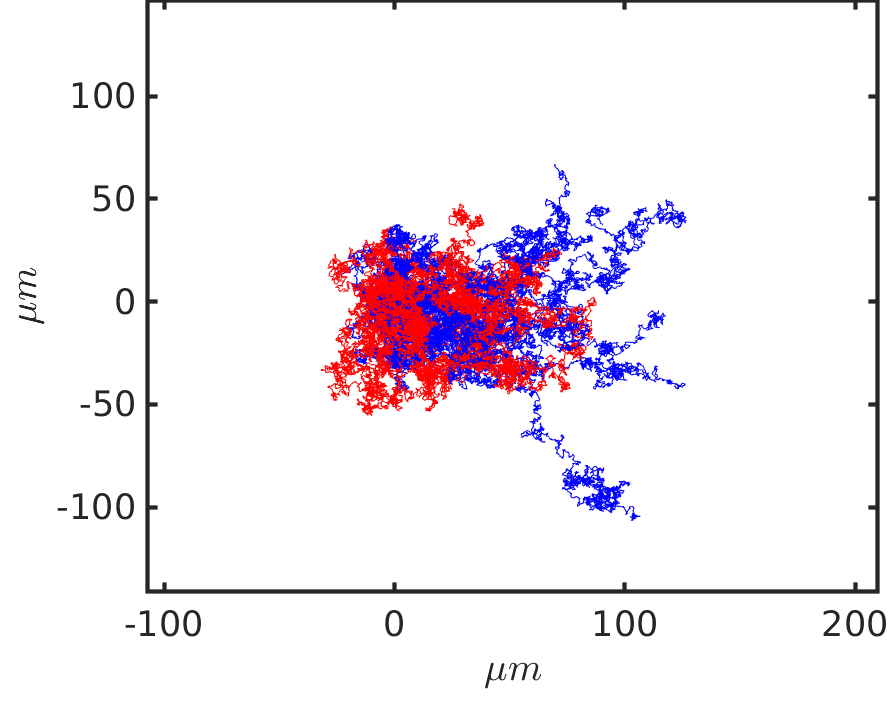}
}
\subfloat[]
{
	\includegraphics[width=43mm,height=39mm]{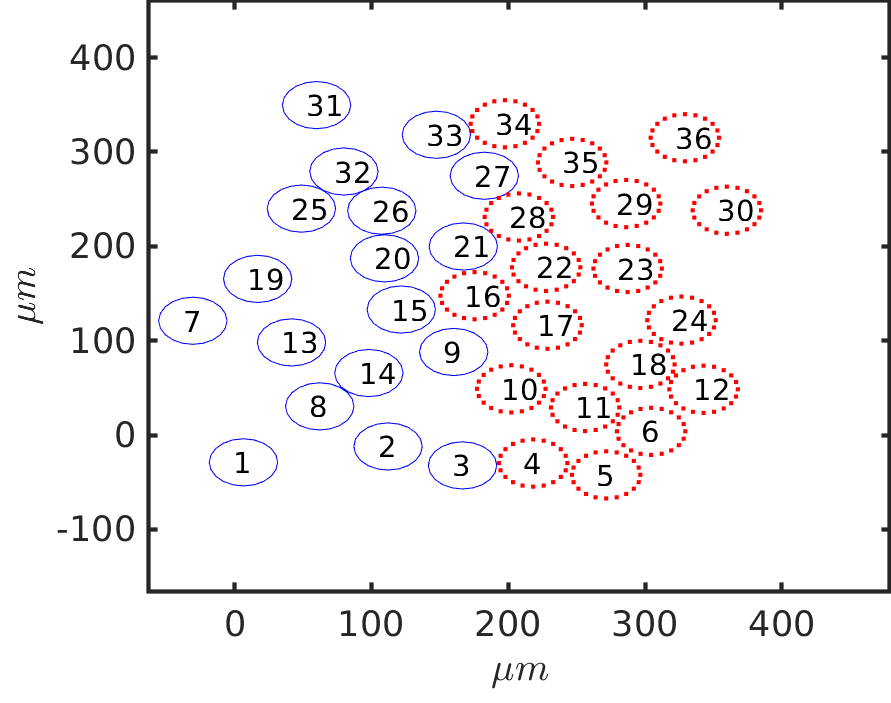}
}
\subfloat[]
{
	\includegraphics[width=43mm,height=39mm]{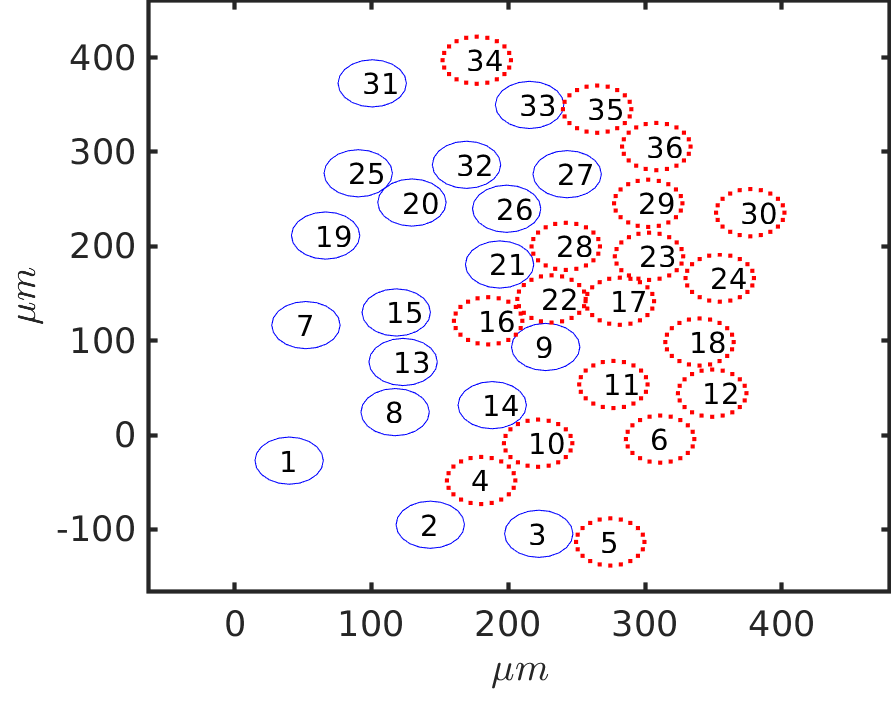}
}
\hfill
\subfloat[]
{
	\includegraphics[width=43mm,height=39mm]{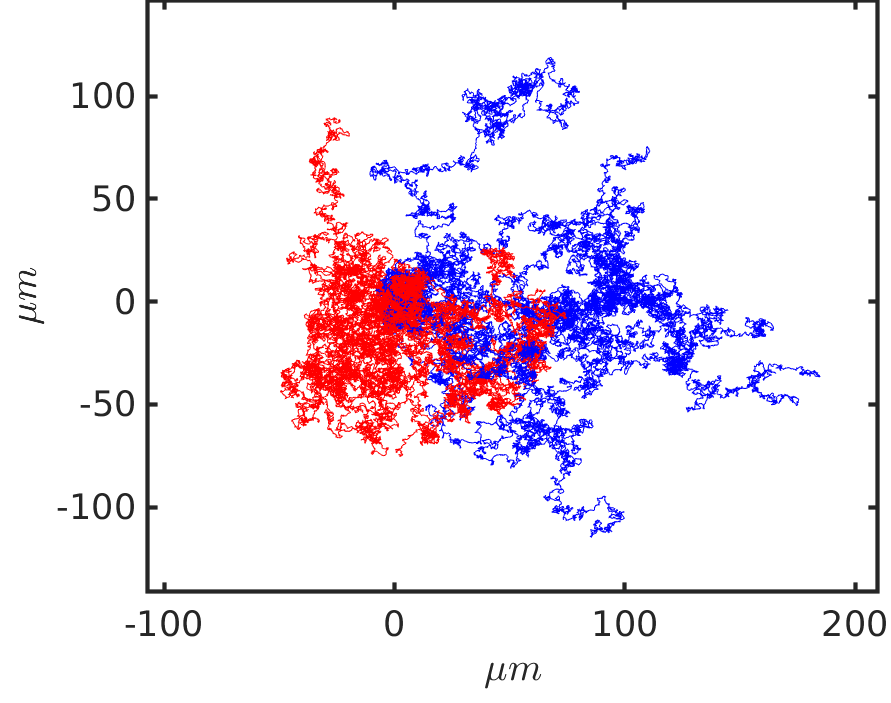}
}
\subfloat[]
{
	\includegraphics[width=43mm,height=39mm]{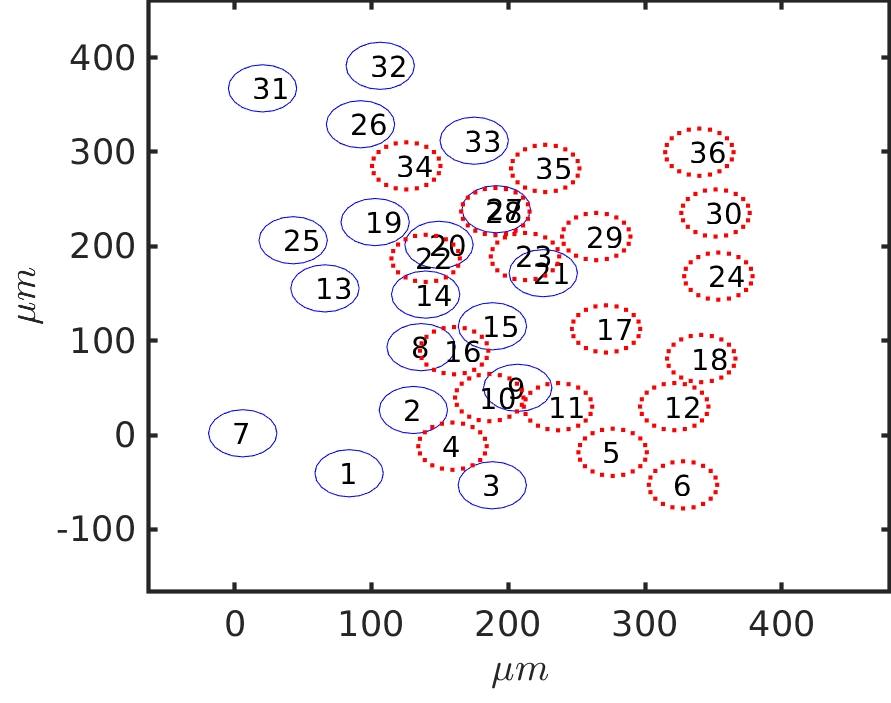}
}
\subfloat[]
{
	\includegraphics[width=43mm,height=39mm]{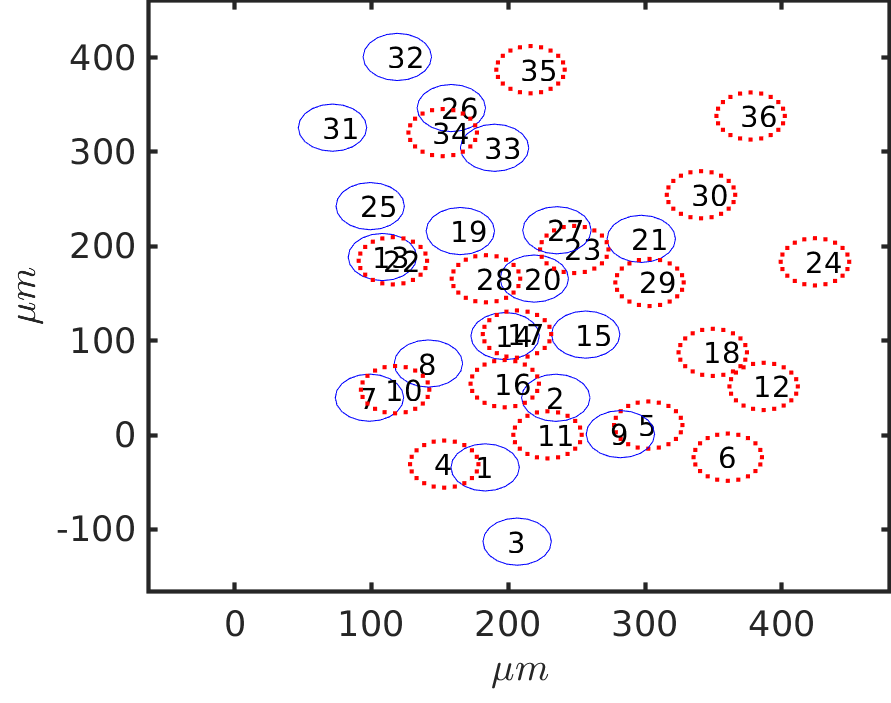}
}
\hfill
\subfloat[]
{
	\includegraphics[width=35mm,height=35mm]{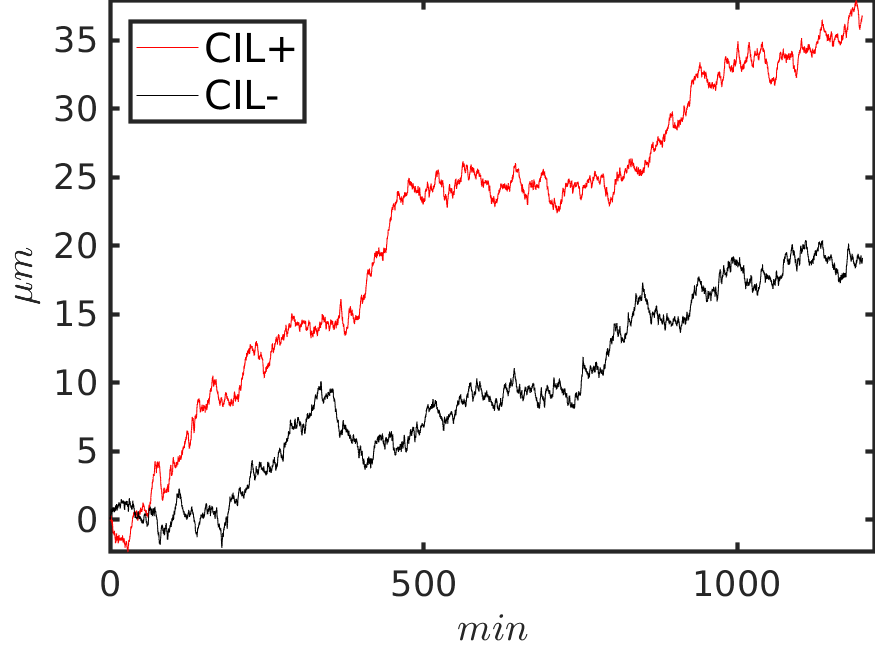}
}
\subfloat[]
{
	\includegraphics[width=35mm,height=35mm]{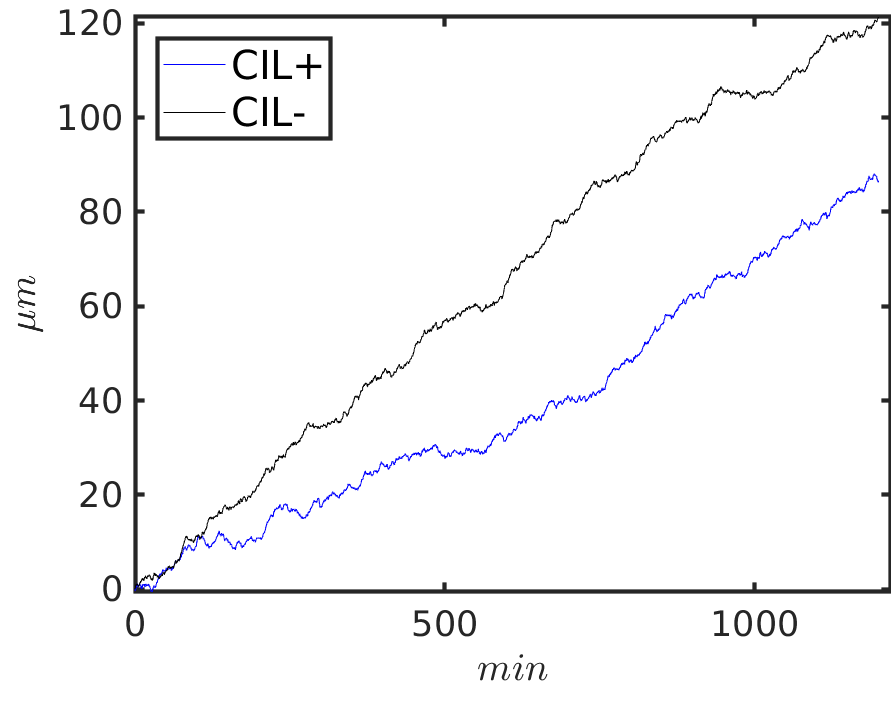}
}
\hfill
\subfloat[]
{
	\includegraphics[width=35mm,height=35mm]{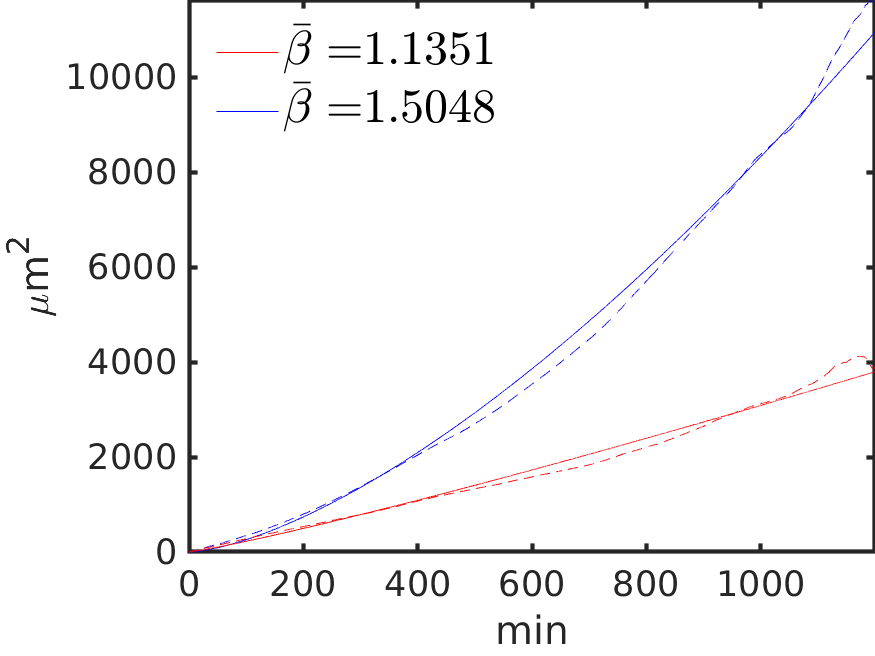}
}
\subfloat[]
{
	\includegraphics[width=35mm,height=35mm]{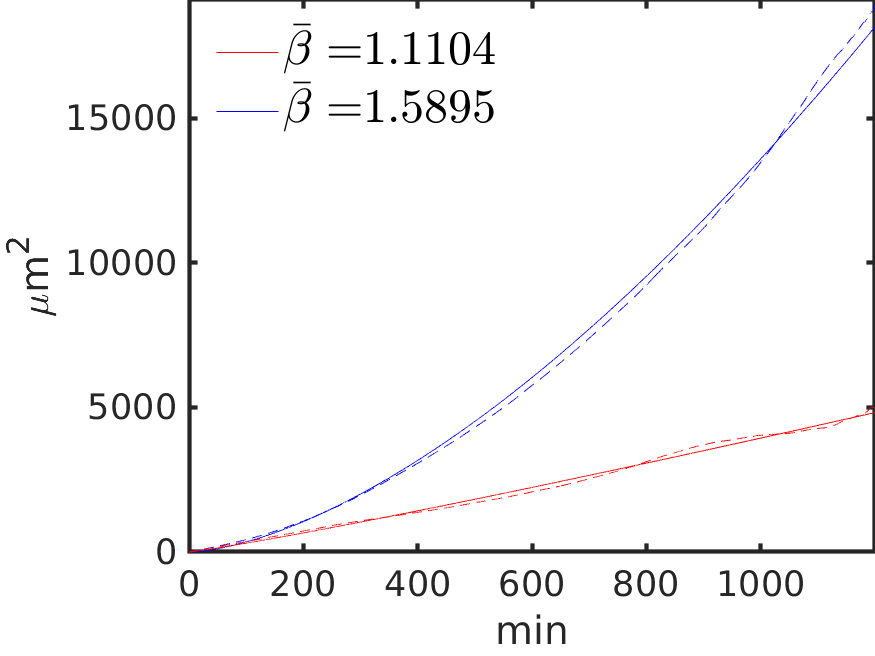}
}
%\vspace{-5mm}
%\subfloat[]
%{
%	\includegraphics[width=40mm,height=39mm]{images/vel_dirs_COM_mixing}
%}
%\subfloat[]
%{
%	\includegraphics[width=48mm,height=44mm]{images/vel_dirs_COM}
%}
\caption{Simulation results for the separated population with (top row) and without (middle row) heterogeneous CIL. Initially, cells are position as in Figure \ref{fig: initial positions populations}c. (a,d) Centered trajectories of 9 chemotaxing (blue) and non-chemotaxing (red) cells. (b,e) and (c,f) Positions of chemotaxing (blue, solid) and non-chemotaxing (red, dotted) cells at $t=600min$ in (b,e) and at $t=1200min$ in (c,f). (g,h) $x$ components of non-chemotaxing (g) and chemotaxing (h) cells' centers of mass with (colored) and without (black) heterotypic CIL. (i,j) Mean squared displacements of chemotaxing (blue) and non-chemotaxing (red) cells with (i) and without (j) heterotypic CIL. }
\label{fig: chase}
\end{figure}

When evenly mixed (Figure \ref{fig: initial positions populations}b), we see that heterotypic CIL does not have a significant impact on chemotaxing or non-chemotaxing cells (Figure \ref{fig: mixing}), as the behavior of each subgroup resembles the case with the corresponding homogeneous populations. This suggests that in a disordered population of cells, homotypic, but not heterotypic CIL facilitates directed migration of cells (as $\bar{\beta}\approx1.34$ in freely chemotaxing cells \cite{2018arXiv181011435U}). Nevertheless, notice that in this unclustered configuration, the chemotaxing cells are able to push their way out, leading to dispersion of the surrounding cells akin to billiard balls (Figure \ref{fig: mixing}a,d): centered trajectories of the non-responsive cells show higher dispersion due to the repulsive interaction with the chemotaxing cells, who must push out the non-responsive cells to achieve the observed directed migration when such interaction is present. Clustering cells according to their responsiveness to an external cue, however, lead to a qualitatively different outcome. If responsive and non-responsive cells are separated as in Figure \ref{fig: initial positions populations}c, we see a cluster-like interaction when heterotypic CIL is present (Figure \ref{fig: chase}): the dividing line between the groups remains discernible for a long time (Figure \ref{fig: chase}b,c), which is not the case when the heterotypic CIL is absent (Figure \ref{fig: chase}e,f). This indicates that the initial clustering (Figure \ref{fig: initial positions populations}c) is conserved due to heterotypic interaction. Unlike the case of evenly mixed cells, we see that the dispersion of the non-chemotaxing cells is not as prominent (Figure \ref{fig: mixing}a vs. Figure \ref{fig: chase}a), and the chemotaxing cells do not push out the non-responsive ones. In fact, we observe that the latter are being displaced in a sheet-like manner by the responsive cells. A similar behavior was observed in \cite{theveneau2013chase}, although in that study the non-chemotaxing cells were themselves the source of a chemoattractant.

Such displacement induces the non-chemotaxing cells to align their motion with the direction of an external cue (Figure \ref{fig: chase}g), although the effect of heterotypic CIL is slight. On the other hand, we see that directed migration of the chemotaxing cells is impeded (Figure \ref{fig: chase}h), which is also reflected in the reduced average displacement (Figure \ref{fig: chase}i). Altogether, these results suggest that the role of heterotypic CIL varies with the distribution of the cell population: it may either facilitate dispersion (Figure \ref{fig: mixing}) or induce directed motion in otherwise randomly migrating cells (Figure \ref{fig: chase}). Its loss, however, is beneficial for tactic migration irrespective of spatial configuration.

%\FloatBarrier
\section{Discussion and Outlook}\label{section: discussion and outlook}
In this paper we extended the single cell migration model from \cite{2018arXiv181011435U} to account for contact inhibition of locomotion arising as a result of cell-cell collisions. Here, the cells, exhibiting CIL response, alter cell-substrate adhesions dynamics and SF contractility following contact with another cell. Mathematically, the model is described by a piecewise deterministic process, whereby collisions occur when some deterministic components (cell-cell distances) reach a corresponding value, and cell motility itself emerges due to mechanochemically mediated stochastic adhesion dynamics. Consequently, the outcome of a collision is also determined stochastically, as reported in \cite{Desai20130717}, \cite{lin2015interplay}, \cite{Scarpa901}.   

Mimicking the experimental setup in \cite{lin2015interplay}, we simulated binary collisions between cells migrating confined to a 1D lane. In this setting, we did not invoke the volume exclusion principle, and showed that a CIL response can be explained solely due to increased cell-substrate adhesion away from the collision site and increased actomyosin contractility in its vicinity. Although cell overlaps occur, we see that by strengthening the CIL response we can reduce its occurrence (Figure \ref{fig: c_distances uniform}b). Our results also show that an external cue can modulate CIL response, in line with \cite{lin2015interplay}. Specifically, typical CIL response can be overridden by chemotaxis (Figure \ref{fig: results taxis}) if post collision velocity is not aligned with the chemotactic gradient.

In an unconfined setting, we simulated the effects of homo- and heterotypic CIL. We found that homotypic CIL leads to increased cell displacement of chemotaxing and non-chemotaxing cells (Figure \ref{fig: homogeneous 2D}h). We also found that the spatial configuration of heterogeneous cells can have an impact on how heterotypic CIL affects migration of cells. In a disordered population it can facilitate the dispersion of randomly migrating cells (Figure \ref{fig: mixing}), while letting directed migration to be unhindered. When separated into groups, our simulations suggest that directed movement can be induced in non-chemotaxing cells (Figure \ref{fig: chase}), as reported in \cite{theveneau2010collective}. Altogether, simulations in the unconfined setting suggest that homotypic, but not heterotypic CIL, is advantageous for dispersive and invasive migration of cells. It has been speculated that such CIL behavior is responsible for the initial spread of cancer cells \cite{MAYOR2010319}, \cite{stramer2017mechanisms}. 

Guided by the study in \cite{Desai20130717}, we assumed that CIL response between two cells is transient and independent of each other. That is, immediately following the collision, cell dynamics and FA event probabilities in both cells are decoupled. However, there is evidence that a mechanical coupling is established prior to repulsion \cite{Stramer681}. Moreover, some cells exhibiting homotypic CIL tend to disperse and reaggregate into small clusters, which increases their chemotactic efficiency \cite{theveneau2010collective}. Thus, addressing mechanical coupling by including cell-cell adhesions represents one of the avenues for future work, whereby collective migration could be investigated further. 

%Encouraged by the fact the essential features of CIL fit well within the modeling framework in \cite{2018arXiv181011435U}, addressing cell-cell adhesions will also allow us to model collective movement of a cell cluster.   

\section*{Acknowledgement}
The author acknowledges support of the German Academic Exchange Service (DAAD).
\clearpage
\begin{appendices}
\section{General CIL model}\label{appendix: general CIL}
In order to construct a motility model with $N$ colliding cells, we proceed as in \cite{2018arXiv181011435U}. In particular, we first provide a formal derivation of the survival function for the next event time and the distribution of the next event index for $N$ cells (the special case of which is given in (\ref{eq: survival collisions})-(\ref{eq: event collisions})). Then we formulate our model as a piecewise deterministic Markov process (see \cite{davis93} for a comprehensive treatment), similarly as in \cite{2018arXiv181011435U}, but including collisions.
\subsection{Preliminaries}\label{appendix: general formulas}
Let $N$ be the number of cells and let $\mu^i,\mathbf{Y}^i$, $i=1,\ldots,N$, be defined as in Section \ref{section: CIL model}, and let $\boldsymbol{\mu}=(\mu^1,\ldots,\mu^N)$. Let $\mathbf{C}^i(t)\in\lbrace0,1\rbrace^N$ denote the collision state of cell $i$ at time $t$ with other cells:
\begin{align*}
C^i_k = 
\begin{cases*}
1, \text{if cell $i$ is in collision state with cell $k$}\\
0, \text{else},
\end{cases*}
\end{align*}
where $k=1,\ldots,N$ and we assume that $C^i_i=0$. Let $\mathbf{\Phi}^i(t)\in[0,2\pi)^N$ denote the vector of collision angles of cell $i$ with other cells, such that $\Phi^i_i(t)=0$. For $N=2$ in Section \ref{section: CIL model}, for example, we have $\mathbf{C}^1(\mathcal{T}_k)=(0,1)$ and $\mathbf{\Phi}^2(\mathcal{T}_k)=(\pi,0)$. For ease of notation, let $\mathbf{X}^i=(\mathbf{x}^i,\mathbf{x}^i_n,\theta^i,\mathbf{\Phi}^i,\mathbf{d}^i)$, where $\mathbf{d}^i$ is defined in \eqref{eq: definition dik}, and  $\mathbf{A}=(\mathbf{A}^1,\ldots,\mathbf{A}^N)$ for $\mathbf{A}\in\lbrace\mathbf{Y},\mathbf{C},\mathbf{X}\rbrace$. 
%Let $\boldsymbol{\mu}\in\lbrace0,1\rbrace^N$ denote the motility

Since there are $N$ cells and $2M$ possible reactions for each cell (binding and unbinding of an FA), then there are $2MN$ possible reactions among all cells. Let $a_{j'}(\bar{\boldsymbol{\mu}},\mathbf{y},\mathbf{c},\mathbf{X}(t))dt$ be the probability, given $\mathbf{X}(t)$ and $\mathbf{A}(t)=\mathbf{a}$, for $\mathbf{A}\in\lbrace\boldsymbol{\mu},\mathbf{Y},\mathbf{C}\rbrace$ and $\mathbf{a}\in\lbrace\bar{\boldsymbol{\mu}},\mathbf{y},\mathbf{c}\rbrace$, that a reaction $j'=1,\ldots,2MN$ will occur in the time interval $[t,t+dt)$.

Finally, let $K_{time}(\tau|t,\bar{\boldsymbol{\mu}},\mathbf{y},\mathbf{c},\mathbf{X}(t))d\tau$ be the probability that a reaction occurs in the time interval $[t+\tau,t+\tau+d\tau)$ and let $K_{index}(j'|t,\tau,\bar{\boldsymbol{\mu}},\mathbf{y},\mathbf{c},\mathbf{X}(t))$ be the probability of reaction $j'$, given that it occurs at time $t+\tau$. Then, we have \cite{2018arXiv181011435U}:
\begin{align}
K_{time}&(\tau|t,\bar{\boldsymbol{\mu}},\mathbf{y},\mathbf{c},\mathbf{X}(t)) \nonumber\\
&=a^N_0(\bar{\boldsymbol{\mu}},\mathbf{y},\mathbf{c},\mathbf{X}(t+\tau))\exp\left(-\int_{t}^{t+\tau}a^N_0(\bar{\boldsymbol{\mu}},\mathbf{y},\mathbf{c},\mathbf{X}(t+\tau'))\tau'\right)
\end{align}
and
\begin{align}
K_{index}(j'|t,\tau,\bar{\boldsymbol{\mu}},\mathbf{y},\mathbf{c},\mathbf{X}(t))=\frac{a_{j'}(\bar{\boldsymbol{\mu}},\mathbf{y},\mathbf{c},\mathbf{X}(t+\tau))}{a^N_0(\bar{\boldsymbol{\mu}},\mathbf{y},\mathbf{c},\mathbf{X}(t+\tau))},
\end{align} 
where $a^N_0 = \sum_{j'=1}^{2MN}a_{j'}$. Here, we adopt the following convention:
\begin{itemize}
\item A reaction $j'$ occurs in cell $i$ if $i=\floor{\frac{j'-1}{2M}}+1$.
\item A reaction $j'$ corresponds to a binding reaction of $j^{\text{th}}$ FA if $2j-1= j'modN$, and to unbinding reaction of $j^{\text{th}}$ FA if $2j=j'modN$.
\end{itemize}
Thus, $a^{+,i}_j=a_{2j-1+2M(i-1)}$ and $a^{-,i}_j=a_{2j+2M(i-1)}$ correspond, respectively, to binding and unbinding probability rates of the $j^{\text{th}}$ FA of cell $i$. For an example utilizing the above, see the special case with $N=2$ in Section \ref{section: CIL model}.
\subsection{PDMP formulation}\label{appendix: pdmp formulation}
Let $A:=\lbrace1,\ldots,2^{N+MN+N^2}\rbrace$ and let $\boldsymbol{\alpha}:A\rightarrow\lbrace0,1\rbrace^N\times\lbrace0,1\rbrace^{MN}\times\lbrace0,1\rbrace^{N^2}$ be a bijection. This is a mapping such that $\boldsymbol{\alpha}(\nu) = (\boldsymbol{\mu},\mathbf{Y},\mathbf{C})$ corresponds to motility, FA, and collision states of $N$ cells.

Let $\nu\in A$ and $\boldsymbol{\alpha}(\nu)= (\boldsymbol{\mu},\mathbf{Y},\mathbf{C})$. Let $D^{i,k}_\nu\subset\mathbb{R}$, $D^i_\nu \subset\mathbb{R}^N$, $D_\nu \subset \mathbb{R}^{N^2}$ be defined as:
\begin{align}\label{eq: definition Dik}
D^{i,k}_\nu&:=
\begin{cases*}
(-\infty,1],\phantom{aaBb} \text{if $C^i_k=0$ and $i,k\in I$}\\
(-\infty,\infty), \phantom{aab} \text{else}
\end{cases*},\\
D^{i}_\nu &:= \prod_{k=1}^{N}D^{i,k}_\nu,\phantom{D} D_\nu := \prod_{i=1}^ND^i_\nu,
\end{align}
where $I\subset\lbrace1,\ldots,N\rbrace$ is the index set of cells exhibiting CIL.
Let $\mathbf{d}^i:[0,\infty)\rightarrow D^i_\nu$, $i=1,\ldots,N$ be defined as:
\begin{align}\label{eq: definition dik}
d^i_k(t) := \exp\left(2R_{cell}^2-\frac{1}{2}\lVert\mathbf{x}^i(t)-\mathbf{x}^k(t)\rVert^2\right),\phantom{a} k=1,\ldots,N,
\end{align}
and let $\mathbf{d}:=\left(\mathbf{d}^1,\ldots,\mathbf{d}^N\right)\in D_\nu$. This particular form of $d^i_k$ is chosen since it satisfies the following requirements, which we impose on $d^i_k$:
\begin{itemize}
\item $d^i_k$ must be a measure of distance between cells $i$ and $k$, such that it attains a unique value when the cells are in contact (in our case the value is one), and such that a certain range of values correspond to the case when the cells overlap. 
\item $d^i_k$ must be bounded and continuously differentiable.
\end{itemize} 
Depending on the form of $d^i_k$, $D^i_k$ must be modified accordingly.

Let  $\mathbf{X}^i:=(\mathbf{x}^i,\mathbf{x}^i_n,\theta^i,\mathbf{\Phi}^i,\mathbf{d}^i)\in\mathbb{R}^2\times\Omega_{cell}\times[0,2\pi)\times D^i_\nu\times[0,2\pi)^N:=E^i_\nu$. For convenience of notation, we define $\boldsymbol{\alpha}^i_{\mathbf{A}}(\nu):=\mathbf{A}^i$, where $\mathbf{A}\in\lbrace\boldsymbol{\mu},\mathbf{Y},\mathbf{C}\rbrace$. We also extend the definition of $u_j$ in \eqref{eq: simple u_j}: 
\begin{align*}
u_j(\mathbf{X}^i,\boldsymbol{\alpha}^i_{\mathbf{C}}(\nu)) := 
\begin{cases*}
1, \phantom{abc}\Phi^i_k-\frac{\pi}{2}\leq\theta^i+(j-1)\frac{2\pi}{M}\leq\Phi^i_k+\frac{\pi}{2}\text{ and }C^i_k=1, \\
0, \phantom{abc}\text{else},
\end{cases*}
\end{align*}
for some $k=1,\ldots,N$ and where $j=1,\ldots,M$. Then, we have:
\begin{align*}
T_j&\rightarrow T_j(1+\delta_{myo}u_j(\mathbf{X}^i,\boldsymbol{\alpha}^i_{\mathbf{C}}(\nu)))\\
\mathbf{F}_j&\rightarrow \mathbf{F}_j (\mathbf{X}^i,\boldsymbol{\alpha}^i_{\mathbf{C}}(\nu))\\
\mathbf{F}&\rightarrow\mathbf{F}(\boldsymbol{\alpha}^i_{\mathbf{Y}}(\nu),\mathbf{X}^i,\boldsymbol{\alpha}^i_{\mathbf{C}}(\nu)).
\end{align*}
Let $E_{\nu}:=\prod_{i=1}^{N}E^i_{\nu}$ and define $\mathbf{H}^i_\nu:E_\nu\rightarrow\mathbb{R}^{5+2N}$ as:
\begin{align}\label{eq: appendix cell ODE system}
\frac{d}{dt}\mathbf{X}^i = 
\begin{pmatrix*}
\boldsymbol{\alpha}^i_{\boldsymbol{\mu}}(\nu)\beta_{ECM}^{-1}\mathbf{F}(\boldsymbol{\alpha}^i_{\mathbf{Y}}(\nu),\mathbf{X}^i,\boldsymbol{\alpha}^i_{\mathbf{C}}(\nu))\cdot\hat{\mathbf{r}}(\mathbf{x}^i_n)\hat{\mathbf{r}}(\mathbf{x}^i_n)\\
\beta_{cell}^{-1}\mathbf{F}(\boldsymbol{\alpha}^i_{\mathbf{Y}}(\nu),\mathbf{X}^i,\boldsymbol{\alpha}^i_{\mathbf{C}}(\nu))\\
\boldsymbol{\alpha}^i_{\boldsymbol{\mu}}(\nu)\beta_{rot}^{-1}\lVert\mathbf{x}^i_n\rVert\mathbf{F}(\boldsymbol{\alpha}^i_{\mathbf{Y}}(\nu),\mathbf{X}^i,\boldsymbol{\alpha}^i_{\mathbf{C}}(\nu))\cdot\hat{\boldsymbol{\varphi}}(\mathbf{x}_n)\\
\mathbf{0}\\
-(\mathbf{x}^i-\mathbf{x}^1)\cdot(\dot{\mathbf{x}}^i-\dot{\mathbf{x}}^1)d^i_1\\
\vdots\\
-(\mathbf{x}^i-\mathbf{x}^N)\cdot(\dot{\mathbf{x}}^i-\dot{\mathbf{x}}^N)d^i_N
\end{pmatrix*}:=
\mathbf{H}^i_\nu(\mathbf{X}).
\end{align}
This is simply an ODE system that governs the evolution of $\mathbf{X}^i$ between events. The equations governing $\mathbf{x}^i,\mathbf{x}^i_n$, and $\theta^i$ were presented in Sections \ref{section: single cell PDMP}-\ref{section: CIL model}. 
Note that $\boldsymbol{\Phi}^i$ changes only when collisions occur, and is constant at all other times. For a collection of $N$ cells, we then have:
\begin{align}\label{eq: appendix N cell ODE system}
\frac{d}{dt}\mathbf{X}_t&=\mathbf{H}_\nu(\mathbf{X}_t)\nonumber\\
\mathbf{X}_0&=\mathbf{Z}\in E_\nu
\end{align}
where $\mathbf{H}_\nu:E_\nu\rightarrow\mathbb{R}^{5N+2N^2}$ and $\mathbf{H}_\nu:=(\mathbf{H}^1_\nu,\ldots,\mathbf{H}^N_\nu)$. One can also show that there exists a unique solution to \eqref{eq: appendix N cell ODE system}, by using the result for a single cell model in $\cite{2018arXiv181011435U}$. 

Let $\phi_\nu:\mathbb{R}_+\times E_\nu\rightarrow E_\nu$ be the flow corresponding to \eqref{eq: appendix N cell ODE system}. Note that a cell $i$ collides with a cell $k$, if $d^i_k\in\partial D_\nu^{i,k}=\lbrace1\rbrace$ for some $\nu\in A$ such that $C_i^k=0$. Thus, the boundary of $E_\nu$ plays an important role in addressing the collisions. Let $\partial E_\nu$ denote the boundary of $E_\nu$, and define $\partial^*E_\nu$, $\Gamma^*$ as:
\begin{align*}
\partial^*E_\nu &:= \lbrace\mathbf{X}\in\partial E_\nu:\phi_\nu(t,\mathbf{Z})=\mathbf{X},\phantom{a} (t,\mathbf{Z})\in\mathbb{R}_+\times E_\nu\rbrace\\
\Gamma^*&:=\lbrace(\nu,\mathbf{X}):\nu\in A,\phantom{a}\mathbf{X}\in \partial^*E_\nu\rbrace.
\end{align*}
Let $E:=\lbrace(\nu,\mathbf{X}):\nu\in A, \mathbf{X}\in E_\nu\rbrace$ and define $t^*:E\rightarrow \mathbb{R}_+$ as:
\begin{align*}
t^*(\nu,\mathbf{X}) = \inf\lbrace t>0: \phi_\nu(t,\mathbf{X})\in\partial^*E_\nu\rbrace.
\end{align*}
Here, $t^*$ is simply the next collision time, given the state of the system $(\nu,\mathbf{X})\in E$. Let $a_0^N:E\rightarrow\mathbb{R}_+$ be defined as above:
\begin{align*}
a_0^N(\nu,\mathbf{X}) = \sum_{j=1}^{2MN}a_j(\nu,\mathbf{X}),
\end{align*}
where for ease of notation we write $a_0^N(\nu,\mathbf{X})=a_0^N(\boldsymbol{\alpha}(\nu),\mathbf{X})$ and $a_j(\nu,\mathbf{X})=a_j(\boldsymbol{\alpha}(\nu),\mathbf{X})$ for $(\nu,\mathbf{X})\in E$, $j=1,\ldots,NM$. We define 
\begin{align*}
\mathcal{E} = \lbrace B: B = \lbrace (\nu,\mathbf{X}):\nu\in A, \phantom{a} \mathbf{X}\in\widetilde{E}_\nu\rbrace, \phantom{a}\widetilde{E}_\nu \in \mathbf{\mathcal{E}}_\nu \rbrace,
\end{align*}
where $\mathbf{\mathcal{E}}_\nu$ denotes the Borel sets of $E_\nu$. Finally, let $(\Omega,\mathcal{F},\mathbb{P})$ be a probability space and define a transition measure $Q:\mathcal{E}\times E\cup\Gamma^*\rightarrow [0,1]$. 

We now have all the ingredients to specify and construct a piecewise deterministic process of cell motility including collisions: 
\begin{itemize}
\item Vector fields $\left(\mathbf{H}_\nu, \nu\in A\right)$, given by \eqref{eq: appendix cell ODE system}, governing the system's evolution between events and such that there exists a unique global solution to \eqref{eq: appendix N cell ODE system}.
\item An intensity function $a^N_0$, determining the arrival times of FA events, such that $s\mapsto a_0^N(\nu,\phi_\nu(s,\mathbf{X}))$ is integrable for $(\nu,\mathbf{X})\in E$. 
\item A transition measure $Q$ (to be specified below), determining the system's state after an event, such that $(\nu,\mathbf{X})\mapsto Q(B,(\nu,\mathbf{X}))$ is measurable for fixed $B\in\mathcal{E}$, and $Q(\cdot,(\nu,\mathbf{X}))$ is a probability measure for $(\nu,\mathbf{X})\in E$.
\end{itemize}
Suppose the process $(\nu_t,\mathbf{X}_t)$ starts at $(\nu_0,\mathbf{X}_0)\in E$. Let the survival function $S$ be defined by
\begin{align}\label{eq: appendix survival}
S(t) = 
\begin{cases*}
\exp\left(-\int_{0}^{t}a^N_0(\nu_0,\phi_{\nu_0}(s,\mathbf{X}_0))ds\right),\phantom{a} t<t^*(\nu_0,\mathbf{X}_0)\\
0, \phantom{aaaaaaaaaaaaaasdasaaaaAaaa} t\geq t^*(\nu_0,\mathbf{X}_0).
\end{cases*}
\end{align}
Then, $\mathbb{P}(\mathcal{T}_1>t) = S(t)$, where $\mathcal{T}_k$ denotes the $k^{\text{th}}$ event time, and the motion of $(\nu_t,\mathbf{X}_t)$ is given by:
\begin{align*}
(\nu_t,\mathbf{X}_t) = 
\begin{cases*}
(\nu_0,\phi_{\nu_0}(t,\mathbf{X}_0)), \phantom{a}t<T_1\\
(\nu_1,\mathbf{X}_1), \phantom{aAAaaa}t=T_1,
\end{cases*}
\end{align*}
where $(\nu_1,\mathbf{X}_1)$ is distributed according to $Q(\cdot,(\nu_0,\phi_{\nu_0}(\mathcal{T}_1,\mathbf{X}_0)))$. At time $t=\mathcal{T}_1$, the next event time $\mathcal{T}_2$ is determined according to $\mathbb{P}(\mathcal{T}_2-\mathcal{T}_1>t)=S(t)$ and the motion continues as above. Note that after an event, the motion of $\mathbf{X}_t$ continues according to \eqref{eq: appendix N cell ODE system} until either an FA event (in \textit{one} of the $N$ cells) or a collision occurs.

Let $(\nu,\mathbf{X})\in E\cup\Gamma^*$. Then, we have:
\begin{align*}
Q(\lbrace\eta\rbrace\times d\mathbf{X}',(\nu,\mathbf{X})) = \mathbb{P}(\lbrace\eta\rbrace\times d\mathbf{X}'\phantom{a}|\phantom{a}(\nu,\mathbf{X})\in\Gamma^*)+\mathbb{P}(\lbrace\eta\rbrace\times d\mathbf{X}'\phantom{a}|\phantom{a}(\nu,\mathbf{X})\notin\Gamma^*).   
\end{align*}
The first and the second terms on the right are, respectively, transition probabilities given that a collision or an FA event occurred. Using our previous results in \cite{2018arXiv181011435U}, we have:
\begin{align*}
\mathbb{P}(\lbrace\eta\rbrace\times &d\mathbf{X}'\phantom{a}|\phantom{a}(\nu,\mathbf{X})\notin\Gamma^*) \\
&= \delta_{\mathbf{X}}(d\mathbf{X}')\times\sum_{i=1}^{N}\left[\sum_{j=1}^{M}\delta_{\bm{\alpha}^i_{\boldsymbol{\mu}}(\eta),0}\frac{a^{+,i}_j(\nu,\mathbf{X})}{a^N_0(\nu,\mathbf{X})}\delta_{\bm{\alpha}^i_{\mathbf{Y}}(\eta)_j,1}\prod_{j'\neq j}^{M}\delta_{\bm{\alpha}^i_{\mathbf{Y}}(\eta)_{j'},\bm{\alpha}^i_{\mathbf{Y}}(\nu)_{j'}}\nonumber\right.\\
&\phantom{abAAAAAAAAAaa}\left.+\delta_{\bm{\alpha}^i_{\boldsymbol{\mu}}(\eta),1}\frac{a^{-,i}_j(\nu,\mathbf{X})}{a^N_0(\nu,\mathbf{X})}\delta_{\bm{\alpha}^i_{\mathbf{Y}}(\eta)_j,0}\prod_{k\neq j}^{M}\delta_{\bm{\alpha}^i_{\mathbf{Y}}(\eta)_k,\bm{\alpha}^i_{\mathbf{Y}}(\nu)_k}\vphantom{\sum_{j=1}^{M}}\right]\\
&\phantom{abAAAAAAAa}\times\prod_{k\neq i}^{N}\delta_{\bm{\alpha_\mu}^k(\eta),\bm{\alpha_\mu}^k(\nu)}\prod_{j'=1}^{M}\delta_{\bm{\alpha}^k_{\mathbf{Y}}(\eta)_{j'},\bm{\alpha}^k_{\mathbf{Y}}(\nu)_{j'}}\prod_{l=1}^{N}\delta_{\bm{\alpha_C}^k(\eta)_l,\bm{\alpha_C}^k(\nu)_l}\\
&\phantom{abAAAAAAAa}\times\prod_{k\neq i}^{N} \left[\delta_{\bm{\alpha_C}^i(\eta)_k,1}\mathbf{1}_{\mathbb{R}^+\text{\textbackslash} \lbrace 0\rbrace}(d^i_k-1)+\delta_{\bm{\alpha_C}^i(\eta)_k,0}\mathbf{1}_{\mathbb{R}^-\cup \lbrace 0\rbrace}(d^i_k-1)\right].
\end{align*}
The first line indicates that components of $\mathbf{X}$ do not jump at an FA event time. The next two lines reflect the fact that an FA event changes the motility state and the state of one adhesion site. The fourth line corresponds to the fact that an FA event in a cell does not affect other cells. The last line indicates that the collision state of a cell is determined according to cell-cell distances at the time of an FA event. 

Define the following for $(\nu,\mathbf{X})$:
\begin{align*}
B_{(\nu,\mathbf{X})} &:= \lbrace(m,l)\in\lbrace 1,\ldots,N\rbrace^2: d^m_l=d^l_m=1, \bm{\alpha}^m_{\mathbf{C}}(\nu)_l=\bm{\alpha}^l_{\mathbf{C}}(\nu)_m=0\rbrace\\
B^c_{(\nu,\mathbf{X})} &:= \lbrace 1,\ldots,N\rbrace^2\text{\textbackslash}B_{(\nu,\mathbf{X})},
\end{align*}     
i.e. tuples of cell indices that have collided, and the remaining pairs, respectively.
Let $\mathbf{b}:\Gamma^*\rightarrow\mathbb{R}^{5N+2N^2}$ and $\hat{\bm{\Phi}}:E\rightarrow[0,2\pi)^N$ be given by:
\begin{align*}
b_i(\nu,\mathbf{X}) &:= (\mathbf{x}^i,\mathbf{x}_n^i,\theta^i,\hat{\bm{\Phi}}^i(\nu,\mathbf{X}),\mathbf{d}^i)\\
\hat{\Phi}_k^i(\nu,\mathbf{X}) &:= 
\begin{cases*}
\Phi^i_k,\phantom{AAAAa}\text{if $(i,k)\in B^c(\nu,\mathbf{X})$}\\
\hat{\varphi}(\mathbf{x}^i,\mathbf{x}^k),\phantom{A} \text{else},
\end{cases*}
\end{align*}
where $\hat{\varphi}(\mathbf{x}^i,\mathbf{x}^k)$ is the polar angle at which a contact between cells $i$ and $k$ occurred. Then, we have:
\begin{align*}
\mathbb{P}(\lbrace\eta\rbrace\times d\mathbf{X}'|(\nu,\mathbf{X})\in\Gamma^*)&=\delta_{\mathbf{b}(\nu,\mathbf{X})}(d\mathbf{X}')\prod_{(m,l)\in B(\nu,\mathbf{X})}\delta_{\bm{\alpha_C}^m(\eta)_l,1}\delta_{\bm{\alpha_\mu}^m(\eta),0}\\
&\phantom{a}\times\prod_{k=1}^{NM}\delta_{\bm{\alpha_Y}(\eta)_l,\bm{\alpha_Y}(\nu)_l}\prod_{(m,l)\in B^c_{(\nu,\mathbf{X})}}\delta_{\bm{\alpha_C}^m(\eta)_l,\bm{\alpha_C}^m(\nu)_l}\delta_{\bm{\alpha_\mu}^m(\eta),\bm{\alpha_\mu}^m(\nu)}.
\end{align*}
The first line on the right reflects that at the time of collision, the contact angles, collision, and motility states jump to new values. The second line indicates that the FA, collision, and motility states of other cells are unaffected.

If the process hits the boundary (i.e. there is a collision), the post jump location is necessarily in $E$. That is, $\mathbb{P}(\lbrace\eta\rbrace\times d\mathbf{X}'\phantom{a}|\phantom{a}(\nu,\mathbf{X})\in\Gamma^*)=0$ if $\lbrace\eta\rbrace\times d\mathbf{X}'\not\subset E$. This implies that expected number of events in a finite time is finite, and $\mathcal{T}_k\rightarrow\infty$ almost surely (see Chapter 2 in \cite{davis93}).
\subsubsection{Homotypic and heterotypic CIL}\label{appendix: homo and hetero CIL}
In order to take into account mixed populations with different CIL response, we only need to slightly modify the definition of $D^i_k$ in \eqref{eq: definition Dik}. Let $I_1\subset\lbrace1,\ldots,N\rbrace$, $I_2\subset\lbrace1,\ldots,N\rbrace$, be index sets of cells with and without CIL, respectively, such that $I_1\cap I_2=\emptyset$. Then:
\begin{align*}
D^{i,k}_\nu&:=
\begin{cases*}
(-\infty,1],\phantom{aaBb} \text{if $C^i_k=0$, $i,k\in I_1$ or $i,k\in I_2$ }  \\
(-\infty,\infty), \phantom{abA} \text{else},
\end{cases*}\\
D^{i}_\nu &:= \prod_{k=1}^{N}D^{i,k}_\nu,\phantom{D} D_\nu := \prod_{i=1}^ND^i_\nu.
\end{align*}
Thus, only members of the same group undergo CIL. Here, in the absence of heterotypic CIL we effectively rule out collisions between members of different groups. 

%\subsubsection{Heterogeneous populations}
%In order to take into account the case when only some cells exhibit CIL, we need to slightly modify the definition of $D^i_k$ in \eqref{eq: definition Dik}. Let $I\in\lbrace1,\ldots,N$ be an index set of cells undergoing CIL. Then:
%\begin{align}\label{eq: modified definition Dik}
%D^{i,k}_\nu&:=
%\begin{cases*}
%(-\infty,1],\phantom{aaBb} \text{if $C^i_k=0$} \\
%(-\infty,\infty), \phantom{ab} \text{else}
%\end{cases*},\\
%D^{i}_\nu &:= \prod_{k=1}^{N}D^{i,k}_\nu,\phantom{D} D_\nu := \prod_{i=1}^ND^i_\nu.
%\end{align}
 
\subsection{Simulation method} 
To simulate the constructed process we employ Algorithm \ref{algorithm: method} presented below. 
\begin{algorithm}[h]
\begin{enumerate}
\item Set $(\nu_0,\mathbf{X}_0)\in E$ and $t=\mathcal{T}_0=0$, $k=0$.

\item Generate interarrival time $\hat{\Delta}_k$ using Algorithm 2 in \cite{2018arXiv181011435U} applied to the ODE system \eqref{eq: appendix N cell ODE system} and the survival function \eqref{eq: appendix survival}.
\item Find $\mathbf{X}_{\mathcal{T}_k+\hat{\Delta}_k}=\phi_{\nu_{\mathcal{T}_k}}(\hat{{\Delta}}_k,\mathbf{X}_{\mathcal{T}_k})$ and $\hat{B}_{\left(\nu_{\mathcal{T}_k},\mathbf{X}_{\mathcal{T}_k+\hat{\Delta}_k}\right)}$. Set $\Delta_k=\hat{\Delta}_k$.
\item If $\hat{B}_{\left(\nu_{\mathcal{T}_k},\mathbf{X}_{\mathcal{T}_k+\hat{\Delta}_k}\right)}\neq\emptyset$ (Collision)\\
$\phantom{AB}\Delta_k=\min\left\{ s>0: d^m_l(\mathcal{T}_k+s)=d^l_m(\mathcal{T}_k+s)=1, (m,l)\in\hat{B}_{\left(\nu_{\mathcal{T}_k},\mathbf{X}_{\mathcal{T}_k+\hat{\Delta}_k}\right)}\right\}$\\
\item Set $\mathcal{T}_{k+1} := \mathcal{T}_k+\Delta_k$\\
$\phantom{AB}(\nu_{\mathcal{T}_{k+1}},\mathbf{X}_{\mathcal{T}_{k+1}})\sim Q(\cdot,(\nu_{\mathcal{T}_{k}},\mathbf{X}_{\mathcal{T}_k+\Delta_k}))$\\
$\phantom{AB}k:=k+1$
%$\phantom{AB}\mathcal{T}_{k+1} := \mathcal{T}_{k}+\hat{\Delta}_k$.\\
%$\phantom{AB}$Set $(\nu_{\mathcal{T}_{k+1}},\mathbf{X}_{\mathcal{T}_{k+1}})\sim Q(\cdot,(\nu_{\mathcal{T}_{k}},\mathbf{X}_{\mathcal{T}_k+\hat{\Delta}_k}))$.\\
%Else\\
%$\phantom{AB}\Delta_k:=\min\left\{ s>0: d^m_l(\mathcal{T}_k+s)=d^l_m(\mathcal{T}_k+s)=1, (m,l)\in\hat{B}_{\left(\nu_k,\mathbf{X}_{\mathcal{T}_k+\hat{\Delta}_k}\right)}\vphantom{\prod_{i=1}^{n}}\right\}$\\
%$\phantom{AB}\mathcal{T}_{k+1} := \mathcal{T}_{k}+\Delta_k$.\\
%$\phantom{AB}$Set $(\nu_{\mathcal{T}_{k+1}},\mathbf{X}_{\mathcal{T}_{k+1}})\sim Q(\cdot,(\nu_{\mathcal{T}_{k}},\mathbf{X}_{\mathcal{T}_k+\Delta_k}))$.
\end{enumerate}
\caption{Simulation of the PDMP}
\label{algorithm: method}
\end{algorithm}

Here, we use our previously developed method in \cite{2018arXiv181011435U} to simulate a general piecewise deterministic process. However, we now need to take into account collisions as well. To do so, we define
\begin{align}\label{eq: appendix set B}
\hat{B}_{\left(\nu,\mathbf{X}\right)}:= \lbrace(m,l)\in\lbrace 1,\ldots,N\rbrace^2: d^m_l=d^l_m\geq1, \phantom{a}\bm{\alpha}^m_{\mathbf{C}}(\nu)_l=\bm{\alpha}^l_{\mathbf{C}}(\nu)_m=0\rbrace.
\end{align}
Note that if $(m,l)\notin \hat{B}_{\left(\nu_t,\mathbf{X}_t\right)}$ and $(m,l)\in \hat{B}_{\left(\nu_t,\mathbf{X}_{t+s}\right)}$, then this implies that a collision between cells $m$ and $l$ occurred in the time interval $[t,t+s]$. 

After initialization in Step 1 of the algorithm below, we find the interarrival time $\hat{\Delta}_k$ of the next FA event in Step 2 using the method described in \cite{2018arXiv181011435U}. Then, in Step 3 we evolve the ODE system \eqref{eq: appendix N cell ODE system} and identify the cells, which collided in this time period. For each colliding pair, we find their collision time $s$, and their minimum in Step 4. The collision time $s\in(0,\hat{\Delta}_k]$ for $(m,l)\in\hat{B}_{\left(\nu_{\mathcal{T}_k},\mathbf{X}_{\mathcal{T}_k+\hat{\Delta}_k}\right)}$ is the root of
\begin{align}\label{eq: f(s)}
f(s)=f(\mathbf{X}_{\mathcal{T}_k+s})=d^m_l(\mathcal{T}_k+s)-1=0.
\end{align}
Note that after Step 3, the solution $\mathbf{X}_t$ of the ODE system \eqref{eq: appendix N cell ODE system} is available at the time points $t=\mathcal{T}_k+s_i$, where $i=0,\ldots,n$ and $s_n=\hat{\Delta}_k$. Thus, 
\begin{align*}
f(s)= f(\mathbf{X}_{\mathcal{T}_k+s}) = f(\phi_{\nu_{\mathcal{T}_k}}(s-s_i,\mathbf{X}_{\mathcal{T}_k+s_i})).
\end{align*}
Therefore, evaluation of \eqref{eq: f(s)} needed for a root finding method amounts to advancing the ODE system for a single time step of size $s-s_i$. This way, the amount of extra computations needed to find the collision time is minimized, which yields increasing computational savings as the number of cells $N$ increases. Finally, in Step 5 we set the time of the next event $\mathcal{T}_{k+1}$ and update the system according to the event occurred. 

This method can be used to efficiently simulate an arbitrary PDMP, where solving an ODE system is expensive and the boundary hitting time is finite. 
%\subsection{A measure of directionality}\label{appendix: measure of directionality}
%Let $V=\lbrace v\in\mathbb{R}:v=x_1(t_i)-x_1(t_{i-1})\rbrace$ and $V^{+}=\lbrace v>0:v\in V\rbrace$, $V^{-}=\lbrace v<0:v\in V\rbrace$, where $x_1$ is the $x-component$ of $\mathbf   
%
%To measure directionality of cell movement along $x$-axis, 

\end{appendices}

%\clearpage
%\newpage
\providecommand{\noopsort}[1]{}

\bibliographystyle{abbrv}
\end{document}